\documentclass[journal,comsoc]{IEEEtran}%
\makeatletter\if@twocolumn\PassOptionsToPackage{switch}{lineno}\else\fi\makeatother

      \makeatletter
\usepackage{wrapfig}
\usepackage{multirow}
\usepackage[table,xcdraw]{xcolor}
\usepackage{cite}
\usepackage{hyperref}
\usepackage{float}

\newcounter{aubio}

\long\def\bioItem{%
\@ifnextchar[{\@bioItem}{\@@bioItem}}

\long\def\@bioItem[#1]#2#3{
 \stepcounter{aubio}
 \expandafter\gdef\csname authorImage\theaubio\endcsname{#1}
 \expandafter\gdef\csname authorName\theaubio\endcsname{#2}
 \expandafter\gdef\csname authorDetails\theaubio\endcsname{#3}
}

\long\def\@@bioItem#1#2{
 \stepcounter{aubio}
 \expandafter\gdef\csname authorName\theaubio\endcsname{#1}
 \expandafter\gdef\csname authorDetails\theaubio\endcsname{#2}
}

\newcommand{\checkheight}[1]{%
  \par \penalty-100\begingroup%
  \setbox8=\hbox{#1}%
  \setlength{\dimen@}{\ht8}%
  \dimen@ii\pagegoal \advance\dimen@ii-\pagetotal
  \ifdim \dimen@>\dimen@ii
    \break
  \fi\endgroup}

\def\printBio{%
  \@tempcnta=0
   \loop
     \advance \@tempcnta by 1
     \def\aubioCnt{\the\@tempcnta}
     \setlength{\intextsep}{0pt}%
     \setlength{\columnsep}{10pt}%
     \expandafter\ifx\csname authorImage\aubioCnt\endcsname\relax%
      \else%
       \checkheight{\includegraphics[height=1.25in,width=1in,keepaspectratio]{\csname authorImage\aubioCnt\endcsname}}
        \begin{wrapfigure}{l}{25mm}
         \includegraphics[height=1.25in,width=1in,keepaspectratio]{\csname authorImage\aubioCnt\endcsname}
        \end{wrapfigure}\par
      \fi
     \noindent\textbf{\csname authorName\aubioCnt\endcsname}\csname authorDetails\aubioCnt\endcsname \par\bigskip
      \ifnum\@tempcnta < \theaubio
   \repeat
   }
\makeatother

\usepackage{bm}
\usepackage{graphicx}
\usepackage[T1]{fontenc}
\usepackage{mathtools}

\usepackage{url,multirow,morefloats,floatflt,cancel,tfrupee}
\makeatletter

\AtBeginDocument{\@ifpackageloaded{textcomp}{}{\usepackage{textcomp}}}
\makeatother
\usepackage{colortbl}
\usepackage{xcolor}
\usepackage{pifont}
\usepackage[nointegrals]{wasysym}
\makeatletter

\def\mcWidth#1{\csname TY@F#1\endcsname+\tabcolsep}

\def\cAlignHack{\rightskip\@flushglue\leftskip\@flushglue\parindent\z@\parfillskip\z@skip}
\def\rAlignHack{\rightskip\z@skip\leftskip\@flushglue \parindent\z@\parfillskip\z@skip}
\usepackage{array}

\usepackage{ifxetex}
\ifxetex\else\if@twocolumn\usepackage{dblfloatfix}\fi\fi

\AtBeginDocument{
\expandafter\ifx\csname eqalign\endcsname\relax
\def\eqalign#1{\null\vcenter{\def\\{\cr}\openup\jot\m@th
  \ialign{\strut$\displaystyle{##}$\hfil&$\displaystyle{{}##}$\hfil
      \crcr#1\crcr}}\,}
\fi
}

\AtBeginDocument{%
  \@ifpackageloaded{endfloat}%
   {\renewcommand\efloat@iwrite[1]{\immediate\expandafter\protected@write\csname efloat@post#1\endcsname{}}}{\newif\ifefloat@tables}%
}%

\def\BreakURLText#1{\@tfor\brk@tempa:=#1\do{\brk@tempa\hskip0pt}}
\let\lt=<
\let\gt=>
\def\processVert{\ifmmode|\else\textbar\fi}

\@ifundefined{subparagraph}{
\def\subparagraph{\@startsection{paragraph}{5}{2\parindent}{0ex plus 0.1ex minus 0.1ex}%
{0ex}{\normalfont\small\itshape}}%
}{}

\newcommand\role[1]{\unskip}
\newcommand\aucollab[1]{\unskip}

\@ifundefined{tsGraphicsScaleX}{\gdef\tsGraphicsScaleX{1}}{}
\@ifundefined{tsGraphicsScaleY}{\gdef\tsGraphicsScaleY{.9}}{}
\def\checkGraphicsWidth{\ifdim\Gin@nat@width>\linewidth
	\tsGraphicsScaleX\linewidth\else\Gin@nat@width\fi}

\def\checkGraphicsHeight{\ifdim\Gin@nat@height>.9\textheight
	\tsGraphicsScaleY\textheight\else\Gin@nat@height\fi}

\def\fixFloatSize#1{}
\let\ts@includegraphics\includegraphics

\def\inlinegraphic[#1]#2{{\edef\@tempa{#1}\edef\baseline@shift{\ifx\@tempa\@empty0\else#1\fi}\edef\tempZ{\the\numexpr(\numexpr(\baseline@shift*\f@size/100))}\protect\raisebox{\tempZ pt}{\ts@includegraphics{#2}}}}

\AtBeginDocument{\def\includegraphics{\@ifnextchar[{\ts@includegraphics}{\ts@includegraphics[width=\checkGraphicsWidth,height=\checkGraphicsHeight,keepaspectratio]}}}

\DeclareMathAlphabet{\mathpzc}{OT1}{pzc}{m}{it}

\def\URL#1#2{\@ifundefined{href}{#2}{\href{#1}{#2}}}

\def\UrlOrds{\do\*\do\-\do\~\do\'\do\"\do\-}%
\g@addto@macro{\UrlBreaks}{\UrlOrds}

\@ifundefined{quoteAttrib}
	{}
	{}

\@ifundefined{titlequoteAttrib}
	{}{}

\newenvironment{title-quote}
	{\list{}{\fontsize{10pt}{12pt}\selectfont\leftmargin.5in\itshape\rightmargin\leftmargin}%
  \item\relax}
  {\endlist}

\makeatother

\makeatletter
\AtBeginDocument{\@ifpackageloaded{longtable}{%
\def\LT@makecaption#1#2#3{%
  \LT@mcol\LT@cols c{\hbox to\z@{\hss\parbox[t]\LTcapwidth{%
    \sbox\@tempboxa{#1{#2: } #3}%
    \ifdim\wd\@tempboxa>\hsize
      #1{#2: }\textsc{#3}%
    \else
      \hbox to\hsize{\hfil\box\@tempboxa\hfil}%
    \fi
    \endgraf\vskip\baselineskip}%
  \hss}}}
}{}}
\makeatother
\usepackage[utf8]{inputenc}
\usepackage[T1]{fontenc}
\usepackage{subfigure}
\usepackage{tikz}
\usepackage{fancyhdr}

\usetikzlibrary{arrows, calc, decorations.markings, positioning}

\makeatletter
\newenvironment{timeline}[6]{%

    \newcommand{\startyear}{#1}
    \newcommand{\tlendyear}{#2}

    \newcommand{\yearcolumnwidth}{#3}
    \newcommand{\rulecolumnwidth}{#4}
    \newcommand{\entrycolumnwidth}{#5}
    \newcommand{\timelineheight}{#6}

    \newcommand{\templength}{}

    \newcommand{\entrycounter}{0}

    \long\def\ifnodedefined##1##2##3{%
        \@ifundefined{pgf@sh@ns@##1}{##3}{##2}%
    }

    \newcommand{\ifnodeundefined}[2]{%
        \ifnodedefined{##1}{}{##2}
    }

    \newcommand{\drawtimeline}{%
        \draw[timelinerule] (\yearcolumnwidth+5pt, 0pt) -- (\yearcolumnwidth+5pt, -\timelineheight);
        \draw (\yearcolumnwidth+0pt, -10pt) -- (\yearcolumnwidth+10pt, -10pt);
        \draw (\yearcolumnwidth+0pt, -\timelineheight+15pt) -- (\yearcolumnwidth+10pt, -\timelineheight+15pt);

        \pgfmathsetlengthmacro{\templength}{neg(add(multiply(subtract(\startyear, \startyear), divide(subtract(\timelineheight, 25), subtract(\tlendyear, \startyear))), 10))}
        \node[year] (year-\startyear) at (\yearcolumnwidth, \templength) {\startyear};

        \pgfmathsetlengthmacro{\templength}{neg(add(multiply(subtract(\tlendyear, \startyear), divide(subtract(\timelineheight, 25), subtract(\tlendyear, \startyear))), 10))}
        \node[year] (year-\tlendyear) at (\yearcolumnwidth, \templength) {\tlendyear};
    }

    \newcommand{\entry}[2]{%

        \pgfmathtruncatemacro{\lastentrycount}{\entrycounter}
        \pgfmathtruncatemacro{\entrycounter}{\entrycounter + 1}

        \ifdim \lastentrycount pt > 0 pt%
            \node[entry] (entry-\entrycounter) [below of=entry-\lastentrycount] {##2};
        \else%
            \pgfmathsetlengthmacro{\templength}{neg(add(multiply(subtract(\startyear, \startyear), divide(subtract(\timelineheight, 25), subtract(\tlendyear, \startyear))), 10))}
            \node[entry] (entry-\entrycounter) at (\yearcolumnwidth+\rulecolumnwidth+10pt, \templength) {##2};
        \fi

        \ifnodeundefined{year-##1}{%
            \pgfmathsetlengthmacro{\templength}{neg(add(multiply(subtract(##1, \startyear), divide(subtract(\timelineheight, 25), subtract(\tlendyear, \startyear))), 10))}
            \draw (\yearcolumnwidth+2.5pt, \templength) -- (\yearcolumnwidth+7.5pt, \templength);
            \node[year] (year-##1) at (\yearcolumnwidth, \templength) {##1};
        }

        \draw ($(year-##1.east)+(2.5pt, 0pt)$) -- ($(year-##1.east)+(7.5pt, 0pt)$) -- ($(entry-\entrycounter.west)-(5pt,0)$) -- (entry-\entrycounter.west);
    }

    \newcommand{\plainentry}[2]{

        \pgfmathtruncatemacro{\lastentrycount}{\entrycounter}
        \pgfmathtruncatemacro{\entrycounter}{\entrycounter + 1}

        \ifdim \lastentrycount pt > 0 pt%
            \node[entry] (entry-\entrycounter) [below of=entry-\lastentrycount] {##2};
        \else%
            \pgfmathsetlengthmacro{\templength}{neg(add(multiply(subtract(\startyear, \startyear), divide(subtract(\timelineheight, 25), subtract(\tlendyear, \startyear))), 10))}
            \node[entry] (entry-\entrycounter) at (\yearcolumnwidth+\rulecolumnwidth+10pt, \templength) {##2};
        \fi

        \ifnodeundefined{invisible-year-##1}{%
            \pgfmathsetlengthmacro{\templength}{neg(add(multiply(subtract(##1, \startyear), divide(subtract(\timelineheight, 25), subtract(\tlendyear, \startyear))), 10))}
            \draw (\yearcolumnwidth+2.5pt, \templength) -- (\yearcolumnwidth+7.5pt, \templength);
            \node[year] (invisible-year-##1) at (\yearcolumnwidth, \templength) {};
        }

        \draw ($(invisible-year-##1.east)+(2.5pt, 0pt)$) -- ($(invisible-year-##1.east)+(7.5pt, 0pt)$) -- ($(entry-\entrycounter.west)-(5pt,0)$) -- (entry-\entrycounter.west);
    }

    \begin{tikzpicture}
        \tikzstyle{entry} = [%
            align=left,%
            text width=\entrycolumnwidth,%
            node distance=5mm,%
            anchor=west]
        \tikzstyle{year} = [anchor=east]
        \tikzstyle{timelinerule} = [%
            draw,%
            decoration={markings, mark=at position 1 with {\arrow[scale=1.5]{latex'}}},%
            postaction={decorate},%
            shorten >=0.4pt]

        \drawtimeline
}
{
    \end{tikzpicture}
    \let\startyear\@undefined
    \let\tlendyear\@undefined
    \let\yearcolumnwidth\@undefined
    \let\rulecolumnwidth\@undefined
    \let\entrycolumnwidth\@undefined
    \let\timelineheight\@undefined
    \let\entrycounter\@undefined
    \let\ifnodedefined\@undefined
    \let\ifnodeundefined\@undefined
    \let\drawtimeline\@undefined
    \let\entry\@undefined
}
\makeatother

\usepackage{multirow}
\usepackage{setspace}
\usepackage{bm}
\usepackage{mathrsfs}
\usepackage{amsmath}
\usepackage{forest}
\usepackage{supertabular}
\usepackage{siunitx}
\usepackage{graphicx}
\usepackage{ulem}

\begin{document}

\title{The Evolution of Quantum Secure Direct Communication: On the Road to the Qinternet}

\author{Dong Pan, \IEEEmembership{Member, IEEE}, Gui-Lu Long, \IEEEmembership{Member, IEEE}, Liuguo Yin, \IEEEmembership{Senior Member, IEEE,} Yu-Bo Sheng, \\Dong Ruan, Soon Xin Ng, \IEEEmembership{Senior Member, IEEE}, Jianhua Lu, \IEEEmembership{Fellow, IEEE}, Lajos Hanzo, \IEEEmembership{Life Fellow, IEEE}
	\thanks{Dong Pan acknowledges support from the National Natural Science Foundation of China under Grant No. 12205011; Dong Pan and Gui-Lu Long acknowledge support from the Open Research Fund Program of the State Key Laboratory of Low-Dimensional Quantum Physics under Grant No. KF202205; Gui-Lu Long acknowledges support from the National Natural Science Foundation of China under Grant No. 11974205, the Key R\&D Program of Guangdong province under Grant No. 2018B030325002, Beijing Advanced Innovation Center for Future Chip (ICFC), and Tsinghua University Initiative Scientific Research Program. Lajos Hanzo would like to acknowledge the financial support of the Engineering and Physical Sciences Research Council projects EP/W016605/1, EP/X01228X/1 and EP/Y026721/1 as well as of the European  Research Council's Advanced Fellow Grant QuantCom (Grant No. 789028). Liuguo Yin acknowledges support from the National Natural Science Foundation of China under Grant 62025110. Dong Ruan acknowledges support from the National Natural Science Foundation of China under Grant 62131002.} 
  \thanks{Dong Pan and Gui-Lu Long are with the Beijing Academy of Quantum Information Sciences, Beijing 100193, China (e-mail: pandong@baqis.ac.cn; gllong@tsinghua.edu.cn). Dong Ruan and Gui-Lu Long are with the State Key Laboratory of Low-dimensional Quantum Physics and Department of Physics, Tsinghua University, Beijing 100084, China (e-mail: dongruan@tsinghua.edu.cn). Yu-Bo Sheng is with the College of Electronic and Optical Engineering and College of Flexible Electronics (Future
Technology), Institute of Quantum Information and Technology, Nanjing University of Posts and Telecommunications, Nanjing 210003, China (email: shengyb@njupt.edu.cn). Gui-Lu Long, Liuguo Yin, and Jianhua Lu are with the  Beijing National Research Center for Information Science and Technology, Frontier Science Center for Quantum Information, Beijing 100084, China. Liuguo Yin and Jianhua Lu are with the School of Information Science and Technology, Tsinghua University, Beijing 100084, China (e-mail: $\{$yinlg, lhh-dee$\}$@tsinghua.edu.cn). Soon Xin Ng and Lajos Hanzo are with the School of Electronics and Computer Science, University of Southampton, Southampton SO17 1BJ, United Kingdom (e-mail: $\{$sxn, lh$\}$@ecs.soton.ac.uk). This paper was drafted while Dong Pan was a visiting researcher in School of Electronics and Computer Science, University of Southampton, UK.}
\thanks{Corresponding author: Gui-Lu Long and Lajos Hanzo.}
}

\maketitle
\IEEEpeerreviewmaketitle

\date{\today}

\begin{abstract}
Communication security has to evolve to a higher plane in the face of the threat from the massive computing power of the emerging quantum computers. Quantum secure direct communication (QSDC) constitutes a promising branch of quantum communication, which is provably secure and overcomes the threat of quantum computing, whilst conveying secret messages directly via the quantum channel. In this survey, we highlight the motivation and the status of QSDC research with special emphasis on its theoretical basis and experimental verification. We will detail the associated point-to-point communication protocols and show how information is protected and transmitted. Finally, we discuss the open challenges as well as the future trends of QSDC networks, emphasizing again that QSDC is not a pure quantum key distribution (QKD) protocol, but a fully-fledged secure communication scheme.
\end{abstract}


\section*{Abbreviations}
\tablefirsthead{}
\tablehead{}
\tabletail{}
\tablelasttail{}
\begin{supertabular}{p{15mm}p{65mm}}
AAG & Adjustable Air Gap\\
AM & Amplitude Modulator\\
AMZI & Asymmetric Mach-Zehnder Interferometer\\
ASE & Amplified Spontaneous Emission\\
Att, Attn & Attenuator\\
BB84 & Bennett-Brassard 1984\\
BD & Beam Displacer\\
BPSK & Binary Phase Shift Keying\\
BS & Beam Splitter\\
BSM & Bell-State Measurement\\
CIR & Circulator \\
CM & Control Mode\\
CW & Continuous-Wave\\
CWDM & Coarse Wavelength-Division Multiplexer\\
CNOT & Controlled-NOT\\
DCF & Dispersion-Compensating Fiber\\
\end{supertabular}
\begin{supertabular}{p{15mm}p{65mm}}
CV & Continuous Variable\\
DI   & Device-Independent \\
DL & Delay Line\\
DL04 & Deng-Long 2004\\
DM & Dichroic Mirror\\
DQKD & Deterministic Quantum Key Distribution\\
DSF & Dispersion Shifted Fiber\\
DSQC & Deterministic Secure Quantum Communication\\
DWDM & Dense Wavelength Division Multiplexing\\
DV & Discrete Variable \\
EDFA & Erbium-Doped Fiber Amplifier\\
ENT & A utility for evaluating random number sequences \\
EOM & Electro-Optical Modulator\\
EPC & Electronic Polarization Controller\\
EPR & Einstein-Podolsky-Rosen\\
FC & Fiber Coupler\\
FEC & Forward Error Correction\\
FM & Faraday Mirror\\
FPGA & Field Programmable Gate Array\\
FR & Faraday Rotator\\
FSO & Free-Space Optical\\
GHZ & Greenberger-Horne-Zeilinger\\
HD, Hom & Homodyne Detector\\
HWP & Half Wave Plate\\ 
ILP & In-Line Polarizer\\
IM & Intensity Modulator\\
ISO & Isolator\\
ITU & International Telecommunication Union \\
LD & Laser Diode\\
LO & Local Oscillator\\
MDI & Measurement Device Independent\\
MOT & Magneto-Optical Trap\\
MS06 & Marino-Stroud 2006\\
MZI & Mach-Zehnder Interferometer\\
N.A. & Not Available\\
NOPA &Nondegenerate Optical Parametric Amplifier\\
OAM & Orbital Angular Momentum\\
OC & Optical Circulator\\
ODL & Optical Delay Line\\
OPA & Optical Parametric Amplifier\\
PBS & Polarization Beam Splitter\\
PBML08 & Pirandola-Braunstein-Mancini-Lloyd 2008\\
PC & Polarization Controller\\
PM & Phase Modulation\\
PMCIR & Polarization-Maintaining Circulator\\
PMFC & Polarization Maintaining Filter Coupler\\
Pol & Polarizer\\
PPLN & Periodically Poled Lithium Niobate\\
PQC & Post-Quantum Cryptography\\
PQKD & Probabilistic Quantum Key Distribution\\
PS & Phase Shifter\\
PZT & PieZoelecTric\\
QBER & Quantum Bit Error Rate\\
QC & Quantum Communication\\
QDL & Quantum Data Locking\\
QI & Quantum Illumination\\
Qinternet & Quantum Internet\\
QKD & Quantum Key Distribution\\
\end{supertabular}
\begin{supertabular}{p{15mm}p{65mm}}
QKPC & Quantum Keyless Private Communication \\
QLPI & Quantum Low Probability of Intercept \\
QM & Quantum Mechanics\\
QSDC & Quantum Secure Direct Communication\\
QSS & Quantum Secret Sharing\\
QT & Quantum Teleportation\\
RECON & REpeatable Classical ONe-time-pad \\
Rx & Receiver \\
SHA & Secure Hash Algorithm\\
SMZI & Sagnac-Mach-Zehnder Interferometers\\
SNSPD & Superconducting Nanowire Single Photon Detector\\ 
SPD & Single-Photon Detector\\
SPDC & Spontaneous Parametric DownConverter\\
SR & Secure Repeater\\
TCSPC & Time-Correlated Single Photon Counting\\
TEC & ThermoElectric Cooler\\
TFOC & Triplet Fiber-Optic Collimator\\
TMSS & Two-Mode Squeezed State\\
Tx  & Transmitter \\
WDM & Wavelength Division Multiplexing\\
VOA & Variable Optical Attenuator\\
WP  & Wave Plate \\
ZXFZ & Zhu-Xia-Fan-Zhang\\
\end{supertabular}

\section{Introduction}
\label{sec:intro}
The widespread application of communicating computers and mobile devices has changed our daily life beyond recognition. However, retaining the confidentiality of sensitive information is of crucial importance for individuals, enterprises, and governments in our information age. Cryptography is used for maintaining the confidentiality, integrity, and authenticity of information for the authorized users in the face of the malicious activities of third-party adversaries~\cite{menezes2018handbook,lindell2014introduction}. The operational cryptographic systems tend to rely on carrying out mathematical operations, that are hard to `decypher' with the aid of state-of-the-art computers. In other words, practical computer science offers what is termed as computational security, which is practically unbreakable within a relatively short space of time by using practical computational resources. However, this conventional cryptography faces challenges imposed by the evolution of ever more powerful computing hardware.

\subsection{Unveiling Quantum Computing}
\label{sec:Unveiling quantum computing}
\begin{figure}[!h]
\begin{center}
\includegraphics[width=7cm,angle=0]{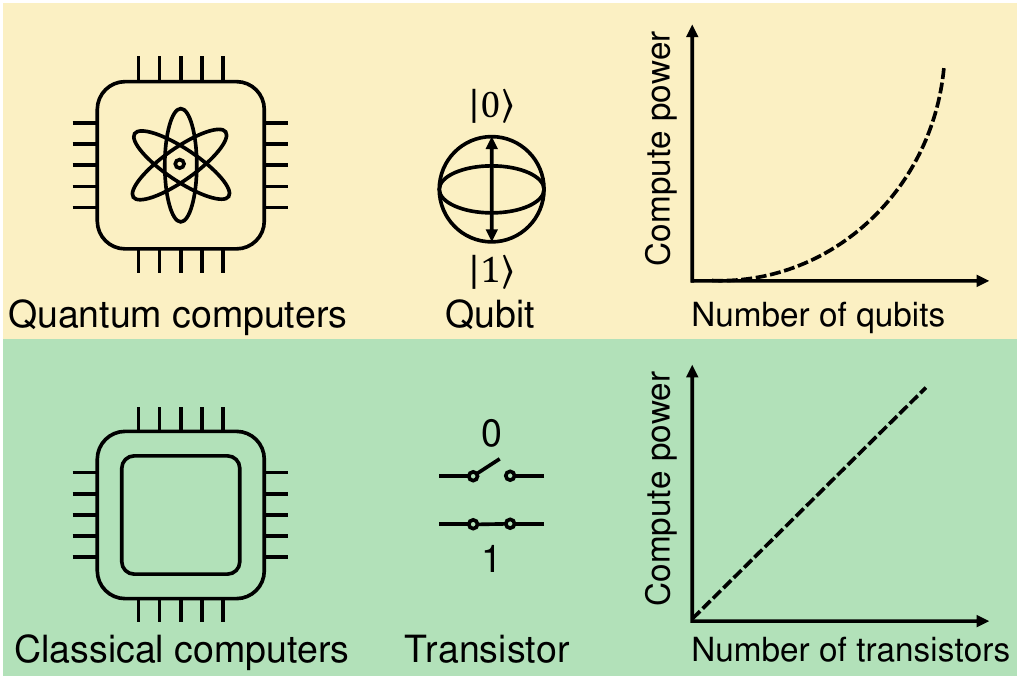}
\caption{Comparison of computational power between quantum computers and classical computers. The symbol `$|\cdot\rangle$' is the notation for quantum states, which will be elaborated on in further detail in Section~\ref{sec:Quantum-states-and-qubits}.}
\label{fig:QCVsCC}
\end{center}
\end{figure}

The concept of quantum computing was introduced in the 1980s~\cite{benioff1980computer,yuri1980manin,feynman1982simulating,deutsch1985quantum}. It is a revolutionary paradigm that harnesses the principles of quantum mechanics to perform computations at speeds unattainable by classical computers. In contrast to classical computers, which tend to rely on Boolean logic and on exploiting that the classical bits can only be in one of two states, namely 0 or 1, the basic variable of quantum computers - namely the qubit - exhibits entirely different properties. It is not confined to the two states, but can exist in their superposition~\cite{nielsen2002quantum,li2001quantum,botsinis2018quantum}. Information encoding, storing and processing may be carried out more efficiently in quantum computers than in its classical counterpart, when relying on superimposed quantum states. Explicitly, massive parallel computations are facilitated by quantum computers, resulting in an exponential increase in computational efficiency for certain tasks. For example, a quantum computer with $N$ qubits can be likened to $2^N$ classical computers running in parallel. Fig.~\ref{fig:QCVsCC} illustrates a brief comparison between quantum computing and classical computing. Note that the concept of qubits will be elaborated upon in Section~\ref{sec:Quantum-states-and-qubits}.

The computational power of a quantum computer increases exponentially with the number of qubits, while the computational power of a classical computer increases in a 1:1 linear relationship with the number of transistors. Currently, quantum computers demonstrate acceleration advantages for specific problems such as cryptography cracking, quantum simulation, and quantum optimization~\cite{bova2021commercial}. By contrast, classical computers excel in handling routine computational tasks.

\subsection{Cryptosystems in the Quantum World}
\label{sec:Cryptosystems-in-the-quantum world}
Asymmetric cryptography\footnote{Asymmetric cryptography, also known as public-key cryptography, relies on a pair of keys - a public key and a private key - to encrypt and decrypt confidential plaintext information, respectively. If data is encrypted with the public key, it can only be decrypted using the corresponding private key. The public key is made public and can be used by anyone, while the private key must be kept confidential. This ensures that plaintext messages remain secure from unauthorized third parties during transmission, since it is computationally infeasible to compute private key from public key by using practical computational resources~\cite{menezes2018handbook}. By contrast, symmetric encryption typically utilizes a single pair of identical keys for both encryption and decryption. It can also involve the use of two keys that can be easily deduced from each other. The keys involved are kept secret~\cite{menezes2018handbook}.} primarily relies on excessive complexity calculations. For example, the well-known Rivest-Shamir-Adleman cryptosystems~\cite{rivest1978method} are designed based on the problem of factoring large numbers into their prime factors. However, at the time of writing, the security of asymmetric cryptography is facing increasing threats from quantum computers, which are capable of factoring large numbers. Shor's algorithm~\cite{shor1994algorithms} shows that a quantum computer is capable of efficient factorization of large prime numbers and of solving elliptic curve based problems. Similarly, quantum annealing computers are equally powerful \cite{peng2019factoring}. The noisy intermediate-scale quantum computers may challenge the Rivest-Shamir-Adleman cryptosystems by using a hybrid quantum-classical algorithm~\cite{yan2022factoring}. Thus, the above-mentioned commonly used cryptographic algorithms are no longer secure in the quantum era \cite{cheng2017securing}.

Another well-known algorithm running on a quantum computer is Grover's search algorithm~\cite{grover1996fast,grover1997quantum,grover1998quantum,long1999phase,long2001grover}, which is capable of finding a known entry in unsorted databases~\cite{botsinis2013quantum,botsinis2015iterative}. Grover's algorithm is eminently suitable for analyzing symmetric encryption systems, such as the Data Encryption Standard and the Advanced Encryption Standard~\cite{lindell2014introduction,wang2022variational}. These cryptographic protocols are generally analyzed using `brute force' search in the space of all legitimate keys. Specifically, the 56-bit keys of the Data Encryption Standard~\cite{lindell2014introduction} may be cracked by Grover's algorithm, which will use on the order of hundred million search steps, which is much lower than that of the number of operations to be carried out by a classical computer \cite{brassard1997searching}. Furthermore, the hash function SHA-2 (Secure Hash Algorithm) and SHA-3 \cite{lindell2014introduction} are facing the same threats in symmetric encryption systems, since their security relies on the difficulty of finding two different messages that map to the same fixed length. However, Grover's algorithm provides an improvement to this problem~\cite{brassard1998quantum}, although an exponential speed-up in the database search problem is infeasible even for Grover's optimal algorithm~\cite{bennett1997strengths,long2001grover,toyama2013quantum,castagnoli2016highlighting}. Cryptologists believe that the Advanced Encryption Standard, SHA-2, and SHA-3 are relatively secure even in a quantum world, but defense mechanisms should nonetheless be conceived to guard against quantum search attacks \cite{buchanan2017will}. It is believed that doubling the key size of a symmetric encryption system or that of the output length of a hash function is urgently needed.

Large-scale quantum computers are expected to have far-reaching influence on the existing cryptographic solutions as reported by the National Institute of Standards and Technology \cite{chen2016report}. Table \ref{table:ImpactofQC} summarizes these impacts, indicating that the progress of quantum computation tends to threaten the security of modern cryptosystems. Although it has been argued that public-key cryptosystems will only be broken when practical quantum computers have been built for handing thousands quantum bits and can perform thousands of quantum gate operations~\cite{proos2003shor,roetteler2017quantum,gidney2019factor}, it is time for the research community to conceive new cryptosystems, which remain secure in the quantum era.

Ensuring the privacy and security of our communication in the quantum era is the major task of cryptography. There are two different alternative candidate families, namely post-quantum cryptography and quantum cryptography or quantum communication. Post-quantum cryptography is also based on solving challenging mathematical problems, but they rely on other problems than the factoring of large numbers. Instead, they rely on hash-based cryptography~\cite{bernstein2009introduction}, on multivariate-quadratic-equations cryptography~\cite{bernstein2009introduction}, on lattice-based cryptography \cite{micciancio2009lattice}, and on code-based cryptography \cite{sendrier2017code}. Quantum cryptography or quantum communication is another secure-communication solution that exploits the properties of quantum mechanics itself, which additionally has the capability of detecting eavesdropping.

\begin{table}[]
\begin{footnotesize}
\begin{center}
\caption{Impact of large-scale quantum computers on current cryptographic algorithms.}
\begin{tabular}{|m{1.9cm}|m{1.8cm}|m{1.8cm}|m{1.5cm}|}
\hline
Cryptosystems     & Purposes                                     & Threats        & Security                                     \\ \hline

Rivest-Shamir-Adleman               & Key establishment and signature             & Shor's algorithm   & Insecure                                   \\ \cline{1-2} \cline{4-4}
Elliptic Curve Digital Signature Algorithm, Elliptic Curve Diffie-Hellman (Elliptic-Curve Cryptography) & Key exchange and signature &                           & Insecure                                   \\ \cline{1-2} \cline{4-4}
Digital Signature Algorithm (Finite-Field Cryptography) & Key exchange and signature            &                             & Insecure                                   \\ \cline{1-2} \cline{4-4}
Diffie-Hellman                              & Key exchange                                &                                     & Insecure                                   \\ \hline
Advanced Encryption Standard                             & Symmetric encryption                        & Grover's algorithm & Relatively secure, but large keys are needed \\
                                       \cline{1-2} \cline{4-4}
SHA-2, SHA-3                     & Hash functions                              &                                     & Relatively secure, but larger output needed    \\ \hline
\end{tabular}
\label{table:ImpactofQC}
\end{center}
\end{footnotesize}
\end{table}

\subsection{From Quantum Key Distribution to Quantum Secure Direct Communication}
\label{sec:patterns-for-commu}
The roots of quantum cryptography or quantum communication can be traced back to the idea of Wiesner in the late 1960s, who proposed the concept of unforgeable quantum money by relying on quantum physics. Similarly to many other radical concepts, he had difficulty in publishing his paper, but finally his much-delayed paper was published in 1983~\cite{wiesner1983conjugate}, where he described how information may be stored and conveyed with the aid of polarized photons. In 1984~\cite{bennett1984quantum}, enlightened by Wiesner's idea, Bennett and Brassard discovered that a pair of communicating parties can generate a cryptographic key over an insecure channel by using appropriately polarized single photons~\cite{bennett1992experimental}. It is what we know today as the Bennett-Brassard 1984 (BB84) quantum key distribution (QKD) protocol, marking the beginning of quantum cryptography. The security of quantum-domain cryptosystems is based on the laws of quantum mechanics rather than on conceiving mathematically challenging problems, which enables the legitimate communicating parties to have unconditionally secure links. As a benefit, communication systems become secure even in the presence of an eavesdropper who has unlimited computational power, which is an explicit benefit of exploiting the laws of physics.

\begin{figure}[!h]
\begin{center}
\includegraphics[width=8cm,angle=0]{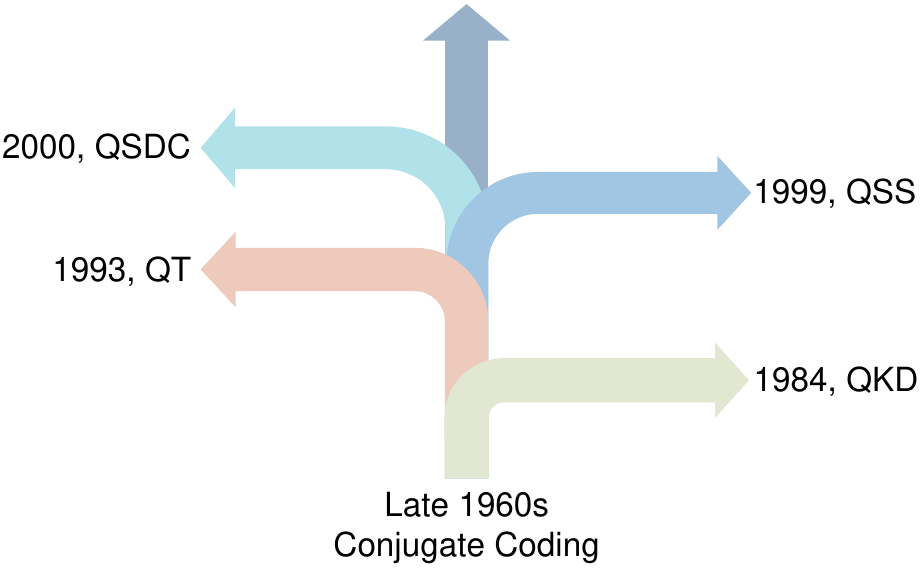}
\caption{The main branches of quantum communication. QKD, quantum key distribution; QT, quantum teleportation; QSS, quantum secret sharing; QSDC, quantum secure direct communication.}
\label{fig:QCbranches}
\end{center}
\end{figure}

Hence numerous quantum cryptographic or quantum communication protocols have been proposed, which can be classified into four main branches: QKD~\cite{bennett1984quantum}, quantum teleportation (QT)~\cite{bennett1993teleporting}, quantum secret sharing (QSS)~\cite{hillery1999quantum}, and quantum secure direct communication (QSDC) \cite{long2000theoretical,long2002theoretically}, as shown in Fig.~\ref{fig:QCbranches}. Bennett \textit{et al}. \cite{bennett1993teleporting} introduced quantum teleportation in 1993, showing how to send an unknown quantum state to a remote receiver, with the assistance of classical communication and pre-shared entangled photons. Hillery \textit{et al}. \cite{hillery1999quantum} proposed quantum secret sharing in 1999, which is a scheme using entangled quantum states for sharing a random bit among several parties so that no subset of them is able to reconstruct a shared ramdom bit - all of them have to work together~\cite{xiao2004efficient}. As a further development, in 2000, a QSDC protocol was proposed by Long and Liu \cite{long2000theoretical,long2002theoretically} for transmiting a predetermined information. QSDC is a beneficial secure communication technique, where secret information can be transmitted directly through the quantum channel without a pre-distributed cryptographic key.

\begin{figure}[!h]
\begin{center}
\includegraphics[width=\columnwidth,angle=0]{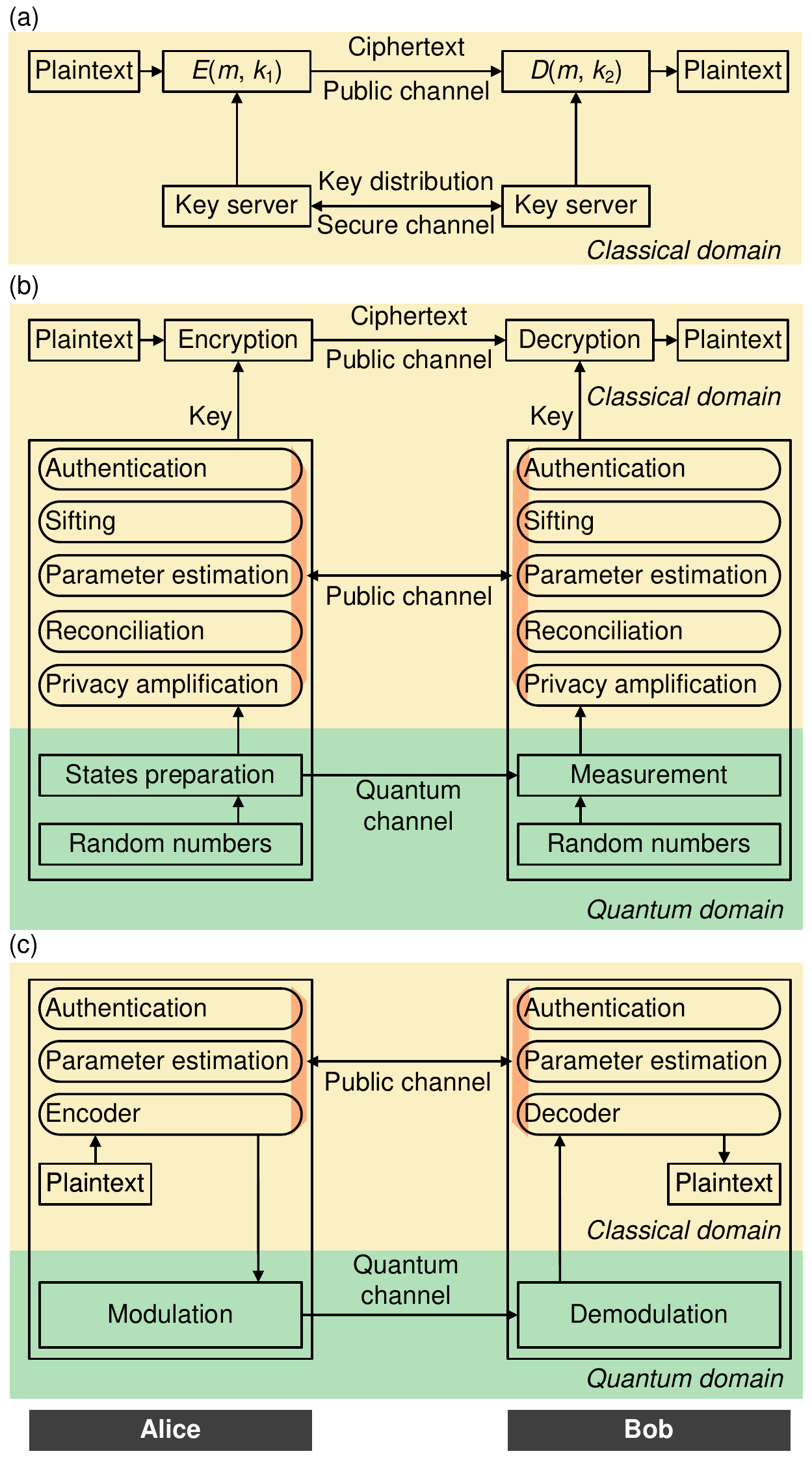}
\caption{Different models of communication systems. \textbf{(a) classical cryptosystem.} \textbf{(b) QKD system.} Key agreement is carried out by QKD, and information is transmitted via classical communication, the same as in (a).  \textbf{(c) QSDC system.} No key distribution, no key management and no ciphertext are required in QSDC.}
\label{fig:Commsys}
\end{center}
\end{figure}

Fig.~\ref{fig:Commsys} portrays the different models of secure communications. Fig.~\ref{fig:Commsys} (a) highlights the secure communication structure commonly used at the time of writing. It relies on a pair of channels: the ciphertext channel and the key distribution channel. The transmitter of Alice first transforms a plaintext $m$ into a ciphertext using a secret key $k_1$ and an encryption algorithm $E(m,~k_1)$, and she sends the ciphertext to the receiver - namely to Bob - through the ciphertext channel. Bob uses the secret key $k_2$ and the decryption algorithm $D(m,~k_2)$ to recover the ciphertext for accessing the plaintext $m$ upon receipt \cite{shannon1949communication}. Depending on the keys used by the communicating parties in Fig.~3 (a), the classical secure communication systems can be divided into two broad categories. One being symmetric cryptosystems for $k_1=k_2$ or for the case that it is straightforward to determine $k_2$ when only $k_1$ is known and $k_1$ can be determined from $k_2$~\cite{menezes2018handbook}. The other being asymmetric cryptosystems, where $k_1$ is publicly available while $k_2$ is kept secret~\cite{menezes2018handbook}. The ciphertext is communicated through a public channel (ciphertext channel), which is usually insecure. Hence the ciphertext can in principle be intercepted by an eavesdropper, Eve, without detection. The key is distributed through another classical channel, for example, using the asymmetric Rivest-Shamir-Adleman scheme \cite{lindell2014introduction,rivest1978method}. During the key distribution, the key is also encrypted and the ciphertext representing the encrypted key can also be intercepted by Eve. Key management is at the heart of this secure communication infrastructure, because the cryptographic key must be generated, exchanged, stored, and finally disposed of in a secure manner. It is clear that the adversary, Eve, can readily steal the ciphertext representing the message during its transmission over the public channel, and the ciphertext conveying the key during the key exchange process without being detected by either of the legitimate communication parties. If the key were stolen, Eve would be able to decrypt all the communications between the legitimate users. 

At the time of writing, only the one-time-pad has been proven to be perfectly secure \cite{shannon1949communication}. The one-time-pad protocol applies the exclusive OR logic operation between the plaintext message and a pre-shared key to generate the ciphertext. The length of the random key string has to be at least as high as that of the plaintext, and should never be reused. Even though this long key represents a 100\% transmission overhead, no other cryptographic protocols relying on shorter keys have been proven to be perfectly secure. Thus, for practical applications, the repeated use of the cryptographic key in a cryptographic system should be avoided, and the length of the key should be set sufficiently high - depending on the computational power available.

The model of an unconditionally secure end-to-end cryptosystem can be constructed by combining QKD and the classical one-time pad, which is shown in Fig.~\ref{fig:Commsys} (b). QKD allows two parties to agree on a secret key by exchanging qubits over a quantum channel~\cite{bennett1984quantum,ekert1991quantum,bennett1992quantum}. An authenticated public channel is also required in support of the associated sifting, parameter estimation, reconciliation, privacy amplification. More explicitly, at least one additional classical information bit is required for each qubit for key \textit{sifting} in QKD. The malicious action of Eve would perturb the state of the qubits, hence the communicating parties can discover Eve through \textit{parameter estimation}, for example estimating quantum bit error rate (QBER). Any potential errors imposed by imperfections of hardware and channel are mitigated by \textit{reconciliation}. Furthermore, \textit{privacy amplification} is used for ensuring that Eve has only negligible information about the final secret key which is achieved by compressing the key~\cite{bennett1995generalized}. Eve has almost no information about the secret key shared between two legitimate users via QKD. In the following process, the encryption, transmission and decryption of data are identical to the aforementioned classical secure communication systems. 

It is observed from Fig. \ref{fig:Commsys} (c) that QSDC constitutes a new secure communication paradigm, which provides a complete confidential near-instantaneous communication solution by means of transmitting the actual messages directly over a quantum channel, rather than only managing the negotiation of secret keys, as in QKD. More explicitly, the plaintext messages are mapped to quantum bits at Alice's station before transmitting them to Bob. No additional classical information is required for decoding the data. Therefore, QSDC does not use cryptographic keys, encryption algorithms, and decryption algorithms. Hence it does not have ciphertext either. Nonetheless, a classical authenticated channel is also needed in QSDC, but only for parameter estimation of detecting eavesdroppers as well as for encoder and decoder, which includes the required service communication for forward error correction and secure coding based on the  universal hashing families~\cite{pan2023free}. The encoder is composed of secure coding encoder, error correction code encoder, and anti-loss encoder, while the decoder includes secure coding decoder, error correction code decoder, and anti-loss decoder~\cite{pan2023free}. Note that the authentication seen both in Fig. \ref{fig:Commsys} (b) and Fig. \ref{fig:Commsys} (c) can be moved from the classical domain to the quantum domain, which has been an important research topic since its seminal source appeared \cite{duvsek1999quantum}. This will be discussed in Section \ref{sec:The cryptographic applications of point-to-point QSDC protocols}.

In general, the theoretically unbreakable classical cryptosystem developed by Shannon is based on the utilization of secret keys \cite{shannon1949communication}. However, conceiving efficient key management is a challenging task. QKD provides a way for a pair of communicating parties to rely on a common secret key for supporting unconditional security, but only the key establishment takes place in the quantum domain, while the messages are sent by a conventional technique over the classical domain. A common feature of both the classical cryptosystems and of QKD seen in Fig. \ref{fig:Commsys} (a) and (b), is that both of them face a security problem in terms of potential key leakage to malicious insiders and outside hackers. QSDC offers an entirely new way of solving all privacy problems. These benefits accrue from QSDC, because the process of communication takes place in the quantum domain, where the actual messages are transmitted through the quantum channel between end users. Hence no information leakage is possible, as guaranteed by applying the laws of quantum mechanics. Without relying on key encryption, no resources are required at all for key management~\cite{dhillon2021qsdc}.

\subsection{Motivation and Contributions}
\label{sec:Motivation and contributions}
The high security of quantum communication has received widespread attention, and the establishment of the quantum internet (Qinternet) will lead to many new applications that have no corresponding classical counterparts~\cite{wehner2018quantum,liu2023quantum}. In recent years, a substantial body of literature has summarized the development of quantum communication, with representative surveys highlighted as follows.
\begin{itemize}
\item Starting from the fundamentals of quantum mechanics (QM), Gisin \textit{et al}.~\cite{gisin2002quantum} provided an overview of the basic concepts of QKD, various protocols, and the progress in experimental implementations. They presented the security analysis of QKD under various attacks.
\item Long \textit{et al}.~\cite{long2007quantum} reviewed the historical development of QSDC theory, detailed some QSDC and deterministic secure quantum communication (DSQC) schemes. They also presented solutions for implementing multi-user QSDC networking. These QSDC schemes belong to the discrete variable (DV) domain.
\item Shenoy-Hejamadi \textit{et al}.~\cite{shenoy2017quantum} surveyed various quantum communication schemes, providing examples of QSDC protocols. Additionally, they discussed the relevant technical issues in implementing quantum communication.
\item Hosseinidehaj \textit{et al}.~\cite{hosseinidehaj2018satellite} elucidated the basic concepts of continuous-variable (CV) QKD and summarized the research progress. Specifically, they discussed CV QKD in the context of free-space communication and introduced its application in satellite-ground communication scenarios.
\item Zawadzki~\cite{zawadzki2021advances} compared several typical QSDC protocols and reported their security analysis. These QSDC schemes also belong to the DV domain.
\item Singh \textit{et al}.~\cite{singh2021quantum} described the fundamental elements of the Qinternet, including qubits, entanglement, quantum teleportation, QKD, and they introduced the  potential applications brought about by the Qinternet.
\item Cao \textit{et al}.~\cite{cao2022evolution} presented the evolution of QKD networks and reported on the progress of QKD standardization.
\item Yang \textit{et al}.~\cite{yang2023survey} offered a review of the key milestones and recent advances in quantum computing and quantum communication (QC). They organized the discussion of various research questions into four major sections: quantum computers, quantum networks, quantum cryptography, and quantum machine learning.
\item Li \textit{et al}.~\cite{li2023entanglement} discussed the basic principles and enabling technologies of entanglement-assisted quantum networks.
\item Pan \textit{et al}.~\cite{pan2023free} introduced the typical protocols of  QSDC, with a focus on progress in their implementation in free-space scenarios. They also anticipated the feasibility of satellite-based QSDC.
\end{itemize}

As shown in Fig.~\ref{fig:Commsys}, QSDC directly supports information transmission in the quantum domain, which is completely different from the motivation of most surveys. Hence, there is a lack of a tutorial and survey literature on QSDC. 

Table~\ref{tab:survey comparison} compares our work to other relevant surveys. By contrast, our survey bridges the gap between classical and quantum domains in a more basic low-paced tutorial manner. Grounded in the relevant quantum mechanics and experimental techniques, it comprehensively reviews the research and applications of QSDC. Importantly, many aspects presented here are introduced for the first time in a survey style.

\begin{table*}[!htbp]
\begin{centering}
\caption{Comparison of this survey with other relevant survey papers. The symbol $\bigtriangleup$ indicates that this paper covers part of this topic.}
\begin{footnotesize}
\begin{tabular}{|m{0.45cm}|m{0.4cm}|m{0.9cm}|m{0.8cm}|m{0.8cm}|m{0.7cm}|m{0.4cm}|m{0.8cm}|m{0.6cm}|m{0.6cm}|m{0.8cm}|m{1.5cm}|m{0.8cm}|m{0.8cm}|m{1.1cm}|}
\hline
Refs & Year & Subject & Security of cryptosystems& Secure communication models & History of QSDC &QM & QC experimental techniques & DV QSDC protocols & CV QSDC protocols & Security analysis of QSDC & Cryptographic applications of QSDC & QSDC network schemes & QSDC Experiments & Challenges of QSDC \\ \hline
 \cite{gisin2002quantum}& 2002 & QKD& \checkmark &  &  & \checkmark & \checkmark &  &  &  &  &  &  &  \\ \hline
 \cite{long2007quantum}& 2007 &QSDC \& DSQC &  &  & \checkmark &  &  & \checkmark &  &  &  & \checkmark &  &  \\ \hline
\cite{shenoy2017quantum}& 2017 & QC schemes &  &  &  &  & \checkmark & \checkmark &  &  &  &  &  &  \\ \hline
 \cite{hosseinidehaj2018satellite}& 2018 & CV QKD &  &  &  & \checkmark & \checkmark &  &  &  &  &  &  &  \\ \hline
 \cite{zawadzki2021advances}& 2021 & QSDC &  &  &  &  &  & \checkmark &  & \checkmark &  &  &  &  \\ \hline
 \cite{singh2021quantum}& 2021 & Qinternet &  &  &  & \checkmark & \checkmark &  &  &  &  &  &  &  \\ \hline
 \cite{cao2022evolution}& 2022 & Qinternet & \checkmark & $\bigtriangleup$ &  &  & $\bigtriangleup$ &  &  &  &  &  &  &  \\ \hline
 \cite{yang2023survey}& 2023 & QC \& quantum computing &  &  &  & \checkmark &  &  &  &  &  &  &  &  \\ \hline
 \cite{li2023entanglement}& 2023 & Qinternet &  &  &  & \checkmark & $\bigtriangleup$ &  &  &  &  &  &  &  \\ \hline
 \cite{pan2023free}& 2023 & FSO QSDC &  &  &  &  & $\bigtriangleup$ & \checkmark & \checkmark &  &  &  &$\bigtriangleup$ & \checkmark \\ \hline
This work &  & \checkmark & \checkmark & \checkmark & \checkmark & \checkmark & \checkmark & \checkmark & \checkmark & \checkmark & \checkmark & \checkmark & \checkmark & \checkmark \\ \hline
\end{tabular}
\end{footnotesize}
\label{tab:survey comparison}
\end{centering}
\end{table*}

The main contributions of our survey are outlined as follows:
\begin{itemize}
\item (1) We conducted a comparative analysis of classical secure communication, secure communication based on QKD, and QSDC in terms of communication modes. This highlights that information transmission in QSDC is truly within the quantum domain. Highlighting its unique advantages, we reviewed over two decades of milestones in its development, indicating that quantum-enabled technology has transitioned from theory to experimentation with a solid foundation for practical applications.
\item (2) We provided an accessible portrayal of the quantum information resources required for understanding QSDC, including qubits, entanglement, operations for mapping information onto quantum states, measurement, superdense coding, quantum teleportation, entanglement swapping, and entanglement purification. We also summarized the fundamental quantum physical principles that underpin quantum-native security in QSDC.
\item (3) We presented the technological foundations for implementing QSDC, covering its essential elements such as light sources, modulation, channels, and detectors.
\item (4) We systematically introduced the family of various QSDC protocols, encompassing both DV and CV protocols, and compared their characteristics. Additionally, we discussed the transition from point-to-point communication to multi-user networking.
\item (5) We conducted an in-depth investigation into the design of other quantum cryptographic protocols relying on QSDC.
\item (6) Regarding the security and performance metrics of QSDC, we presented the  advances in terms of security proofs for QSDC. Critical to improving the QSDC performance attained, we reported on the current state of experimental advances in QSDC.
\item (7) Finally, we identified open questions in the evolution of QSDC and highlighted promising research directions. Specifically, we discussed its protocol design, security proofs, and its experimental implementations. We proposed a hybrid approach combining QSDC with classical secure communication networks in support of secure next-generation communication.
\end{itemize}

\subsection{Article Structure}
\label{sec:paperstructure}
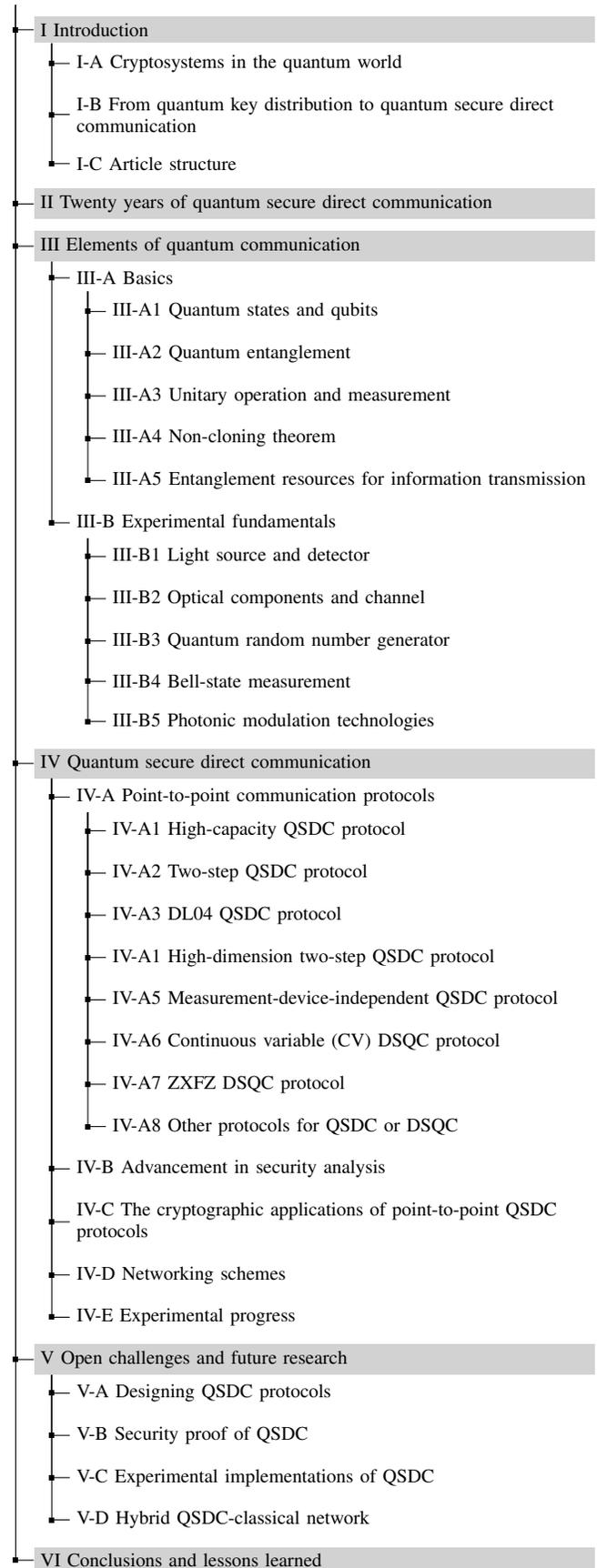
\begin{figure}[!htbp]
\begin{footnotesize}
\begin{forest}
  for tree={
    grow'=0,
    child anchor=west,
    parent anchor=south,
    anchor=west,
    calign=first,
    edge path={
      \noexpand\path [draw, \forestoption{edge}]
      (!u.south west) +(7.5pt,0) |- node[fill,inner sep=1pt] {} (.child anchor)\forestoption{edge label};
    },
    before typesetting nodes={
      if n=1
        {insert before={[,phantom]}}
        {}
    },
    fit=band,
    s sep=4.0pt,
    before computing xy={l=15pt},
  }
[[\ref{sec:intro} Introduction,fill=gray!35, text width=8cm
    [\ref{sec:Unveiling quantum computing} Unveiling Quantum Computing]
    [\ref{sec:Cryptosystems-in-the-quantum world} Cryptosystems in the Quantum World]
    [\ref{sec:patterns-for-commu} From Quantum Key Distribution to Quantum Secure Direct Communication, text width=8cm]
    [\ref{sec:Motivation and contributions} Motivation and Contributions]
    [\ref{sec:paperstructure} Article Structure]
  ]
  [\ref{sec:history of QSDC} Twenty Years of Quantum Secure Direct Communication,fill=gray!35, text width=8cm
  ]
  [\ref{sec:Preliminaries} Preliminaries of Quantum Communication,fill=gray!35, text width=8cm
    [\ref{sec:Theoretical foundations} Theoretical Foundations
      [\ref{sec:Quantum-states-and-qubits} Quantum States and Qubits]
      [\ref{sec:Quantum-entanglement} Quantum Entanglement]
      [\ref{sec:Unitary-operation-and-measurement} Unitary Operation and Measurement]
      [\ref{sec:No-cloning theorem} Non-Cloning Theorem]
      [\ref{subsec:EntRes} Entanglement Resources for Information Transmission]
    ]
    [\ref{sec:Experimental-fundamentals} Experimental Fundamentals
      [\ref{subsec:LD} Light Source and Detector]
      [\ref{sec:componentsandchannel} Optical Components and Channel]
      [\ref{sec:QRNG} Quantum Random Number Generator]
      [\ref{sec:BSM} Bell-State Measurement]
      [\ref{sec:encodingtech} Photonic Modulation Technologies]
    ]
  ]
  [\ref{sec:QSDC} Quantum Secure Direct Communication,fill=gray!35, text width=8cm
    [\ref{sec:PtoP} Point-to-Point Communication Protocols
      [\ref{sec:High-capacity QSDC protocol} High-Capacity QSDC Protocol]
      [\ref{sec:Two-step QSDC protocol} Two-Step QSDC Protocol]
      [\ref{sec:DL04 QSDC protocol} Deng-Long 2004 (DL04) QSDC Protocol]
      [\ref{sec:High-capacity QSDC protocol} High-Dimension Two-Step QSDC Protocol]
      [\ref{sec:Measurement-device-independent QSDC protocol} Measurement-Device-Independent QSDC Protocol]
      [\ref{sec:MS06 DSQC protocol} Marino-Stroud 2006 (MS06) DSQC Protocol]
      [\ref{sec:ZXFZ DSQC protocol} ZXFZ DSQC Protocol]
      [\ref{sec:Other protocols for QSDC or DSQC} Other Protocols for QSDC or DSQC]
    ]
    [\ref{sec:Advancement in security analysis} Advancement in Security Analysis]
    [\ref{sec:The cryptographic applications of point-to-point QSDC protocols} The Cryptographic Applications of Point-to-Point QSDC Protocols, text width=8cm]
    [\ref{sec:Networking schemes} Networking Schemes]
    [\ref{sec:Experimental progress} Experimental Progress]
  ]
  [\ref{sec:challenges and future works} Open Challenges and Future Research,fill=gray!35, text width=8cm
    [\ref{sec:Designing QSDC protocol} Designing QSDC Protocols]
    [\ref{sec:Security proof of QSDC} Security Proof of QSDC]
    [\ref{sec:QSDC experimental implementation} Experimental Implementations of QSDC]
    [\ref{sec:Hybrid QSDC-classical network} Hybrid QSDC-Classical Network]
  ]
  [\ref{sec:Con} Conclusions and Lessons Learned, fill=gray!35, text width=8cm]
]
\end{forest}
\end{footnotesize}
\caption{The structure of this survey article.}
\label{fig:paperstructure}
\end{figure}

This article is organized as seen in Fig. \ref{fig:paperstructure}. As we have discussed in Section \ref{sec:intro}, quantum communication guarantees cryptographic security against quantum computation attacks, even when quantum computing becomes the norm in the post-quantum era. QSDC, as one of the prominent branches of quantum communication supports the confidential transmission of information via quantum channels. In Section \ref{sec:history of QSDC}, the development of QSDC is surveyed by presenting its milestones. In Section \ref{sec:Preliminaries}, we will introduce the fundamental theory and experimental techniques of quantum communication. We will only rely on modest theoretical background for understanding QSDC, followed by some practical guidelines. In Section \ref{sec:QSDC}, the salient QSDC protocols are introduced step by step to present the basic principles of QSDC, highlighting how the secret messages can be directly transmitted over the quantum channel. We will consider both point-to-point scenarios and networking issues. Topics such as their cryptographic applications, security proof and recent experimental advances are also covered. The long-term evolution of QSDC, its research directions and challenges are discussed in Section \ref{sec:challenges and future works}. Finally, we conclude in Section \ref{sec:Con}.

\section{Twenty years of quantum secure direct communication}
\label{sec:history of QSDC}
Again, QSDC was originally proposed by Long and Liu in 2000~\cite{long2000theoretical,long2002theoretically} and in these seminal contributions, it was pointed out that a key was produced by Alice before transmission took place \cite{long2000theoretical} (versions 1 and 2). The terminology of `quantum secure direct communication' was introduced for the first time in the two-step QSDC paper \cite{deng2003two}, where the definition of QSDC was also put forward. Since its conception in 2000, it has evolved into a fully-fledged communications protocol, as seen in Fig. \ref{fig:Time}. The early research of QSDC has been focused on the construction of physical schemes, and many QSDC protocols were invented for different information carriers, such as entangled states \cite{long2002theoretically,deng2003two}, single photons \cite{deng2004secure}, and Greenberger-Horne-Zeilinger (GHZ) states \cite{wang2005multi,jin2006three}. Recently significant breakthroughs have been made in the experimental demonstration of QSDC protocols \cite{hu2016experimental,zhang2017quantum}, which paved the way for their practical application \cite{zhu2017experimental,qi2019implementation}. 

\begin{figure*}
\begin{centering}
\begin{footnotesize}
\begin{timeline}{2000}{2023}{1cm}{1.5cm}{15cm}{23cm}
\entry{2000}{Long and Liu~\cite{long2000theoretical} proposed an effiicent QSDC protocol using the block transmission of Einstein-Podolsky-Rosen (EPR) pairs.}
\entry{2003}{Deng \textit{et al}.~\cite{deng2003two} constructed the two-step QSDC protocol and the standard criterion of QSDC was discussed.}
\entry{2004}{Deng and Long~\cite{deng2004secure} presented the DL04 QSDC protocol with single photons.}
\entry{2005}{Wang \textit{et al}.~\cite{wang2005quantum} utilized the source of $d$-dimension Bell states to realize a high-dimension two-step QSDC protocol.}
\plainentry{2005}{Wang \textit{et al}. \cite{wang2005multi} created the multi-step QSDC scheme by using GHZ-state dense coding.}
\plainentry{2005}{Zhang \textit{et al}.~\cite{zhang2005multiparty1} proposed a scheme of multiparty quantum secret sharing based on the DL04 QSDC protocol~\cite{deng2004secure}.}
\entry{2006}{Jin \textit{et al}.~\cite{jin2006three} conceived the three-party simultaneous QSDC based on GHZ states.}
\plainentry{2006}{Lee \textit{et al}.~\cite{lee2006quantum} constructed a QSDC protocol with authentication relying on GHZ states.}
\plainentry{2006}{Deng \textit{et al}.~\cite{deng2006quantum} proposed a QSDC network enabling any one of the authorized users can communicate another one on the network.}
\entry{2007}{Deng \textit{et al}.~\cite{deng2007quantum} proposed the QSDC network with entanglement and decoy photons.}
\entry{2008}{Lin \textit{et al}.~\cite{lin2008quantum} designed QSDC protocol using $\chi$-state.}
\plainentry{2008}{Pirandola \textit{et al}.~\cite{pirandola2008quantum} constructed a continuous-variable quantum direct communication protocol.}
\entry{2009}{Qin \textit{et al}.~\cite{qin2009quantum} studied QSDC over collective amplitude damping channel.}
\entry{2010}{Gao \textit{et al}.~\cite{gao2010cryptanalysis} analyzed the security of a multi-party controlled QSDC protocol based on GHZ state.}
\entry{2011}{Lu \textit{et al}.~\cite{lu2011unconditional} provided the unconditional security proof of the four-state quantum communication which is suit for DL04 protocol.}
\entry{2012}{Sun \textit{et al}.~\cite{sun2012quantum} proposed a QSDC protocol with cluster state.}
\entry{2013}{Chang \textit{et al}.~\cite{chang2013quantum} proposed a QSDC protocol with authentication using single photons.}
\entry{2014}{Yadav \textit{et al}.~\cite{yadav2014two} proposed a two-step QSDC with the help of order rearrangement.}
\plainentry{2014}{Shapiro \textit{et al}.~\cite{shapiro2014secure,zhuang2015ultrabroadband} proposed a QSDC protocol based on quantum illumination.}
\entry{2015}{Farouk \textit{et al}.~\cite{farouk2015generalized} designed a $N$-party QSDC protocol with authentication.}
\entry{2016}{Hu \textit{et al}.~\cite{hu2016experimental} designed and experimentally demonstrated a single-photon frequency coded DL04 QSDC protocol.}
\plainentry{2016}{Lum \textit{et al}.~\cite{lum2016quantum} showed that quantum data locking can be applied in QSDC.}
\entry{2017}{Zhang \textit{et al}.~\cite{zhang2017quantum} employed the state-of-art atomic quantum memory and demonstrated the principles of the efficient QSDC and the two-step QSDC protocol for the first time.}
\plainentry{2017}{Zhu \textit{et al}.~\cite{zhu2017experimental} realized long-distance QSDC using the efficient QSDC and the two-step QSDC protocols.}
\entry{2018}{Zhou \textit{et al}.~\cite{zhou2020measurement} reported a measurement-device-independent (MDI) QSDC scheme of single photons.}
\plainentry{2018}{Niu \textit{et al}.~\cite{niu2018measurement} proposed MDI QSDC scheme of EPR pairs.}
\plainentry{2018}{Sun \textit{et al}.~\cite{sun2018design} designed a QSDC protocol that does not require quantum mempery, overcoming a bottleneck obstacle of practical QSDC.}
\entry{2019}{Qi \textit{et al}.~\cite{qi2019implementation} implemented a practical QSDC system with the security analysis of the Wyner wiretap channel theory.}
\plainentry{2019}{Shapiro \textit{et al}.~\cite{shapiro2019quantum} proposed that the quantum low probability of perception protocol can be viewed as an example of QSDC.}
\plainentry{2019}{Zhou \textit{et al}.~\cite{zhou2020device} proposed the device-independent (DI) QSDC protocol.}
\plainentry{2019}{Massa \textit{et al}.~\cite{massa2019experimental} experimentally demonstrated two-way QSDC~\cite{Del2018two}.}
\entry{2020}{Sun \textit{et al}.~\cite{sun2020toward} proposed a quantum-memory-free DL04 QSDC protocol using coding theory.}
\plainentry{2020}{Pan \textit{et al}.~\cite{pan2020experimental} reported a free-space QSDC.}
\plainentry{2020}{Lindsey \textit{et al}.~\cite{lindsey2020transmission} suggested a spectrum approach to achieve QSDC over the noisy quantum channels.}
\entry{2021}{Qi \textit{et al}.~\cite{qi202115} demonstrated a 15-user QSDC network based on entanglement distribution.}
\plainentry{2021}{V{\'a}zquez-Castro \textit{et al}.~\cite{vazquez2021quantum} utilized a quantum version of on-off keying modulation to directly transmit confidential information over a quantum channel.}
\entry{2022}{Zhang \textit{et al}.~\cite{zhang2022realization} declared breakthrough in \SI{100}{km} fiber-based QSDC.}
\plainentry{2022}{Wu \textit{et al}.~\cite{wu2022quantum} proved the security of QSDC considering the finite-size effect.}
\plainentry{2022}{Long \textit{et al}.~\cite{long2022evolutionary} proposed a secure repeater network and experimental demonstrated it.}
\plainentry{2022}{Liu \textit{et al}.~\cite{liu2022fiber} reported a proof-of-principle QSDC experiment over a \SI{5}{km} fiber channel.}
\plainentry{2022}{Sheng \textit{et al}.~\cite{sheng2022one} relaxed the requirement of two-way transmission of photons in the two-step QSDC scheme, proposing a one-way transmission to achieve QSDC.}
\entry{2023}{Panda \textit{et al}.~\cite{panda2023quantum} presented a QSDC protocol by utilizing quantum walks on orbital angular momentum (OAM) states.}
\plainentry{2023}{Zhou \textit{et al}.~\cite{zhou2023device} designed a DI QSDC scheme relying on single-photon sources.}
\plainentry{2023}{Li \textit{et al}.~\cite{li2023single1,li2023single2} propounded a single-photon-memory MDI QSDC protocol and deduced its secrecy capacity.} 
\plainentry{2023}{Sun \textit{et al}.~\cite{sun2023one} relaxed the requirement for state preparation in MDI QSDC and determined the practical secrecy capacity of the protocol by integrating customized decoy-state methods.}
\plainentry{2023}{Xu \textit{et al}.~\cite{xu2023when} proposed a quantum blockchain scheme relying on QSDC.}
\end{timeline}
\caption{Timeline of important milestones in quantum secure direct communication.}
\label{fig:Time}
\end{footnotesize}
\end{centering}
\end{figure*}

The lack of practical quantum memory has been a serious obstacle for the evolution of quantum communication, because the employment of relaying requires memory. Hence, practical QKD has remained confined to intra-city distances. For inter-city applications of QKD, so-called trusted relays are used as temporary replacements for perfectly reliable quantum repeaters. Satellite relays provide a promising technique for global quantum key distribution~\cite{bacsardi2013way,liao2017satellite}. But for QSDC even over short distances, quantum memory has remained indispensable, until quite recently, because the associated block-based transmission requires the quantum states to be stored before their security is assured. However, quite recently, a quantum-memory-free QSDC protocol~\cite{sun2018design,sun2020toward} was designed, where the plaintext was encrypted using a pre-shared random key and then the ciphertext was used to distill secret keys for encrypting later message blocks~\cite{pan2020single}. This development has finally made QSDC applicable for intercity distances~\cite{zhang2022realization}. Furthermore, combining QSDC and post-quantum cryptography enables the construction of a secure repeater, which can establish a large-scale quantum communication network using existing technology~\cite{long2022evolutionary}. This approach avoids the security risks that arise from relying on trusted relays.

Many QKD protocols~\cite{bennett1984quantum,ekert1991quantum,bennett1992quantum,huttner1995quantum} have a probabilistic nature, since an uncontrolled key sequence is established between two users, who randomly choose their so-called rectilinear or diagonal quantum basis to measure the qubits and the key is produced on the basis of random instances, where the pair of communicating users choose the same bases on a probabilistic basis. Here we emphasize that this probabilistic random sequence should NOT contain any meaningful information, just a sequence of random bits, not a key. It is essentially used for eavesdropper detection and if an eavesdopper is detected during its transmission, this random sequence can be discarded. There are several examples of deterministic quantum key distribution (DQKD) protocols. Briefly, DQKD is a protocol designed for handing over the above-mentioned deterministic key to the intended receiver and no basis reconciliation is required for decoding. To exemplify the DQKD protocols, Goldenberg and Vaidman proposed such a scheme in 1995 using a Mach-Zehnder interferometer~\cite{goldenberg1995quantum}, while Boström \textit{et al}.~\cite{bostrom2002deterministic} conceived a QKD protocol using EPR pairs, which has the fond connotation of the Ping-Pong protocol. The two-way QKD protocols of~\cite{deng2004bidirectional,lucamarini2005secure,pirandola2008continuous} are also prominent examples of DQKD. An essential feature of QKD is that the transmitted data may become fully or partially leaked to Eve. Fortunately, QKD is capable of detecting eavesdropping, but it cannot prevent the leakage of the transmitted data. This is why QKD has to resort to the transmission of meaningless random sequences for Eve-detection, and again, if eavesdropping is detected, the transmitted data will be discarded. It is apparent that the Ping-Pong protocol~\cite{bostrom2002deterministic} cannot convey secret messages over the quantum channel due to its QKD nature. Although the original Ping-pong protocol of~\cite{bostrom2002deterministic} is insecure even for QKD, later it has been rendered secure and has been generalized to numerous applications~\cite{wojcik2003eavesdropping,zhang2004improving,cai2004ping,pavivcic2013quantum,li2013security}. As a result, a simple way of distinguishing a QSDC protocol from a DQKD protocol~\cite{long2007quantum} is to check whether the transmitted data would or would not be leaked to Eve. For QSDC, the transmitted confidential information would not be leaked, because the eavesdropper would only be able to acquire completely random information, whereas the data transmitted in DQKD would be partially or completely leaked to Eve. It is worth pointing out that classical communications cannot detect eavesdropping.

There is a particular variant of DQKD, which is also mistakenly referred to as QSDC by some authors. To elaborate a little further, normally QKD is performed first to establish a shared key between Alice and Bob. The key is then used for encrypting the message into the related ciphertext, which is then transmitted through a classical channel. For DQKD, the procedure can be appropriately modified, where Alice can choose a random sequence as her key to encrypt her message into the ciphertext, which is then transmitted to Bob through a quantum channel. Then they assess the grade of security during the ciphertext transmission, for example by estimating the error rate. If they are sure that the security has not been compromised, implying that tempering by Eve has not affected the ciphertext, then Alice sends the key through a classical channel to Bob. This variant of DQKD is usually termed as DSQC~\cite{long2007quantum,long2010quantum}. A simple rule to judge whether a protocol belongs to the family of DSQC is to ask the question: is there any need for classical communication for announcing the key rather than for basis choice reconciliation and eavesdropping detection in the protocol? If the answer is affirmative, we can conclude that it is indeed a DSQC protocol. In the DSQC protocol the receiver cannot directly read the secret message, unless it receives one bit of additional classical information from the transmitter for reading the secret message. In 2001, Beige \textit{et al}.~\cite{beige2001secure} proposed a DSQC protocol, which also needs additional classical communication. Their protocol became insecure, when an adversary acquired the secret information by applying a so-called quantum non-demolition measurement. As a further advance, DSQC has also been extended to  entanglement distribution~\cite{yan2004scheme,zhu2006secure} and to continuous variable based implementations~\cite{marino2006deterministic}. Pan \textit{et al}.~\cite{pan2023free} identified the essential characteristics that a point-to-point QSDC protocol should possess. If a protocol fails to satisfy these criterias, but nevertheless transmits information directly through a quantum channel, it is referred to as a quasi-QSDC protocol in their work.

Thus based on the above paragraph, it should be born in mind that the above statements regarding QSDC are not applicable to the DSQC family, whose members are essentially of DQKD nature. The nature of some of the popular communication protocols is summarized at a glance in Table \ref{t2-cmmunicationmodels}. It is clear that only QSDC is capable of avoiding the leakage of transmitted data, when Eve intercepts the transmission. 

\begin{table*}[h!]
\centering
\begin{footnotesize}
\caption{Comparisons of Different Communication Patterns. PQKD, Probabilistic QKD; Leakage, Leakage of transmitted data if Eve intercepts.}
\label{t2-cmmunicationmodels}
\begin{tabular}{ccccc}\hline
	Protocol                 & Deterministic         & Eve detection         & Leakage       & Examples  \\ \hline
	Classical communication           & Yes           & No                    & Yes                             & Any\\
	PQKD                    & No            & Yes                   & Yes                            &\cite{bennett1984quantum} \\
	DQKD                     & Yes           & Yes                   & Yes                            & \cite{goldenberg1995quantum}\\
	DSQC                     & Yes           & Yes                   & Yes                            &\cite{beige2001secure}\\
	QSDC                     & Yes           &Yes                    & No                             & \cite{long2000theoretical}\\ \hline
\end{tabular}
\end{footnotesize}
\end{table*}

\section{Preliminaries of quantum communication}
\label{sec:Preliminaries}
QSDC involves mapping private information onto quantum states for transmission. To introduce this communication paradigm, we begin with a mathematical description of its information carrier - the quantum state, we then introduce the commonly used communication facilitator - the entangled state. Successful information transmission requires mapping the information onto quantum states, followed by demodulation. Hence, we will present the unitary operations used for mapping the information and for its measurement i.e. the retrieval. A distinctive feature of QSDC is its high security, rooted in the no-cloning theorem. Additionally, this section will also introduce the elements commonly involved in QSDC protocols, such as superdense coding, quantum teleportation, and entanglement swapping.

As mentioned in Section~\ref{sec:history of QSDC}, QSDC has been around for two decades, and researchers have moved on to experimental demonstrations, showing significant potential for practical applications. To this end, we will discuss the progress in experiments, hoping to inspire experimentalists to further enhance its performance. Here, we will introduce the fundamental experiments, which primarily include its basic components, signal modulation techniques, signal detection, and so on.

\subsection{Theoretical Foundations}
\label{sec:Theoretical foundations}
\subsubsection{Quantum States and Qubits}
\label{sec:Quantum-states-and-qubits}
In quantum mechanics, the state of a physical system is described by a state vector in the Hilbert space $\mathscr{H}$ \cite{imre2005quantum}. The Hilbert space is a complex-valued vector space encompassing every conceivable state a quantum system may inhabit. The so-called `ket' notation describing a vector in the complex number space $\mathbb{C}^{n}$ representing a pure quantum state is mathematically denoted as
\begin{eqnarray}
|\psi\rangle=\begin{pmatrix}
\alpha_{1}\\
\alpha_{2}\\
\alpha_{3}\\
\vdots\\ 
\alpha_{n}\\
\end{pmatrix},
\end{eqnarray}
where the symbol `$|\cdot\rangle$' is the Dirac notation \cite{dirac1981principles} of a symbol $\psi$ for a vector and we have $\alpha_{n}\in\mathbb{C}^{n}$ associated with $\sum_{i=1}^{n}\left | \alpha_{i} \right |^{2}=1$. The `ket' notation simply represents the second half of the word `Dirac-ket'. This unit vector can also be written in the form of the superposition $|\psi\rangle=\sum_{i=1}^{n}\alpha_{i}|\psi_{i}\rangle$, where $\alpha_{i}$ is the amplitude of the vector in an orthonormal basis $|\psi_{i}\rangle$. The orthonormal basis obeys $\langle\psi_{i}|\psi_{i}\rangle=1$ and $\langle\psi_{i}|\psi_{j}\rangle=0$ when $i\ne j$. The conjugate transpose of $|\psi\rangle$ is $\langle\psi|$, which is called a bra vector, formulated as:
\begin{eqnarray}
\langle\psi|=\left(|\psi\rangle\right)^{\dagger}=
\begin{pmatrix}
\alpha_{1}^{\ast}&\alpha_{2}^{\ast}&\alpha_{3}^{\ast}&\cdots&\alpha_{n}^{\ast}
\end{pmatrix}.
\end{eqnarray}
The combined bra and ket notations $\langle\psi|\phi\rangle$, and $|\psi\rangle\langle\phi|$ represent the inner product and the outer product of the vectors, respectively. But in some cases a quantum system cannot be described by a single vector, it is rather described by a probability distribution, which may be in the state $|\phi_{i}\rangle$ with probability $p_{i}$. Such a quantum system is said to be in a mixed state, which corresponds to a probabilistic mixture of pure quantum states. A pure state is a particular case of a mixed state associated with $p_{i}=1$ and $p_{j}=0$ $(i\neq j)$. A commonly adopted way of describing mixed states in quantum mechanics is to use the so-called density matrix, which is the weighted sum of pure states in the form of \cite{jones2012quantum,gyongyosi2018survey}
\begin{eqnarray}
\rho=\sum_{i=1}^{N}p_{i}|\phi_{i}\rangle\langle\phi_{i}|.
\end{eqnarray}

\begin{figure}[!h]
\begin{center}
\includegraphics[width=5cm,angle=0]{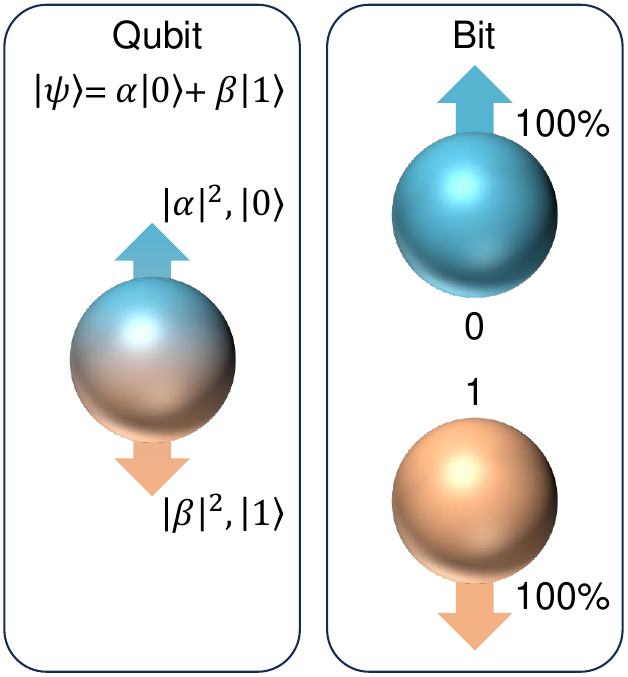}
\caption{Comparison between qubit and classical bit.}
\label{fig:QubitVsbit}
\end{center}
\end{figure}

The basic element of quantum information processing is a qubit formulated as \cite{nielsen2002quantum}
\begin{eqnarray}
|\psi\rangle=\alpha|0\rangle+\beta|1\rangle=\begin{pmatrix}
\alpha\\
\beta
\end{pmatrix},
\end{eqnarray}
where $|0\rangle=\begin{pmatrix}1\\ 0\end{pmatrix}$ and $|1\rangle=\begin{pmatrix}0\\ 1\end{pmatrix}$ are the orthonormal bases, while $\alpha$ and $\beta$ are complex numbers associated with $\left | \alpha \right |^{2}+\left | \beta \right |^{2}=1$. The classical bits either assume a logical `0' or `1' value, while a qubit may be found in an arbitrary superposition of the two basis states $|0\rangle$ and $|1\rangle$, as shown in Fig.~\ref{fig:QubitVsbit}. When measuring or observing the qubit in Fig.~\ref{fig:QubitVsbit}, we will obtain $|0\rangle$ with probability $|\alpha|^2$ and $|1\rangle$ with probability $|\beta|^2$ (which will be explained in detail later). In physically tangible terms this superposition may be interpreted as a coin spinning in a box in the equi-probable superposition of `head and tail'. But when we lift the lid of the box and `observe' the coin, this superposition of states `collapses' back into one of the basis states of `head or tail'. 

\begin{figure}[!h]
\begin{center}
\includegraphics[width=7cm,angle=0]{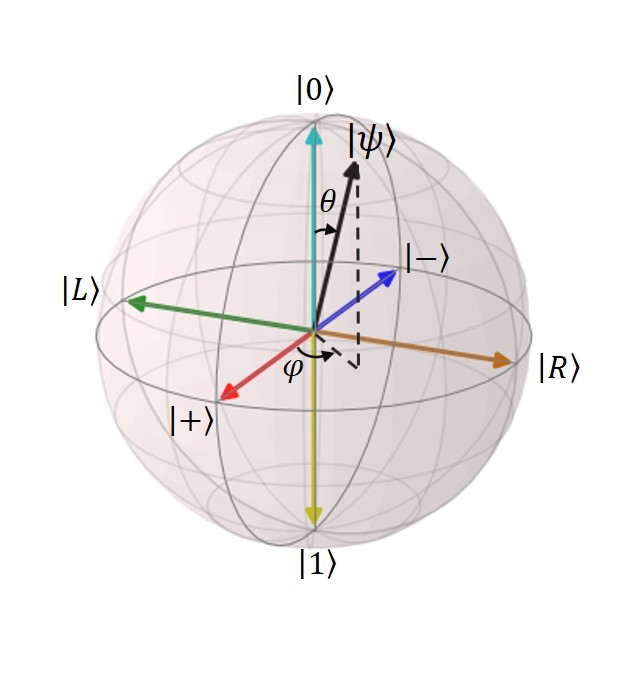}
\caption{Visualization of qubit states in the Bloch sphere.}
\label{fig:BlochSphere}
\end{center}
\end{figure}

If the amplitudes are parameterized by $\alpha=\cos\left(\frac{\theta}{2}\right)$ and $\beta=e^{i\varphi}\sin\left(\frac{\theta}{2}\right)$, then a particularly useful form is 
\begin{equation}
|\psi\rangle=\cos\left(\frac{\theta}{2}\right)|0\rangle+e^{i\varphi }\sin\left(\frac{\theta}{2}\right)|1\rangle.
\end{equation}
The state of a qubit can be geometrically represented by a vector in a Bloch sphere, shown in Fig. \ref{fig:BlochSphere}, where $\theta$ $(0\leqslant\theta\leqslant\pi)$ and $\varphi $ $(0\leqslant\varphi< 2\pi)$, correspond to the polar angle and azimuthal angle, respectively. A pure qubit state can be represented by a point on the surface of the Bloch sphere, while mixed states are the points inside the Bloch sphere. Furthermore, the center of the sphere is the maximally mixed state, such as
\begin{equation}
\frac{1}{2}|0\rangle\langle0|+\frac{1}{2}|1\rangle\langle1|=\frac{1}{2}\begin{pmatrix}
1&0 \\
0&1
\end{pmatrix}=\frac{I}{2}.
\end{equation}
The basis state $|0\rangle$ is located at the North pole, while $|1\rangle$ at the South pole. Observe from Fig. \ref{fig:BlochSphere} that in addition to the $\{|0\rangle, |1\rangle\}$ basis,  two other important states are, 
\begin{eqnarray}
|+\rangle=\frac{1}{\sqrt{2}}\left(|0\rangle+|1\rangle\right), \nonumber\\
|-\rangle=\frac{1}{\sqrt{2}}\left(|0\rangle-|1\rangle\right),
\end{eqnarray}
which are the eigenstates of $\sigma_{x}$, and the states
\begin{eqnarray}
|R\rangle=\frac{1}{\sqrt{2}}\left(|0\rangle+i|1\rangle\right),\; |L\rangle=\frac{1}{\sqrt{2}}\left(|0\rangle-i|1\rangle\right),
\end{eqnarray}
which are the eigenstates of $\sigma_y$. Qubits can be implemented in various physical ways, including the spin of an electron~\cite{stuyck2020integrated,wang2014parallel}, by superconductors~\cite{berggren2004quantum}, ions~\cite{png2022quantum,drmota2023robust}, photons~\cite{Obrien2007optical}, and quantum dots~\cite{Cong2015ultrafast,Giounanlis2019modeling}. Photons are the most common quantum communication carriers.

By contrast, a `qumode' refers to a continuous-variable quantum system, such as the continuous-variable optical field, analogous to the discrete-variable qubit, and it is described mathematically in~\cite{van2011optical,weedbrook2012gaussian,hosseinidehaj2018satellite}.

\subsubsection{Quantum Entanglement}
\label{sec:Quantum-entanglement}
The concept of entanglement originates from the influential argument of Einstein \textit{et al}. \cite{einstein1935can} intended to question the completeness of quantum mechanics. Here, let us define it by using the modern terminology of qubits. Let us consider a composite Hilbert space $\mathscr{H}_{A}\otimes\mathscr{H}_{B}$. Then there is a state of the composite system, which cannot be written as a tensor product\footnote{The tensor product $|\psi\rangle_{A}\otimes|\psi\rangle_{B}$ is often abbreviated to $|\psi\rangle_{A}|\psi\rangle_{B}$ or even more compactly as $|\psi_{A}\psi_{B}\rangle$.} of the states of the individual subsystems,
\begin{eqnarray}
|\psi\rangle_{AB}\neq|\psi\rangle_{A}\otimes|\psi\rangle_{B},\;\forall\,|\psi\rangle_{A}\in\mathscr{H}_{A},\;\forall\,|\psi\rangle_{B}\in\mathscr{H}_{B}.
\end{eqnarray}
Such a state is called an entangled state. For example, the most commonly used entangled states are the four Bell states, also termed as the EPR pairs or EPR states,
\begin{eqnarray}
\label{Eq:entanglement}
|\psi^{+}\rangle_{AB}=\frac{1}{\sqrt{2}}\left(|0\rangle_{A}|1\rangle_{B}+|1\rangle_{A}|0\rangle_{B}\right),\nonumber\\
|\psi^{-}\rangle_{AB}=\frac{1}{\sqrt{2}}\left(|0\rangle_{A}|1\rangle_{B}-|1\rangle_{A}|0\rangle_{B}\right),\nonumber\\
|\phi^{+}\rangle_{AB}=\frac{1}{\sqrt{2}}\left(|0\rangle_{A}|0\rangle_{B}+|1\rangle_{A}|1\rangle_{B}\right),\nonumber\\
|\phi^{-}\rangle_{AB}=\frac{1}{\sqrt{2}}\left(|0\rangle_{A}|0\rangle_{B}-|1\rangle_{A}|1\rangle_{B}\right).
\end{eqnarray}
We have no way of expressing the four Bell states as tensor products. By contrast, the state $\frac{1}{\sqrt{2}}\left(|0\rangle_{A}|0\rangle_{B}+\right|0\rangle_{A}|1\rangle_{B})$ is a separable state, because it can be written in form of the tensor product, $|0\rangle_{A}\otimes\frac{1}{\sqrt{2}}\left(|0\rangle_{B}+|1\rangle_{B}\right)$.

\begin{figure}[!h]
\begin{center}
\includegraphics[width=5.5cm,angle=0]{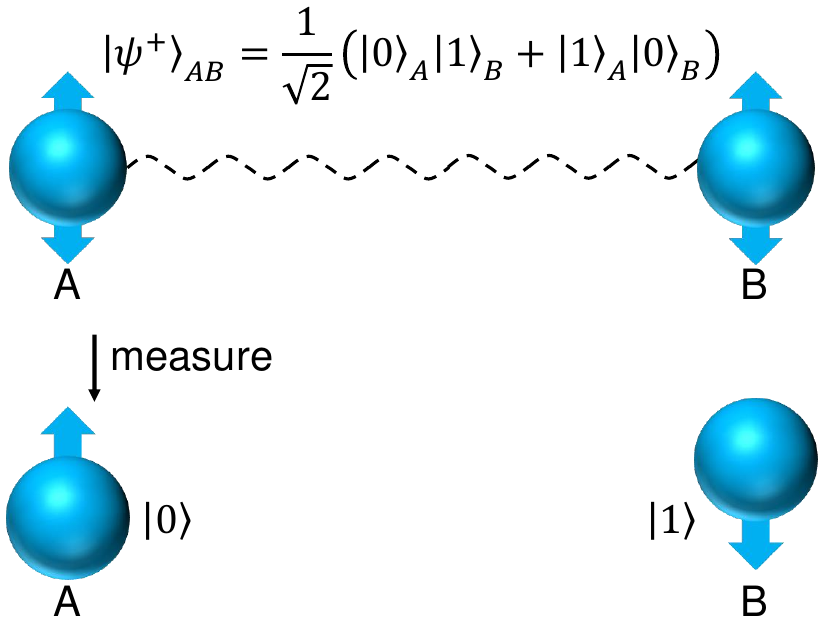}
\caption{Entangled state and its possible measurement outcomes.}
\label{fig:Entanglement}
\end{center}
\end{figure}

Now, the family of entangled states plays a key role both in the protocol design~\cite{chandra2021direct} and in the security proof of quantum communication. Apart from the Bell states of Eq. (\ref{Eq:entanglement}), other members of the entangled state family include the GHZ states \cite{greenberger1989bell,li2016multi} defined as:
\begin{eqnarray}
|\rm GHZ\rangle=\frac{1}{\sqrt{2}}\left(|00\cdots0\rangle\pm|11\cdots1\rangle\right),
\end{eqnarray}
and the W-state \cite{dur2000three}
\begin{eqnarray}
|\rm W\rangle=\frac{1}{\sqrt{N}}(|00\cdots01\rangle+|00\cdots10\rangle+\cdots\nonumber\\
+|01\cdots00\rangle+|10\cdots00\rangle).
\end{eqnarray}
They have been widely used in quantum communication protocols. Loosely speaking, entangled states exhibit `perfect' correlation, which is consistent with the nature of quantum communication, namely sending messages from the transmitter to the receiver is to correlate them~\cite{horodecki2009quantum}. Anecdotally, Einstein referred to the phenomenon of entanglement as a `spooky action at a distance', because measuring one of the entangled particles instantly determines the state of the other particle, regardless of their physical distance. As shown in Fig.~\ref{fig:Entanglement}, assuming an entangled state $|\psi^{+}\rangle_{AB}$ has been established between distant locations A and B, when a measurement at location A yields $|0\rangle$, the state of the particle at location B immediately changes to $|1\rangle$. Having said that, the speed of light cannot be exceeded, because before this `spooky action' can take place, some preparatory classical domain operations are required, which do obey the speed of light.

The two-mode squeezed vacuum state is a type of continuous-variable entangled state, which is mathematically given by~\cite{van2011optical,weedbrook2012gaussian,hosseinidehaj2018satellite}
\begin{equation}
|r\rangle_{\rm EPR}=\sqrt{1-\lambda^2}\sum_{n=0}^{\infty}\left(-\lambda\right)^n|n\rangle_A|n\rangle_B,
\end{equation}
where $\lambda={\rm tanh}r\in\left [0, 1  \right ]$, while $r$ is the parameter related squeezing and the modes $A$ and $B$ constitute an ideal EPR pair when $r\rightarrow \infty$. Furthermore, $n$ is the number of photons.

\subsubsection{Unitary Operation and Measurement}
\label{sec:Unitary-operation-and-measurement}
A quantum system in the process of quantum communication usually undergoes unitary operations and measurement, which carry out information/data encoding and decoding, as well as the observation of the result, respectively. The terminologies of `observation' and `measurement' are used as synonyms and upon measurement a qubit could be converted into a classical bit. A unitary operation is represented by a complex-valued matrix $U$ that satisfies the condition of $U^{\dagger}U=I$, and transforms a state vector into another state vector, which is formulated as:
\begin{equation}
|\psi'\rangle=U|\psi\rangle.
\end{equation}
It is also often written in terms of a sum-of-outer-products given by
\begin{equation}
U=\sum_{ij}M_{ij}|\psi_{i}\rangle\langle\psi_{j}|,
\end{equation}
where $M_{ij}=\langle\psi_{i}|U|\psi_{j}\rangle$ is the matrix element of $U$ between the two basis states.

\begin{figure*}[htbp]
\centering
\subfigure[$|0\rangle\overset{I}{\rightarrow}|0\rangle$.]{
\begin{minipage}[t]{0.3\linewidth}
\centering
\includegraphics[width=2.3in]{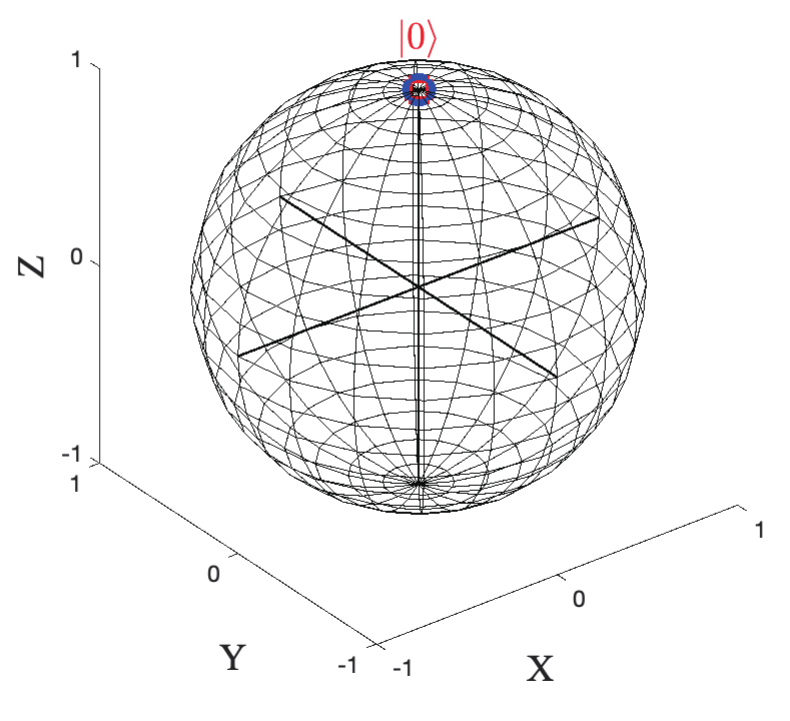}
\end{minipage}%
}%
\subfigure[$|+\rangle\overset{Z}{\rightarrow}|-\rangle$.]{
\begin{minipage}[t]{0.3\linewidth}
\centering
\includegraphics[width=2.3in]{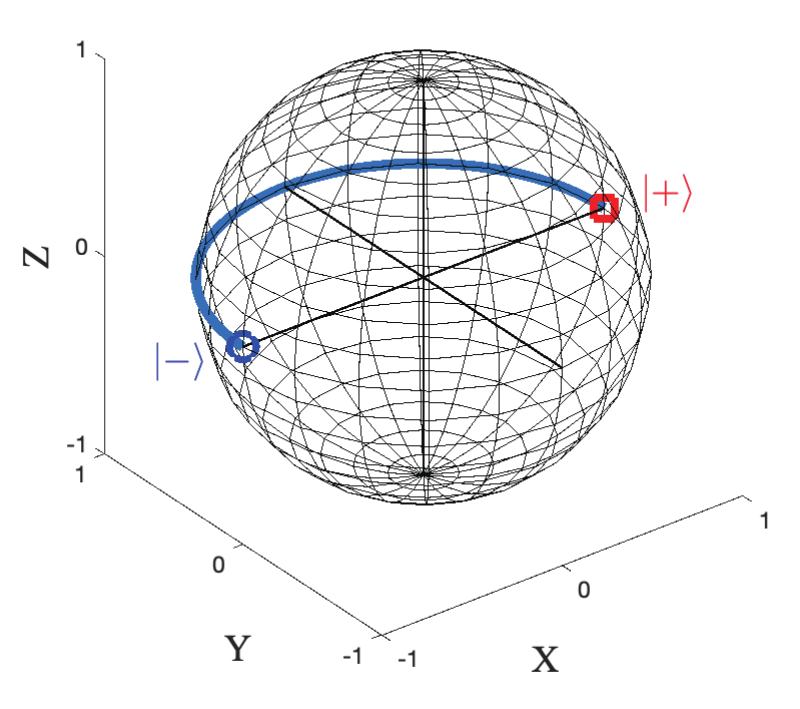}
\end{minipage}%
}%
\subfigure[$|0\rangle\overset{X}{\rightarrow}|1\rangle$.]{
\begin{minipage}[t]{0.3\linewidth}
\centering
\includegraphics[width=2.3in]{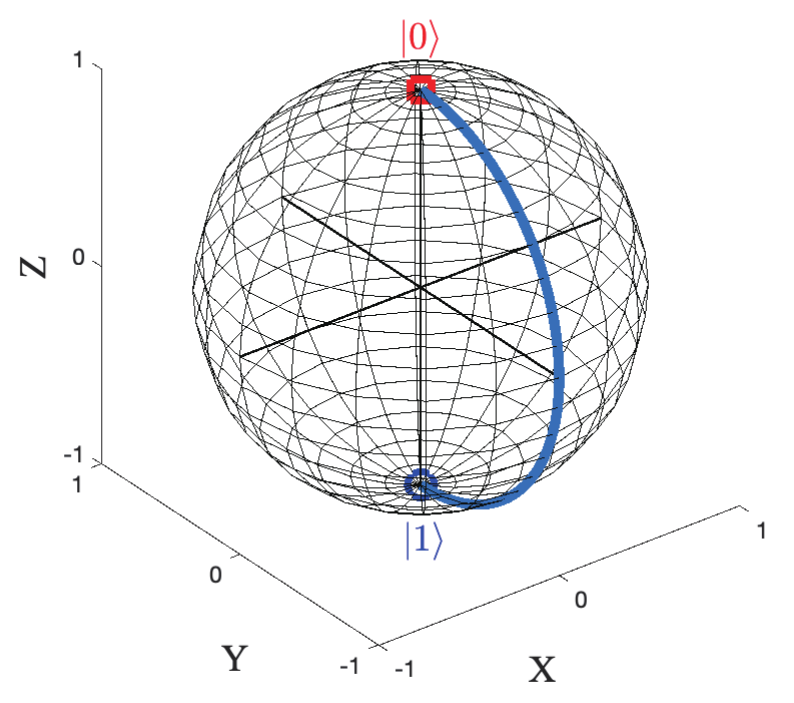}
\end{minipage}
}%

\subfigure[$|+\rangle\overset{Y}{\rightarrow}|-\rangle$.]{
\begin{minipage}[t]{0.3\linewidth}
\centering
\includegraphics[width=2.3in]{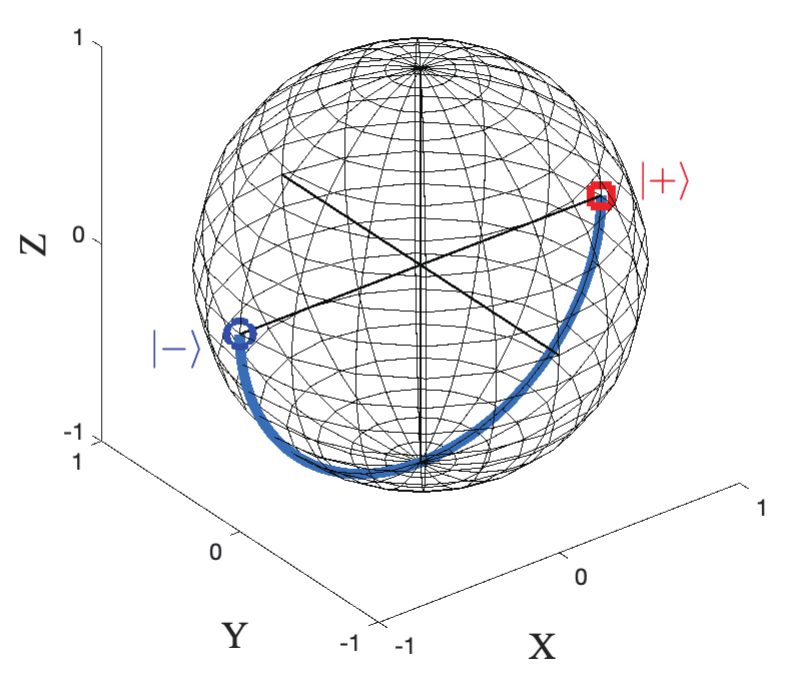}
\end{minipage}
}%
\subfigure[$|0\rangle\overset{H}{\rightarrow}|+\rangle$.]{
\begin{minipage}[t]{0.3\linewidth}
\centering
\includegraphics[width=2.3in]{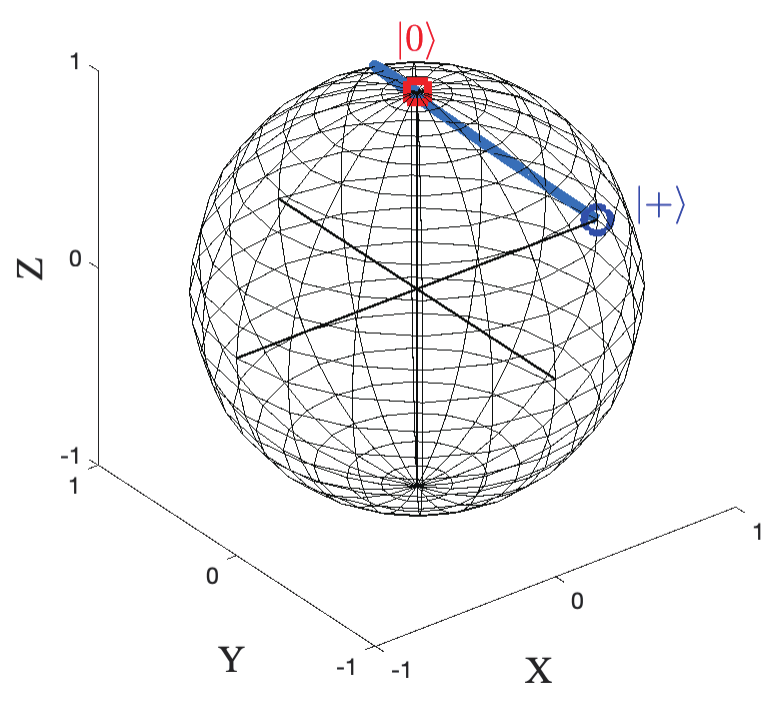}
\end{minipage}
}%
\centering
\caption{The trajectory examples of quantum state under the specific unitary operation on the Bloch sphere. The red squares represent the initial states and the blue circles represent the final states, while the blue solid lines are the trajectories.\label{fig-6:operationBlochSphere}}
\end{figure*}

The most commonly used unitary operations of the quantum communication protocols, which transform single-qubit states are as follows,
\begin{eqnarray}
U_{0}=I=|0\rangle\langle0|+|1\rangle\langle1|=\begin{pmatrix}1 &0 \\ 0& 1\end{pmatrix},\nonumber\\
U_{1}=Z=\sigma_{z}=|0\rangle\langle0|-|1\rangle\langle1|=\begin{pmatrix}1 & 0 \\ 0& -1\end{pmatrix},\nonumber\\
U_{2}=X=\sigma_{x}=|1\rangle\langle0|+|0\rangle\langle1|=\begin{pmatrix}0 & 1 \\ 1& 0\end{pmatrix},\nonumber\\
U_{3}=i\sigma_{y}=|0\rangle\langle1|-|1\rangle\langle0|=\begin{pmatrix}0 & 1 \\ -1& 0\end{pmatrix},\nonumber\\
H=\frac{X+Z}{\sqrt{2}}=|+\rangle\langle0|+|-\rangle\langle1|=\frac{1}{\sqrt{2}}\begin{pmatrix}1 & 1 \\ 1& -1\end{pmatrix}.
\end{eqnarray}
They play quite different roles. Specifically, $U_{0}$ is an identity transform that has no effect on the state, while $U_{1}$ represents the phase flip, which can be written as
\begin{equation}
\label{eq:Unitaryoperation}
U_1|1\rangle=\begin{pmatrix}1 & 0 \\ 0& -1\end{pmatrix}\begin{pmatrix}0\\ 1\end{pmatrix}=\begin{pmatrix}0\\ -1\end{pmatrix}=-|1\rangle.
\end{equation}
Hence, $|0\rangle\overset{U_{1}}{\rightarrow}|0\rangle$,  $|1\rangle\overset{U_{1}}{\rightarrow}-|1\rangle$. Furthermore, using a process similar to Eq.~(\ref{eq:Unitaryoperation}), we can infer that $U_{2}$ is the bit-flip, $|0\rangle\overset{U_{2}}{\rightarrow}|1\rangle$, $|1\rangle\overset{U_{2}}{\rightarrow}|0\rangle$ and finally, $U_{3}$ represents a simultaneous bit-flip and phase-flip, $|1\rangle\overset{U_{3}}{\rightarrow}|0\rangle$, $|0\rangle\overset{U_{3}}{\rightarrow}-|1\rangle$. The operation $H$ is the Hadamard transformation, which may also be represented as, $|0\rangle\overset{H}{\rightarrow}|+\rangle\overset{H}{\rightarrow}|0\rangle$,  $|1\rangle\overset{H}{\rightarrow}|-\rangle\overset{H}{\rightarrow}|1\rangle$. The effects of these unitary operations can be conveniently illustrated on the Bloch Sphere of Fig. \ref{fig-6:operationBlochSphere} for better understanding. 

In entanglement-based quantum communication, we can apply any of the operations $U_{0}$, $U_{1}$, $U_{2}$, or $U_{3}$ to one of the particles of an EPR pair while keeping the other particle untouched, the original Bell state can be turned into another Bell state. The transformation between them is described in Table \ref{table-3:UBellstate}.
\begin{table}
\begin{footnotesize}
\begin{center}
\caption{The relationship between the initial Bell states, the final Bell states and the corresponding unitary operator.}
\begin{tabular}{|m{2cm}||m{1.1cm}<{\centering}|m{1.1cm}<{\centering}|m{1.1cm}<{\centering}|m{1.1cm}<{\centering}|}
\hline
 & \multicolumn{4}{c|}{Initial states} \\ \hline
 Unitary operation
                    & $|\psi^{+}\rangle$ & $|\psi^{-}\rangle$   & $|\phi^{+}\rangle$ & $|\phi^{-}\rangle$ \\ \hline\hline
$U_{0}=I$           & $|\psi^{+}\rangle$ & $|\psi^{-}\rangle$   & $|\phi^{+}\rangle$ & $|\phi^{-}\rangle$ \\ \hline
$U_{1}=\sigma_{z}$  & $|\psi^{-}\rangle$ & $|\psi^{+}\rangle$   & $|\phi^{-}\rangle$ & $|\phi^{+}\rangle$ \\ \hline
$U_{1}=\sigma_{x}$  & $|\phi^{+}\rangle$ & $-|\phi^{-}\rangle$  & $|\psi^{+}\rangle$ & $-|\psi^{-}\rangle$ \\ \hline
$U_{3}=i\sigma_{y}$ & $|\phi^{-}\rangle$ & $-|\phi^{+}\rangle$  & $|\psi^{-}\rangle$ & $-|\psi^{+}\rangle$ \\ \hline
\end{tabular}
\label{table-3:UBellstate}
\end{center}
\end{footnotesize}
\end{table}

Quantum measurements are described by a collection of measurement operators $\{M_{m}\}$ that act on the quantum state, and the index $m$ identifies one of the legitimate outcomes of the measurement. If the system  to be measured is in state $|\psi\rangle$, then the probability of obtaining the result $m$ is given by
\begin{eqnarray}
p\left(m\,|\,|\psi\rangle\right)=\langle\psi|M_{m}^{\dagger}M_{m}|\psi\rangle.\label{e15}
\end{eqnarray}
After measurement, the system collapses to the state
\begin{eqnarray}
\label{Eq:measurement}
|\psi'\rangle=\frac{M_{m}|\psi\rangle}{\sqrt{\langle\psi|M_{m}^{\dagger}M_{m}|\psi\rangle}}.
\end{eqnarray}
The set of measurement operators $\{M_{m}\}$ must satisfy the completeness relationship of $\sum_{m}M_{m}^{\dagger}M_{m}=I$, which results from the fact that the sum of the probabilities $p\left(m\,|\,|\psi\rangle\right)$ is equal to 1.

When we invoke a measurement in the process of quantum communication, we always use sets composed of orthonormal computational bases and the measurement operator is constructed from them. There is a very useful rule of thumb: the measurement operators can be produced in the form of
\begin{eqnarray}
M_{m}=|\psi_{m}\rangle\langle\psi_{m}|,
\end{eqnarray}
according to a set of orthonormal states $\{|\psi_{m}\rangle\}$ \cite{imre2005quantum}. This is a special kind of projective measurement, which is also commonly referred to as the von Neumann measurement~\cite{nielsen2002quantum}. For example, when aiming for determining whether a qubit is in state $|0\rangle$  or $|1\rangle$, the corresponding measurement operators are $\{M_{0}=|0\rangle\langle0|, M_{1}=|1\rangle\langle1|\}$, respectively. According to Eq. (\ref{e15}), the result $|0 \rangle$ will appear with the probability of $p(0\,|\,|\psi\rangle)=\left(\alpha^{*}\langle0|+\beta^{*}\langle1|\right)|0\rangle\langle0|\left(\alpha|0\rangle+\beta|1\rangle\right)=|\alpha|^{2}$, while the result $|1\rangle$ will appear with probability $|\beta|^{2}$.
The state will collapse to the new state of either $|0\rangle$ or $|1\rangle$.

The positive-operator-valued measurement defined in \cite{nielsen2002quantum} is another commonly used notion in quantum information processing, which represents a generalized measurement, because the operators are not necessarily orthogonal. It is described by a set of positive operators $\{E_{m}\}$, formulated as,
\begin{equation}
E_{m}=M_{m}^{\dagger}M_{m},
\end{equation}
which satisfy the completeness condition of $\sum_{m}E_{m}=I$. Therefore, we only care about the probability of getting the specific results $m$, which is given by
\begin{eqnarray}
p\left(m\,|\,|\psi\rangle\right)=\langle\psi|E_{m}|\psi\rangle,
\end{eqnarray}
because we are unable to predict the post-measurement state of the system after carrying out the positive-operator-valued measurement. Fortunately, the post-measurement state is of limited interest in quantum information processing, since most applications are more concerned with the measurement outcomes and with their specific probabilities. 

Let us briefly consider an example of using the positive-operator-valued measurement as a means of distinguishing a pair of nonorthogonal states $|0\rangle$ and $|+\rangle=(|0\rangle+|1\rangle)/\sqrt{2}$. We start by constructing a positive-operator-valued measurement containing three operators
\begin{eqnarray}
E_{1}=\frac{\sqrt{2}}{1+\sqrt{2}}|1\rangle\langle1|, \\
E_{2}=\frac{\sqrt{2}}{1+\sqrt{2}}\frac{(|0\rangle-|1\rangle)(\langle0|-\langle1|)}{2}, \\
E_{3}=I-E_{1}-E_{2}.
\end{eqnarray}
Then $|0\rangle$ is easy to distinguish from $|+\rangle$ with the aid of the measurement outcomes of $\langle0|E_{1}|0\rangle=\langle+|E_{2}|+\rangle=0$ and $\langle+|E_{1}|+\rangle$ as well as $\langle0|E_{2}|0\rangle$ being nonzero: upon measuring a state $|\psi\rangle$ with the aid of $E_1$ and $E_2$, the result will be state $|0\rangle$ if the results are $\langle\psi|E_1|\psi\rangle=0$ and $\langle\psi|E_2|\psi\rangle$ nonzero. By contrast, it will be state $|+\rangle$, if the result of $\langle\psi|E_1|\psi\rangle$ is nonzero and $\langle\psi|E_2|\psi\rangle=0$. To elaborate, if we carry out the positive-operator-valued measurement and obtain a result for $E_1$, the outcome is state $|+\rangle$, while it will be state $|0\rangle$ if the result is obtained in $E_2$. Finally, $E_{3}$ is needed due to the completeness condition, although the measurement result of $E_{3}$ is useless: both $\langle0|E_{3}|0\rangle$ and $\langle+|E_{3}|+\rangle$ are nonzero, hence we cannot tell whether the state is $|0\rangle$ or $|+\rangle$ from the $E_3$ positive-operator-valued measurement. Note that the physical laws tell us that nonorthogonal quantum states cannot be distinguished with perfect reliability, but it is possible to distinguish the states some fraction of the time \cite{huttner1996unambiguous,eldar2001quantum,sun2002optimum}. Therefore, both the legitimate users \cite{B92} and the attackers \cite{slutsky1998security} can employ this tool for exacting information. By applying positive-operator-valued measurements to quantum cryptography, the security of QKD was analyzed in the face of certain eavesdropping strategies related to positive-operator-valued measurement \cite{ekert1994eavesdropping,lutkenhaus1996security,biham2002security}, and the maximum attainable information rate was calculated in \cite{fuchs1996quantum,fuchs1997nonorthogonal}.

Heisenberg formulated the \textit{uncertainty principle} in 1927~\cite{heisenberg1927anschaulichen}, which is a fundamental principle of quantum mechanics. It states that in certain pairs of physical properties, such as position and momentum or energy and time, it is impossible to simultaneously know or measure their exact values. The security of CV quantum communication is precisely based on this fundamental principle of quantum mechanics. For instance, Alice randomly encodes the key onto a pair of physical properties in CV QKD~\cite{ralph1999continuous,hillery2000quantum}, known as the quadrature components, and the uncertainty principle indicates that these two components cannot be simultaneously determined with sufficiently high precision. If Eve eavesdrops on the encoded quantum states of the key, she will inevitably introduce some  perturbation, which will be detected by the communicating parties.

\subsubsection{No-Cloning Theorem}
\label{sec:No-cloning theorem}
In contrast to classical communication where information can in principle be copied perfectly without limits, an eavesdropper is incapable of copying quantum signals in quantum communication, thanks to the no-cloning theorem \cite{wootters1982single,dieks1982communication,milonni1982photons}.

Let us assume that there exists a quantum cloning device capable of perfectly duplicating the arbitrary quantum states $|\psi\rangle$ and $|\phi\rangle$, where the cloning can be realized by a unitary operation 
\begin{eqnarray}
U_{C}\left(|\psi\rangle|0\rangle\right)&=&|\psi\rangle|\psi\rangle,\nonumber\\
U_{C}\left(|\phi\rangle|0\rangle\right)&=&|\phi\rangle|\phi\rangle,
\label{eq:clone}
\end{eqnarray}
where $|0\rangle$ is the initial state of the cloning device. The inner product between the right-hand sides of Eq. (\ref{eq:clone}) is
\begin{eqnarray}
\langle\phi|\langle\phi|\psi\rangle|\psi\rangle=\langle\phi|\psi\rangle\langle\phi|\psi\rangle=\langle\phi|\psi\rangle^{2},
\end{eqnarray}
while for the left-hand side we obtain
\begin{eqnarray}
\langle\phi|\langle0|U_{C}^{\dagger}U_{C}|\psi\rangle|0\rangle=\langle\phi|\psi\rangle\langle0|0\rangle=\langle\phi|\psi\rangle,
\end{eqnarray}
where the relationships of $U_{C}^{\dagger}U_{C}=I$ and $\langle0|0\rangle$=1 are exploited. Finally, we arrive at:
\begin{eqnarray}
\langle\phi|\psi\rangle\left(\langle\phi|\psi\rangle-1\right)=0,
\end{eqnarray}
then $\langle\phi|\psi\rangle=0$ or $\langle\phi|\psi\rangle=1$, which implies that a cloning device can only clone $|\phi\rangle$ that is orthogonal to $|\psi\rangle$. For example, even if a person has a cloning device that can faithfully copy $|0\rangle$ and $|1\rangle$, it cannot perfectly copy $|+\rangle$ and $|-\rangle$, i.e. we have:
\begin{eqnarray}
U_C|+\rangle\!\!\!&=&\!\!\!U_C\frac{1}{\sqrt{2}}(|0\rangle+|1\rangle)\nonumber \\ 
\!\!\!&=&\!\!\!\frac{1}{\sqrt{2}}(|0\rangle|0\rangle+|1\rangle|1\rangle)\nonumber \\ 
\!\!\!&\ne&\!\!\!|+\rangle|+\rangle,
\end{eqnarray}
where $|+\rangle|+\rangle=(|0\rangle|0\rangle+|0\rangle|1\rangle+|1\rangle|0\rangle+|1\rangle|1\rangle)/2$. The security of quantum communication is derived from this property. In the well-known BB84 QKD protocol~\cite{bennett1984quantum}, Alice randomly sends four types of single photons, namely $|0\rangle$, $|1\rangle$, $|+\rangle$ and $|-\rangle$, to Bob to generate a secret key. If an eavesdropper, Eve, wants to successfully eavesdrop on the key, she must obtain a correct copy of each photon and let one copy go to Bob to avoid detection when Alice and Bob later randomly compare the transmitted parts of the quantum states. However, due to the existence of the no-cloning theorem, Eve cannot successfully complete her eavesdropping action.

As a conclusion, it is claimed in~\cite{wootters1982single} that unknown quantum states cannot be cloned (copied) perfectly. This argument has also been extended to mixed states in \cite{barnum1996noncommuting}. However approximate cloning, or probabilistic cloning is possible for an arbitrary state \cite{buvzek1996quantum,gisin1997optimal,werner1998optimal,cerf2000pauli,lemm2017information,duan1998probabilistic}. Therefore, one cannot gain full information about an unknown quantum state without perturbing it.

Here, we can summarize the fundamental quantum physical principles that underlie the high security of quantum communication, which have no classical counterparts.
\begin{itemize}
\item \textit{Entanglement}: The correlation and non-locality of entangled particles dictates that when an eavesdropper alters the state of one of the entangled particles, the state of the other correlated particle changes correspondingly, thereby making entanglement suitable for verifying the security of communication. Furthermore, when an eavesdropper possesses only one of the entangled particles, they do not gain any useful information.
\item \textit{Measurement}: An eavesdropper attempts to glean information by measuring the quantum signal. However, when the eavesdropper observes or measures a quantum signal, the quantum state undergoes an irreversible change as indicated by Eq.~(\ref{Eq:measurement}), which is detectable by both communicating parties.
\item \textit{No-cloning theorem}: An eavesdropper cannot perfectly replicate the quantum state sent by Alice to Bob to infer the transmitted information.
\item \textit{Heisenberg uncertainty principle}: In CV quantum communication, the Heisenberg uncertainty principle prohibits Eve from simultaneously determining two components, preventing her from inferring any useful information.
\end{itemize}
It should be noted that in order to convey an intuitive understanding of the origins of quantum communication security, the above discussions do not consider the interrelationships of these quantum physical principles. For example, the no-cloning theorem can be viewed as a manifestation of the  uncertainty principle.

\subsubsection{Entanglement Resources for Information Transmission}
\label{subsec:EntRes}
Superdense coding \cite{bennett1992communication}, quantum teleportation \cite{bennett1993teleporting,cacciapuoti2020entanglement}, and entanglement swapping \cite{zukowski1993event} are important quantum-domain operations that rely on EPR pairs. These quantum techniques can be employed for the design of quantum communication protocols as exemplified in~\cite{zhan2009quantum,li2009fault,hwang2011quantum,bandyopadhyay2000teleportation,zhou2011quantum,cheng2005quantum}. We will therefore focus our attention on the basic principles of these approaches for highlighting some of the pivotal protocols. Entanglement purification is another salient quantum communications technique conceived for mitigating the degradation of entanglement, thereby guaranteeing security.

\begin{figure}[!h]
\begin{center}
\includegraphics[width=\columnwidth]{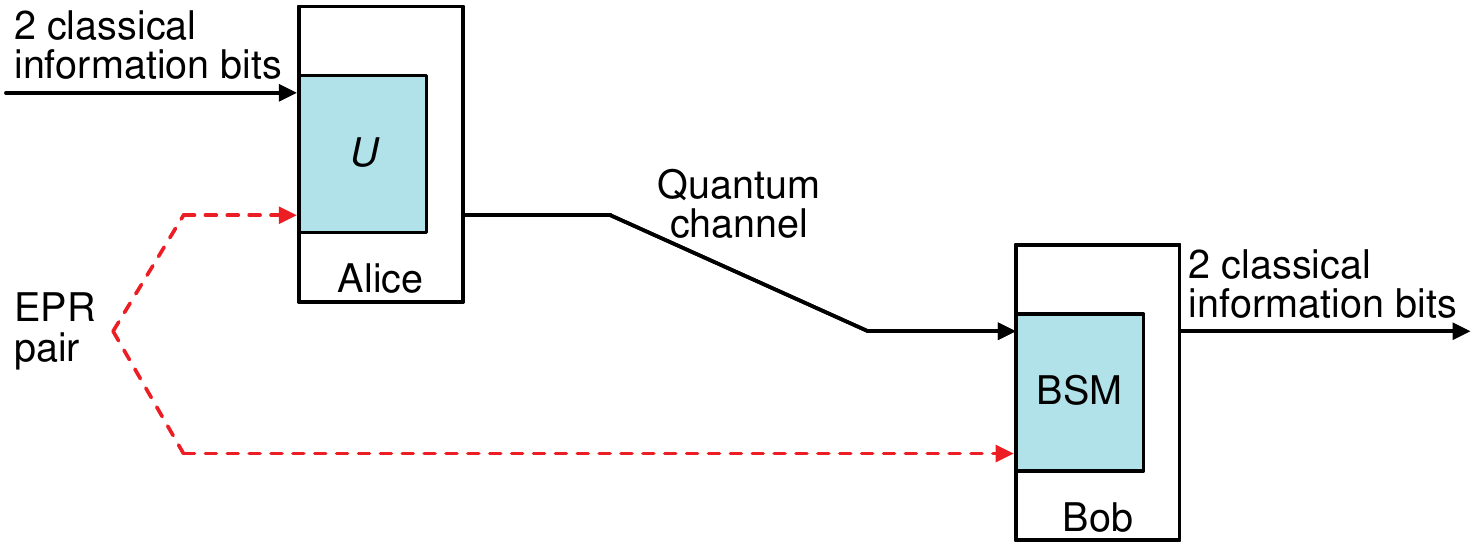}
\caption{Communication model of superdense coding, where the dashed line indicates entanglement and BSM represents the Bell-state measurement.}
\label{fig:SD}
\end{center}
\end{figure}

Fig. \ref{fig:SD} graphically illustrates \textit{superdense coding}, as an effective means of communications, which conveys two bits of classical information from Alice to Bob by sending only a single qubit. 
Let us assume that the EPR pair of Fig. \ref{fig:SD} has been shared by Alice and Bob, so both of them hold one of the particles representing the state $|\psi^{-}\rangle$. Table \ref{table-3:UBellstate} suggests that all four Bell states may be gleaned from either one of them by applying only local operations to one of the particles of Bell states. Consequently, Alice is able to encode two bits of classical information onto a single qubit by applying the unitary operator $U$ of Fig. \ref{fig:SD} according to the following coding rules:
\begin{eqnarray}
|\psi^{-}\rangle_{AB}\xrightarrow [U_{0}=I]{00}|\psi^{-}\rangle_{AB},\nonumber\\
|\psi^{-}\rangle_{AB}\xrightarrow [U_{1}=\sigma_{z}]{01}|\psi^{+}\rangle_{AB},\nonumber\\
|\psi^{-}\rangle_{AB}\xrightarrow [U_{2}=\sigma_{x}]{10}-|\phi^{-}\rangle_{AB},\nonumber\\
|\psi^{-}\rangle_{AB}\xrightarrow [U_{1}=i\sigma_{y}]{11}-|\phi^{+}\rangle_{AB}.
\end{eqnarray}
She then sends her qubit to Bob through the quantum channel of Fig. \ref{fig:SD} and Bob combines the two qubits of the EPR pair considered to perform Bell-state measurement (BSM). The pair of original classical bits of Fig. \ref{fig:SD} are then reconstructed deterministically\footnote{The four Bell states encoding 2 bits of classical information can be distinguished by nonlinear optics based BSM~\cite{kim2001quantum}.}, because the measurement result will unambiguously reveal the state.

\begin{figure}[!h]
\begin{center}
\includegraphics[width=\columnwidth,angle=0]{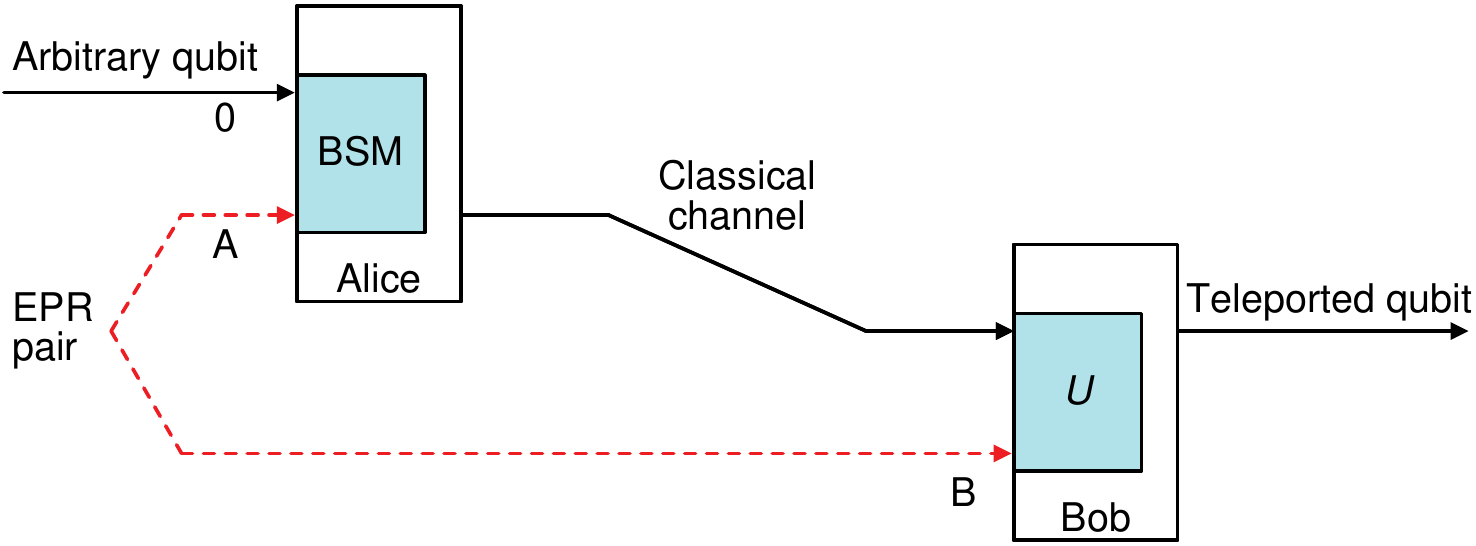}
\caption{Principle of quantum teleportation.}
\label{fig:QT}
\end{center}
\end{figure}

For \textit{quantum teleportation}, the aim is to faithfully deliver an arbitrary qubit $|\psi\rangle_{0}=\alpha|0\rangle+\beta|1\rangle$ between two distant parties, as seen in Fig. \ref{fig:QT}. As in the superdense coding scenario of Fig. \ref{fig:SD}, an EPR pair must be shared by Alice and Bob in advance. Without loss of generality, we assume that the shared EPR state is $|\psi^{-}\rangle_{AB}=1/\sqrt{2}\left(|0\rangle_{A}|1\rangle_{B}-|1\rangle_{A}|0\rangle_{B}\right)$. So the resultant 3-qubit state here is initially $|\psi\rangle_{0}|\psi^{-}\rangle_{AB}$, which can be regrouped and written as
\begin{eqnarray}
|\psi\rangle_{0}|\psi^{-}\rangle_{AB}&=&\!\!\!\!\frac{1}{2}[|\psi^{-}\rangle_{0A}\left(-\alpha|0\rangle-\beta|1\rangle\right)_{B}\nonumber\\
&+&\!\!\!\!|\psi^{+}\rangle_{0A}\left(-\alpha|0\rangle+\beta|1\rangle\right)_{B}\nonumber\\
&+&\!\!\!\!|\phi^{-}\rangle_{0A}\left(\alpha|1\rangle+\beta|0\rangle\right)_{B}\nonumber\\
&+&\!\!\!\!|\phi^{+}\rangle_{0A}\left(\alpha|1\rangle-\beta|0\rangle\right)_{B}].
\end{eqnarray}
If Alice performs a BSM on qubits 0 and $A$ at her side, the measurement will project these two qubits onto one of the four Bell states of Table \ref{table-4:QT} with an equal probability of $1/4$. The measurement outcome is then sent to Bob over a classical channel, hence he will get to know the state of his qubit instantly. Depending on the relationship in Table \ref{table-4:QT}, Bob selects the specific unitary operation that transforms the state of qubit $B$ into the teleported state $|\psi\rangle_{0}$. Thus, the qubit containing the quantum information has been teleported from Alice to Bob. Observe the dual relationship between superdense coding and quantum teleportation by comparing Fig. \ref{fig:SD} and Fig. \ref{fig:QT}. More formally, Werner's proof~\cite{werner2001all} shows that two parties can swap their equipment to convert quantum teleportation into superdense coding under certain conditions, and vice versa.
\begin{table*}[h!]
\begin{footnotesize}
\begin{center}
\caption{Teleporting an arbitrary qubit $|\psi\rangle_{0}=\alpha|0\rangle+\beta|1\rangle$ with EPR state $|\psi^{-}\rangle_{AB}$.}
\begin{tabular}{|l|l|l|l|}
    \hline
    BSM outcome & State of qubit $B$ & Operation & Teleported state \\ \hline
$|\psi^{-}\rangle_{0A}$ & $-\left(\alpha|0\rangle+\beta|1\rangle\right)_{B}$   & $U_{0}=I$   & $-\left(\alpha|0\rangle+\beta|1\rangle\right)_{B}$                \\
$|\psi^{+}\rangle_{0A}$  & $\left(-\alpha|0\rangle+\beta|1\rangle\right)_{B}$  & $U_{1}=\sigma_{z}$  & $-\left(\alpha|0\rangle+\beta|1\rangle\right)_{B}$                \\
$|\phi^{-}\rangle_{0A}$ & $\left(\alpha|1\rangle+\beta|0\rangle\right)_{B}$  & $U_{2}=\sigma_{x}$  & $\left(\alpha|0\rangle+\beta|1\rangle\right)_{B}$                \\
$|\phi^{+}\rangle_{0A}$ & $\left(\alpha|1\rangle-\beta|0\rangle\right)_{B}$              & $U_{3}=i\sigma_{y}$        & $\left(\alpha|0\rangle+\beta|1\rangle\right)_{B}$                \\ \hline
\end{tabular}
\label{table-4:QT}
\end{center}
\end{footnotesize}
\end{table*}

\begin{figure}[!h]
\begin{center}
\includegraphics[width=\columnwidth,angle=0]{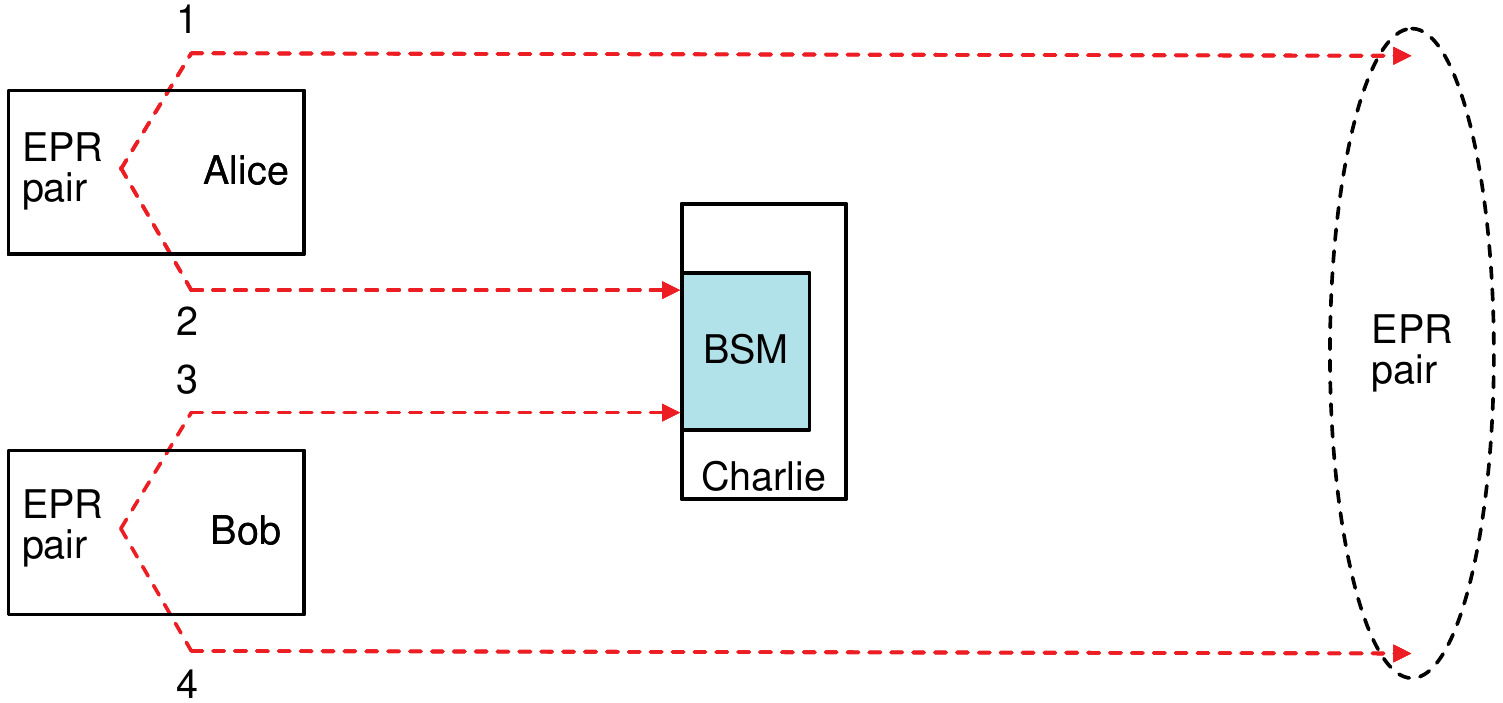}
\caption{Entanglement swapping process. The BSM applied to particles 2 and 3 immediately projects particles 1 and 4 into an EPR pair.}
\label{fig:ES}
\end{center}
\end{figure}

\textit{Entanglement swapping} has the capability of entangling a pair of distant quantum systems that have never been connected in the past\cite{pan1998experimental}. Figure \ref{fig:ES} shows the process of entanglement swapping. Consider a pair of entangled states $|\psi^{-}\rangle_{12}=1/\sqrt{2}\left(|0\rangle_{1}|1\rangle_{2}-|1\rangle_{1}|0\rangle_{2}\right)$ and $|\psi^{-}\rangle_{34}=1/\sqrt{2}\left(|0\rangle_{3}|1\rangle_{4}-|1\rangle_{3}|0\rangle_{4}\right)$, which are generated simultaneously.
We may then pick one photon from each of the two entangled states to make a BSM. As a result, we can immediately see in Fig. \ref{fig:ES} that the measurement 
of particles 2 and 3 projects the original non-entangled particles 1 and 4 into an EPR state. This can be formally expressed as
\begin{eqnarray}
|\psi^{-}\rangle_{12}|\psi^{-}\rangle_{34}\!\!\!\!&=&\!\!\!\!\frac{1}{2}\left(|0\rangle_{1}|1\rangle_{2}-|1\rangle_{1}|0\rangle_{2}\right)\left(|0\rangle_{3}|1\rangle_{4}-|1\rangle_{3}|0\rangle_{4}\right)\nonumber\\
&=&\!\!\!\!\frac{1}{2}(|\psi^{+}\rangle_{14}|\psi^{+}\rangle_{23}-|\psi^{-}\rangle_{14}|\psi^{-}\rangle_{23}\nonumber\\
&+&\!\!\!\!|\phi^{+}\rangle_{14}|\phi^{+}\rangle_{23})-|\phi^{-}\rangle_{14}|\phi^{-}\rangle_{23}).
\end{eqnarray}
Note that the state of newly generated entangled pair is decided by the measurement result, so for example, $|\phi^{+}\rangle_{23}$ yields $|\phi^{+}\rangle_{14}$.

\begin{figure}[!h]
\begin{center}
\includegraphics[width=9cm,angle=0]{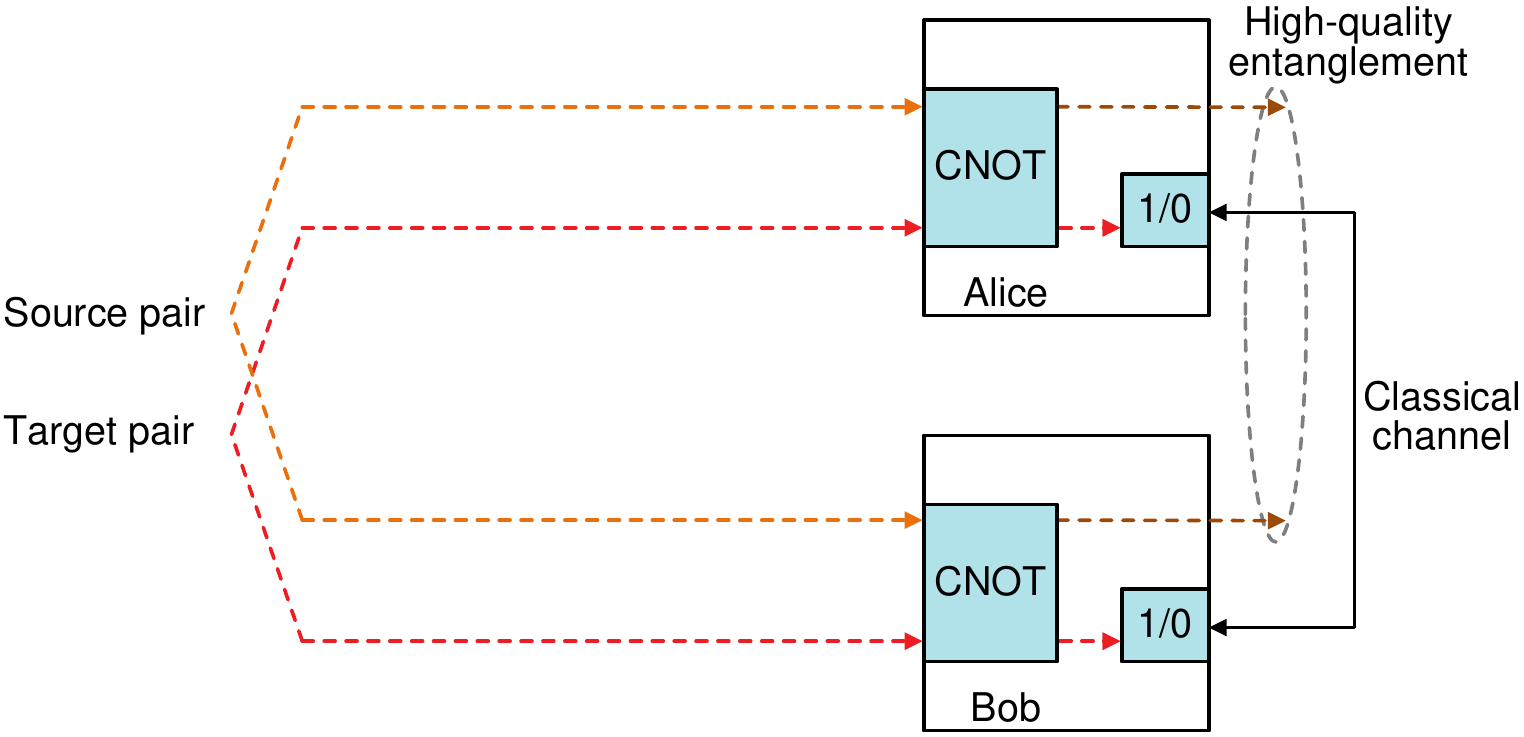}
\caption{The Scheme of entanglement purification by Bennett \textit{et al} \cite{bennett1996purification}. CNOT: Controlled-NOT.}
\label{fig:EP}
\end{center}
\end{figure}

The quality of entangled states decays exponentially upon encountering the unavoidable noise of a quantum channel. This result is absolutely against the original intention of distributing entanglement between a pair of distant nodes without contaminating them. \textit{Entanglement purification} offers a way of mitigating this deleterious effect by extracting a small number of almost perfectly entangled pairs from many poor-quality entangled states with the aid of local operations and classical communications. Figure \ref{fig:EP} shows the original entanglement purification scheme introduced by Bennett \textit{et al}. \cite{bennett1996purification}.
Let us assume having two imperfectly entangled pairs shared by Alice and Bob as seen in Fig. \ref{fig:EP}. One of them is the source pair, which has higher-quality entanglement than the target pair to be purified. Both Alice and Bob apply the Controlled-NOT gate\footnote{In the Controlled-NOT gate the target qubit will be subjected to the NOT operation if and only if the control qubit is in the state $|1\rangle$.} to the particles in their hand. Subsequently, they both measure the qubits of the target pair using the measurement  of Z and compare their measurement outcome by relying on classical communication, as seen in Fig. \ref{fig:EP}. The source pair will be retained as is, because it has a higher degree of entanglement than the original target pair, if their Controlled-NOT outputs shared over the classical channel are the same. Otherwise, the source pair will be discarded. However, since the Controlled-Not gate is difficult to realize experimentally, Pan \textit{et al}. came up with a simpler solution by harnessing a polarizing beam splitter \cite{pan2001entanglement,simon2002polarization}. The benefits of this solution were also demonstrated subsequently in \cite{pan2003experimental}. Some further improved schemes, such as the deterministic entanglement purification protocol were proposed in~\cite{sheng2008efficient,sheng2010deterministic,sheng2012one,zhang2023variational}.

\subsection{Experimental Fundamentals}
\label{sec:Experimental-fundamentals}
\subsubsection{Light Source and Detector}
\label{subsec:LD}
\begin{figure}[!h]
\begin{center}
\includegraphics[width=\columnwidth,angle=0]{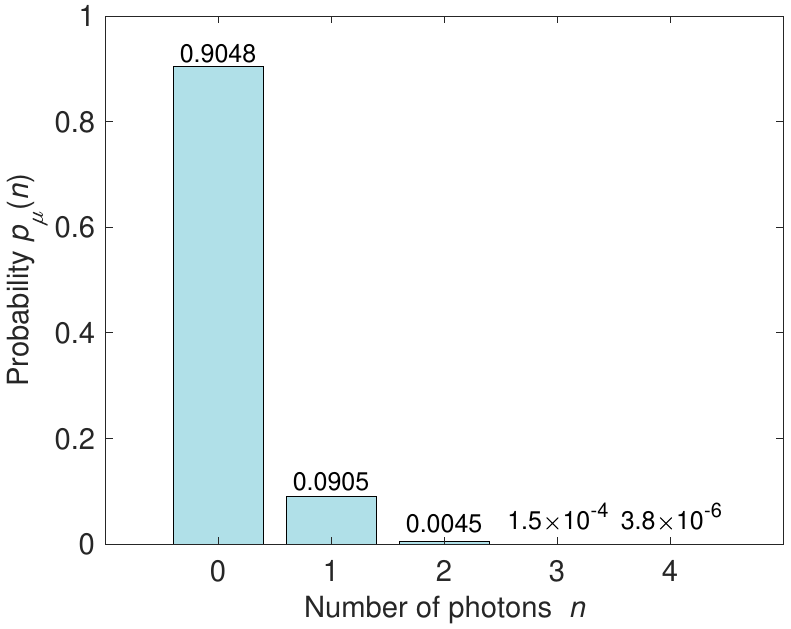}
\caption{Poisson distribution when $\mu=0.1$.}
\label{fig:Poisson}
\end{center}
\end{figure}

In physical implementations, qubits have been realized with the aid of many different systems~\cite{ladd2010quantum}, but for quantum communications the most popular qubit carriers are photons. However, the unconditional security of single-photon based quantum communication requires a \textit{perfect} source~\cite{mayers2001unconditional}, which is difficult to produce. In the experimental implementation, the pragmatic solution is to use a weak coherent pulse as a near-perfect practical single-photon source. The state of a photon emitted by a laser is described by the coherent state $|\alpha\rangle$, which is a superposition of Fock states $|n\rangle$~\cite{walls1994gj},
\begin{eqnarray}
|\alpha\rangle=e^{\frac{-|\alpha|^2}{2}}\sum_{n=0}^{\infty}\frac{\alpha^{n}}{\sqrt{n!}}|n\rangle,
\end{eqnarray}
wherein $\alpha=\sqrt{\mu}e^{i\theta}$, $\sqrt{\mu}$ represents the intensity associated with the average number of photons $\mu$ per pulse and with the phase $\theta$. The probability that a laser pulse contains $n$ photons obeys the Poisson distribution of $p_{\mu}(n)=e^{-|\alpha|^{2}}|\alpha|^{2n}/n!=e^{-\mu}\mu^{n}/n!$~\cite{walls1994gj}. In other words, the practical light source is in a mixture of states $|n\rangle$ with a probability of $p_{\mu}(n)$, rather than obeying the perfect desired single photon Fock state associated with $|n=1\rangle$ (this should not be confused with one of the two states of a qubit). It is plausible that the main components of this state of the laser pulse are the zero-photon vacuum state $|n=0\rangle$ and the single-photon state of $|n=1\rangle$, when the intensity of the laser pulse is attenuated to be sufficiently low. As shown in Fig.~\ref{fig:Poisson}, when the average number of photons is set to $\mu=0.1$, the majority of the pulses emitted by the light source are in the vacuum state $|0\rangle$, while the proportion of single photons $|1\rangle$ is significantly higher than that of multi-photon states $|n\ge2\rangle$.

The weak coherent pulse source has also been proved secure \cite{gottesman2004security}. However, the transmission distances attained remain limited, and the contributions of a fraction of the photons emitted by the laser pulses have to be deleted because they are insecure in the face of the photon number splitting attacks~\cite{gisin2002quantum}. To elaborate a little further, in the photon number splitting attack, Eve captures a photon from each pulse that contains several photons for further eavesdropping action. The attainable distance can be substantially increased by using the so-called decoy state technique \cite{hwang2003quantum,wang2005beating,lo2005decoy}, which improves the overall performance of QKD systems. After the modification, this technique is also applicable to QSDC, enabling resistance to photon number splitting attack and enhancing communication performance~\cite{pan2020experimental,yang2020quantum,liu2021practical,park2022statistical,park2023statistical,sun2023one}. Therefore, combining the weak coherent pulse based and decoy state based solution having the optimal average number of photons $\mu$ constitutes a beneficial practical method for light sources.

At the time of writing, ideal single-photon sources are not yet available \cite{eisaman2011invited}. Alternatively, deterministic single-photon sources and probabilistic single-photon sources are capable of dramatically reducing the relative frequency of vacuum states and multi-photons. Deterministic single-photon sources, as exemplified by quantum dots \cite{fattal2004quantum,bimberg2009quantum,sun2019single} and color centers as detailed in \cite{beveratos2002single,alleaume2004experimental,leifgen2014evaluation}, usually result in a higher energy level first and then emit a single-photon. By contrast, probabilistic single-photon sources, which are also referred to as heralded single-photon sources~\cite{castelletto2008heralded}, generate single photons by measuring one of the photons in an entangled photon pair, which then serves as reference for the generation of a single photon \cite{trifonov2005secure,wang2008experimental}. These two kinds of single-photon sources have the potential of realizing higher information bit rates or longer transmission distances than the weak coherent pulse source~\cite{sun2023one}.

\begin{figure}[!h]
\begin{center}
\includegraphics[width=\columnwidth,angle=0]{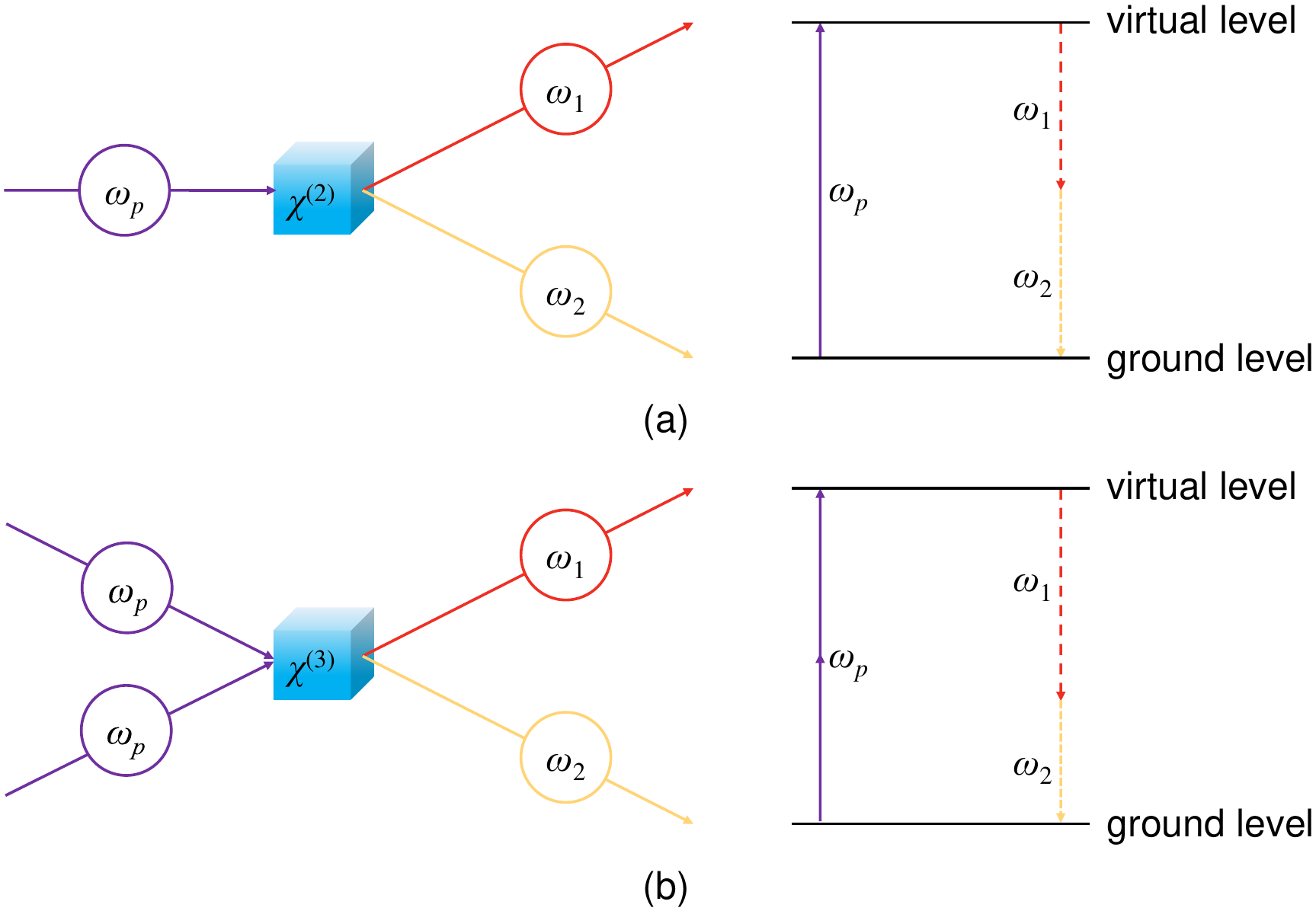}
\caption{Schematic portrayal and energy-level diagram of (a) spontaneous parametric down-conversion and (b) spontaneous four-wave mixing.}
\label{fig:SPDCSFWM}
\end{center}
\end{figure}

The entangled two-photon state is the fundamental resource of many entanglement-based quantum communications techniques. The efforts of generating entanglement originally relied on the so-called atomic system concept of \cite{kocher1967polarization,freedman1972experimental}. As the developments continued, the physical process of spontaneous parametric down-conversion~\cite{migdall2013single} and spontaneous four-wave mixing~\cite{migdall2013single} have been frequently used as the entangled resources. Spontaneous parametric down-conversion is a process of nonlinear interaction, in which a high-frequency pump photon $\omega_{p}$ is converted into a pair of lower-frequency photons $\omega_{1}$ and $\omega_{2}$ (termed as the signal and idler photons), where the pump light illuminates a nonlinear optical crystal characterized by its second-order nonlinear susceptibility $\chi^{(2)}$ as discussed in~\cite{hong1985theory,rubin1994theory,humble2013quantum}, and shown in Fig. \ref{fig:SPDCSFWM} (a). Various types of entanglement, such as OAM entanglement \cite{wang2020satellite}, time-energy entanglement \cite{Steven2016high}, and polarization entanglement~\cite{Ji2017high}, can be generated with the aid of spontaneous parametric down-conversion, which constitute a popular choice for preparing entanglement, since they are relatively simple to construct and hence inexpensive. They have been used as light sources in quantum communications including QKD \cite{ursin2007entanglement} over 140 kilometers, and even for 100+ kilometers for quantum teleportation~\cite{ma2012quantum,ren2017ground}. However, the energy conservation constraints of spontaneous parametric down-conversion formulated as $\hbar\omega_{p}=\hbar\omega_{1}+\hbar\omega_{2}$ and the momentum conservation (also termed as phase-matching) expressed as $\hbar\bm{k}_{p}=\hbar\bm{k}_{1}+\hbar\bm{k}_{2}$ results in the phenomenon that the emitted down-converted photons generated by conventional bulk crystal sources have a cone-shaped spatial multi-mode structure surrounding the pump laser. Hence it is quite challenging to collect and guide the light into single-mode fibers for quantum information processing and transmission \cite{bovino2003effective}. To circumvent this problem, waveguide based spontaneous parametric down-conversion schemes~\cite{banaszek2001generation,sanaka2001new,booth2002counterpropagating} have also been developed.

The spontaneous four-wave mixing sources have also received much attention as another popular candidate for directly generating entangled photons in a single-mode waveguide~\cite{fiorentino2002all,takesue2005generation,fan2007broadband,silverstone2014chip}. As shown in Fig. \ref{fig:SPDCSFWM} (b), a pair of two pump photons is annihilated and a pair of correlated photons is created in the process of spontaneous four-wave mixing by a nonlinear optical medium, characterized by its 3rd-oder nonlinear susceptibility $\chi^{(3)}$ in which both energy conservation $2\hbar\omega_{p}=\hbar\omega_{1}+\hbar\omega_{2}$ and momentum conservation $2\hbar\bm{k}_{p}=\hbar\bm{k}_{1}+\hbar\bm{k}_{2}$ are observed. This process has some distinct advantages over spontaneous parametric down-conversion. Firstly, the resultant entangled pairs are generated in a single spatial mode, making their collection and delivery quite efficient. Hence they can be directly integrated with existing optical-fiber communication networks \cite{fiorentino2002all}. Secondly, spontaneous four-wave mixing exhibits a high brightness facilitated by its long interaction duration and high transverse mode confinement as a benefit of its limited cross-section~\cite{fan2007broadband}. However, spontaneous four-wave mixing requires higher pump power than spontaneous parametric down-conversion because its $\chi^{(3)}$ nonlinearity is weaker than $\chi^{(2)}$. The Raman scattering noise inflicted by its strong pump field must be mitigated in the spontaneous four-wave mixing source, for example by cooling the fiber in liquid nitrogen \cite{takesue20051}. Thus the experimental difficulties increase accordingly.

The single-photon detector constitutes the link between the quantum domain and classical domain, which converts the quantum signals into electrical signals for information detection. At the time writting three popular detectors are used extensively in the experimental implementation of quantum communication: InGaAs/InP avalanche photodiodes, Si avalanche photodiodes, and superconducting single-photon detectors. Six key parameters are routinely used for characterizing the performance of single-photon detectors, including their detection efficiency, dark count rate, dead time, spectral range, time jitter and the ability to distinguish the number of photons~\cite{hadfield2009single}. In a certain spectral range, a perfect single-photon detector must have 100\% detection efficiency, the ability to determine the number of impinging photons, while having all the remaining parameters mentioned above as 0. InGaAs/InP avalanche photodiodes are typically used for detection at telecom wavelengths~\cite{kumar2008two} (typically in 1550 nm and 1310 nm). However, they tend to have a relatively low detection efficiency of around 10\%$\sim20\%$. Naturally, the designer has to strike tradeoffs amongst the key parameters of practical InGaAs/InP avalanche photodiodes. For example, increasing the bias will increase the detection efficiency, but it will also exacerbate the dark count rate \cite{itzler2007single}. Having said that, InGaAs/InP avalanche photodiodes exceeding 50\% detection efficiency are becoming available, which are capable of striking much improved performance tradeoffs \cite{restelli2013single,comandar2015gigahertz}. 

By contrast, Si avalanche photodiodes attain a detection efficiency of more than 60\% at a low dark count rate at specific wavelengths of the visible~\cite{thomas2010efficient} and near-infrared domain~\cite{yang2019spaceborne}, which are eminently suitable for free-space quantum communication \cite{buttler1998practical,hughes2002practical}. Finally, superconducting single-photon detectors also perform well in the visible to mid-infrared wavelength domain. Specifically, they exhibit a high detection efficiency (>90\%) and very low dark count rate (<1 cps), low timing jitter (<100 ps) and short reset time (<100 ns) \cite{marsili2013detecting}. Upon considering each of the above parameters individually an even better performance may be attained~\cite{smirnov2018nbn,shibata2015ultimate,korzh2020demonstration}. Hence they have become one of the most sought after devices for high-performance QSDC~\cite{zhang2022realization}. Most recently, the multipixel superconducting nanowire single-photon detector was shown to support a QKD rate of 100 Mbps~\cite{li2023high,grunenfelder2023fast}. However, superconducting single-photon detector may be deemed excessively costly for conventional applications, because they must be operated at an extremely low temperature of a few Kelvin.

In CV quantum communication, the quadrature components of the optical field can be detected using homodyne and heterodyne detectors, employing a pair of photodiodes as the detection apparatus. These detectors offer the advantage of operating at room temperature, high detection efficiency, and cost-effectiveness. Table~\ref{tab:Comparison of various detectors} provides a comparison among the different detectors.

\begin{table*}[h!]
\begin{center}
\footnotesize
\caption{Comparison of various detectors.}
\label{tab:Comparison of various detectors}
\begin{tabular}{|m{3cm}|c|c|m{3.5cm}|c|}
\hline
\multicolumn{1}{|c|}{Category}        & System          &Typical detection efficiency  & Repetition frequency or bandwidth & Cost   \\ \hline
InGaAs/InP avalanche photodiode        & Discrete variable   & 1550nm@\textgreater{}20\%~\cite{fan2023ultra,he20232}     & \multicolumn{1}{|c|}{GHz}   & Medium \\ \hline
Si avalanche photodiode               & Discrete variable   & 800nm@$\sim$60\%~\cite{yang2019spaceborne}            & \multicolumn{1}{|c|}{GHz} & Medium \\ \hline
Superconducting single photon detector & Discrete variable   & \textgreater{}90\%          & \multicolumn{1}{|c|}{GHz} & High   \\ \hline
Homodyne detector                    & Continuous variable & 1550nm@61.34\%~\cite{zhang2020long}, 1550nm@97.2\%~\cite{tian2022experimental} & \multicolumn{1}{|c|}{GHz~\cite{zhang20181}} & Low   \\ \hline
\end{tabular}
\end{center}
\end{table*}

\subsubsection{Optical Components and Channels}
\label{sec:componentsandchannel}
Optical components, such as beam splitters (BS), polarization beam splitters (PBS), polarization controllers, wave plates, mirrors, Faraday mirrors (FR), phase modulators (PM), attenuators, circulators, intensity modulators, wavelength division multiplexing, and so on, are commonly used in various quantum communication systems \cite{humble2013quantum,qi2019implementation}. The maturity of these devices is conducive for quantum communication.

The quantum channel is a link between Alice and Bob provided for the transmission of quantum information. Basically, there are two popular choices of transmission channels, namely optical fibers and free-space optical (FSO) channels\footnote{It important to note that apart from these practical links, the quantum decoherence effects of quantum signal processing operations are typically modelled by a quantum depolarizing channel. This might be viewed as counterpart of the AWGN channel of classical systems, which characterizes the noise-level in the receiver.}. Optical fiber links have a very low channel loss of about 0.2 dB/km for photons at the 1550 nm wavelength and even 0.16 dB/km~\cite{yin2016measurement,boaron2018secure}, and approximately at 0.3 dB/km for photons at 1310 nm wavelength. QKD transmission has been demonstrated over 421 km of optical fiber \cite{boaron2018secure}, and it is not confined to this distance, when using new mechanisms \cite{lucamarini2018overcoming}. In the absence of perfectly secure quantum repeaters, several hundreds of kilometers are feasible for single photon transmission, which is predominantly limited by the inherent path loss and by the environmental factors of temperature as well as stress.

The FSO transmission is known as a very promising design alternative for quantum communications, which has a wide transmission window in the vicinity of $\sim$800 nm, which conveniently corresponds to the detection range of efficient yet inexpensive Si-avalanche photodiode detectors~\cite{pan2023free}. Moreover, there is only negligible dispersion, but the atmospheric turbulence effects may become hostile, as detailed in~\cite{hosseinidehaj2018satellite}. Further challenges of free-space-based quantum communication are due to the scattered sunlight. To overcome the background noise in FSO scenarios, substantial efforts have been made to use the best wavelength, filtering techniques, and optimized single-photon detection with remarkable results~\cite{liao2017long,restelli2010improved}. 

The total attenuation $\alpha$ of the free-space channel can be evaluated by the contributions of two main effects: diffraction and atmospheric attenuations, including absorption, scattering, and atmospheric turbulence, which may be formulated as $\alpha_{\rm atm}\!\!=\!\!\alpha_{\rm abs}\alpha_{\rm scatt}\alpha_{\rm turb}$. As shown in Table \ref{table-5:FSL}, several theoretical models have been established for calculating the channel loss that results from the above-mentioned effects. Diffraction, also known as geometric loss, will result in beam divergence, hence some fraction of the beam energy cannot be collected by the receiver. A plausible, but costly method of tackling this problem is to increase the telescope aperture \cite{vallone2014free}, which is more realistic for the ground-station than for a compact solar-charged satellite. Nguyen \textit{et al}. \cite{nguyen2017network} applied network coding in a free-space QKD system and studied the diffraction effects. In the process of evaluating the QBER, Shapiro theoretically quantified the diffraction loss~\cite{shapiro2003near}. Absorption and scattering are also inevitable phenomena in the process of FSO propagation through the atmosphere, where they interact with various gasses and particles. Surprisingly, their influence on the channel attenuation tends to remain modest, apart from adverse weather conditions. The atmospheric refractive index tends to fluctuate randomly due to the temperature, pressure, and humidity variations in the air, which is the source of atmospheric turbulence. As a result, typically the probability distribution of transmittance is adopted for characterizing the beam wandering, broadening, and deformation after the laser beam undergoes atmospheric turbulence \cite{vasylyev2012toward,vasylyev2016atmospheric,vasylyev2018theory,trinh2018design}. Then the mean attenuation is given by
\begin{eqnarray}
\alpha_{\rm turb}=\int_{0}^{\eta_{0}}\eta^{2}p(\eta)d\eta,
\end{eqnarray}
where $\eta=\sqrt{\eta_{t}}$, is the intensity transmittance, $\eta_{t}$ is the transmissivity, while the maximum value of $\eta$ is $\eta_{0}$, and $p(\eta)$ is the probability distribution of transmittance. 

\begin{table*}[!h]
\begin{footnotesize}
\begin{center}
\caption{Major propagation phenomena characterizing the free-space channel.}
    \begin{tabular}{|l|l|p{9cm}|}
    \hline
Type of channel loss                       & References                                                                                      & Contributions                                                                                                                                                      \\ \hline
\multirow{2}{*}{Diffraction}               & \cite{shapiro2003near}                                                        & The bounds of the sift and error probabilities of a free-space QKD was deduced from the extended Huygens-Fresnel principle by considering the diffraction effects. \\ \cline{2-3}
                                           & \cite{moli2009performance}                                                    & The additional attenuation caused by diffraction was taken into consideration in the analysis of key-rate performance in earth-satellite QKD.                                            \\ \hline
\multirow{2}{*}{Absorption and scattering} & \cite{kaushal2017optical}, \cite{hosseinidehaj2018satellite} & The propagation loss of adverse weathers conditions was quantified.                                                                                           \\ \cline{2-3}
                                           & \cite{vasylyev2017free}                                                       & A quantum theory of nonclassical light propagation under different weather conditions, including rain or haze, was developed, which agrees well with the data collected from experiments.    \\ \hline
Atmospheric turbulence                     & \cite{vasylyev2018characterization}  & Introducing three models to calculate the probability distribution of transmittance when facing the different propagation distances and optical turbulence strengths.                                                                               \\ \hline
\end{tabular}
    \label{table-5:FSL}
    \end{center}
\end{footnotesize}
\end{table*}

The FSO communication can be rapidly established without end-to-end optical fiber connections, offering mobility and flexibility~\cite{krvzivc2023towards,liu2023high}. However, the exponential decay of quantum states when transmitted in optical fibers poses significant challenges to establishing global quantum communication. Leveraging the low-loss transmission of quantum states in free space, satellite-to-ground quantum communication is currently the preferred approach for establishing global quantum communication. This survey~\cite{pan2023free} summarizes the progress in free-space QSDC and the related engineering challenges.

\subsubsection{Quantum Random Number Generator}
\label{sec:QRNG}
\begin{figure}[!h]
\begin{center}
\includegraphics[width=6.7cm,angle=0]{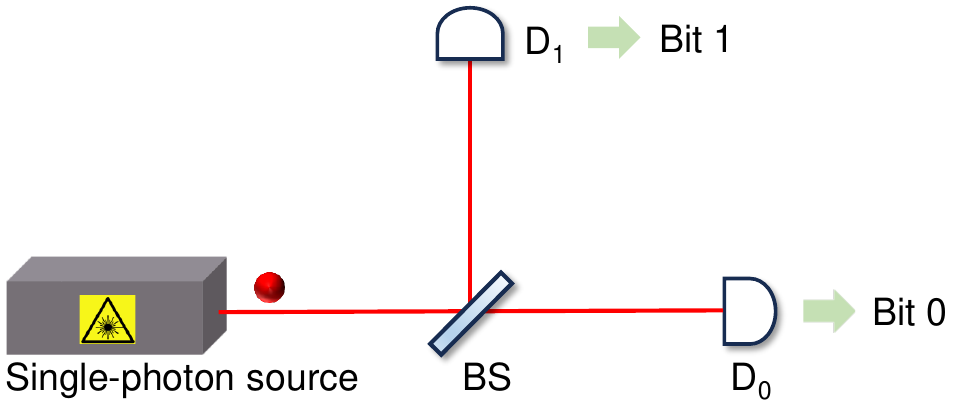}
\caption{Schematic diagram of a random number generator based on a single-photon source. BS, beam splitter; D$_0$, D$_1$, single-photon detector.}
\label{fig:QRNG}
\end{center}
\end{figure}

Traditional pseudo-random number generators are typically algorithm-based, resulting in outputs that are, to some extent, predictable, albeit possessing sufficient complexity to make practical predictions challenging. By contrast, quantum random number generators leverage the inherent uncertainty of quantum mechanics to ensure the generation of entirely unpredictable random numbers. For example, Fig.~\ref{fig:QRNG} illustrates a quantum random number generator based on a single-photon source. A single photon emitted by the single-photon source is subjected to a 50:50 BS. Due to the `indivisibility' of a single photon, it is either randomly transmitted, and then recorded by a single-photon detector $\rm D_0$, generating the bit 0, or alternatively, it is randomly reflected, and then captured by a detector $\rm D_1$, yielding the bit 1. This process results in the generation of a sequence of truly random numbers.

Truely random numbers are considered as an essential part of maintaining the security of quantum communication \cite{trushechkin2018quantum}. As for QSDC systems, the state preparation, state sampling strategies of some protocols, and classical coding~\cite{pan2023free} rely on a random number generator producing samples at a high rate and `high-quality' randomness. Normally, the procedure of designing a quantum random number generator is divided into four steps, as seen in Table \ref{tab-6:QRNG}.

\begin{table*}[!h]
\begin{footnotesize}
\begin{center}
\caption{The design process of quantum random number generator.}
\label{tab-6:QRNG}
\begin{tabular}{lllllllll}
\cline{1-2} \cline{4-4} \cline{6-6} \cline{8-9}
\multicolumn{2}{|l|}{\cellcolor[HTML]{C0C0C0}Source of quantum randomness}                                                                                                                                                                 & \multicolumn{1}{l|}{} & \multicolumn{1}{l|}{\cellcolor[HTML]{C0C0C0}Detection}                                                     & \multicolumn{1}{l|}{} & \multicolumn{1}{m{2cm}|}{\cellcolor[HTML]{C0C0C0}Postprocessing}  & \multicolumn{1}{l|}{} & \multicolumn{2}{l|}{\cellcolor[HTML]{C0C0C0}Randomness test}                                                                        \\ \cline{1-2} \cline{4-4} \cline{6-6} \cline{8-9}
\multicolumn{2}{l}{}                                                                                                                                                                                                                       &                       &                                                                                                            &                       &                                                              &                       & \multicolumn{2}{l}{}                                                                                                                \\ \cline{1-2} \cline{4-4} \cline{6-6} \cline{8-9}
\multicolumn{1}{|p{1.2cm}|}{Discrete}   & \multicolumn{1}{l|}{\begin{tabular}[c]{@{}p{4.4cm}@{}}Spatial randomness of single photons \cite{jennewein2000fast,wang2006scheme,soares2014quantum}\\ Time resolution randomness of single photons \cite{wayne2009photon,furst2010high,nie2014practical}\\ Attenuated coherent light \cite{wei2009bias,ren2011quantum}\\ Quantum tunneling effect \cite{zhou2017quantum,zhou2019quantum}\\
Device-independent self-testing \cite{liu2018device,bierhorst2018experimentally}\end{tabular}} & \multicolumn{1}{l|}{} & \multicolumn{1}{l|}{}                                                                                      & \multicolumn{1}{l|}{} & \multicolumn{1}{l|}{}                                        & \multicolumn{1}{l|}{} & \multicolumn{1}{m{1.4cm}|}{Statistical analysis}   & \multicolumn{1}{p{1.4cm}|}{\begin{tabular}[c]{@{}l@{}}ENT, \\ Diehard, \\ National Institute \\of Standards and \\Technology \\statistical test \\suite~\cite{liu2017117,wang2015robust}\end{tabular}} \\ \cline{1-2} \cline{8-9}
\multicolumn{1}{|p{1.5cm}|}{Continuous} & \multicolumn{1}{l|}{\begin{tabular}[c]{@{}p{4.3cm}@{}}Laser phase fluctuations \cite{guo2010truly,nie2015generation},\\Vacuum fluctuation\cite{zhou2019practical}\\ Shot noise \cite{gabriel2010generator,shen2010practical}\\ Super-luminescent diode \cite{liu2013implementation}\end{tabular}}                                                                            & \multicolumn{1}{l|}{} & \multicolumn{1}{l|}{\multirow{-7}{2.2cm}{Detection varies with different quantum randomness source}} & \multicolumn{1}{l|}{} & \multicolumn{1}{l|}{\multirow{-7}{1.8cm}{Randomness extractors \cite{ma2013postprocessing}}} & \multicolumn{1}{l|}{} & \multicolumn{1}{m{1.4cm}|}{Physical certification} & \multicolumn{1}{m{2.5cm}|}{Bell's theorem \cite{pironio2010random}}                                                   \\ \cline{1-2} \cline{4-4} \cline{6-6} \cline{8-9}
\end{tabular}
\label{tab:QRNG}
\end{center}
\end{footnotesize}
\end{table*}

First the specific source of quantum randomness will be selected, from which we can extract the raw random number sequence using measurement. 
The resultant raw random number sequence then undergoes postprocessing, a step of distillation, to remove the classical sources of contamination (typically appears as bias and redundancy in the sequence) that originates from the imperfection of the devices. Finally, the quality of the random number sequence will be tested either with the aid of statistical analysis or physical certification. The source of quantum randomness falls into two basic categories: discrete and continuous, which depends on the specific source of randomness. The discrete source of quantum randomness has a simple model, but its output rate is usually low. By contrast, the continuous source has a high output rate, but the inevitable classical contamination has to be carefully mitigated.

\subsubsection{Bell-State Measurement}
\label{sec:BSM}
\begin{figure}[!h]
\begin{center}
\includegraphics[width=8.9cm,angle=0]{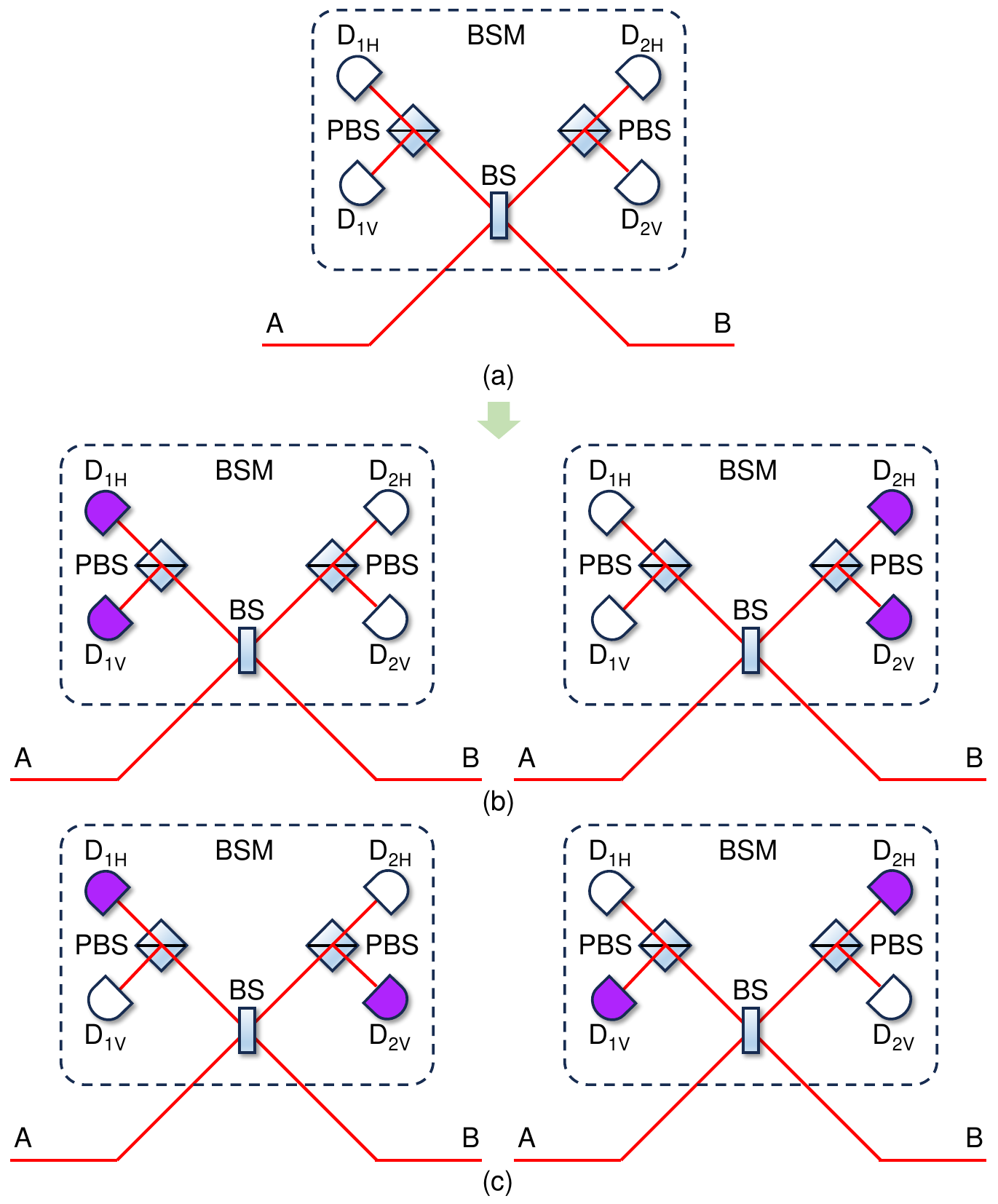}
\caption{(a) Setup to perform BSM. (b) The click patterns of Bell state $|\psi^{+}\rangle_{AB}$. (c) The click patterns of Bell state $|\psi^{-}\rangle_{AB}$. BS: beam splitters; PBS: polarizing beam  splitters;  D: single-photon detectors; the A and B represent two inputs. The single-photon detectors labeled by purple means that this detector is clicked by photons.}
\label{fig:BSM}
\end{center}
\end{figure}

This is one of the key ingredients in quantum information processing, as surveyed in Section \ref{subsec:EntRes}. Entanglement generation was detailed in Section \ref{subsec:LD}. Again, the BSM allows us to distinguish the four Bell states of Eq.~(\ref{Eq:entanglement}) when we want to exploit entangled resources. If we are restricted to using only linear optical devices, such as beam splitters, polarization beam splitters and single-photon detectors, the BSM will be unable to distinguish each of the four Bell-states with 100\% certainty, as shown in~\cite{lutkenhaus1999bell,bayerbach2023bell}. More explicitly, in this case, only two of the four Bell states $|\psi^{\pm}\rangle$ can be reliably discriminated~\cite{lutkenhaus1999bell}. In other words, the success rate of BSM using linear optics is limited to 50\%. Although the four Bell states cannot be distinguished unambiguously by using linear optical devices, these are easier to implement than a complete BSM relying on nonlinear optical components~\cite{kim2001quantum}. Hence, some QSDC protocols~\cite{zhou2020measurement,pan2020single} tend to only choose the pair Bell state $|\psi ^\pm \rangle_{AB}$ of Eq. (\ref{Eq:entanglement}) to transmit information for the sake of simplifying the measurement system. Let us rewrite the Bell states $|\psi ^\pm \rangle_{AB}$ of Eq. (\ref{Eq:entanglement}) in the form of polarization-entangled photonic states, namely,
\begin{eqnarray}
\label{Eq:entanglement-polarization}
|\psi^{+}\rangle_{AB}=\frac{1}{\sqrt{2}}\left(|H\rangle_{A}|V\rangle_{B}+|V\rangle_{A}|H\rangle_{B}\right),\nonumber\\
|\psi^{-}\rangle_{AB}=\frac{1}{\sqrt{2}}\left(|H\rangle_{A}|V\rangle_{B}-|V\rangle_{A}|H\rangle_{B}\right).
\end{eqnarray}
This BSM, which purely relies on linear optical components for discriminating these two polarization-entangled photonic states is shown in Fig. \ref{fig:BSM} (a).
These two devices allow us to identify the incoming Bell state by the so-called click patterns of the single-photon detectors~\cite{da2013long,valivarthi2014efficient,yu2015two}, as shown in Fig. \ref{fig:BSM} (b) and (c). 
If two clicks happen in $\rm D_{1H}$ and $\rm D_{1V}$ or $\rm D_{2H}$ and $\rm D_{2V}$, the measurement result is $|\psi^{+}\rangle_{AB}$. These two click patterns are shown in Fig. \ref{fig:BSM} (b). Furthermore, the measurement result is $|\psi^{-}\rangle_{AB}$ if two clicks $\rm D_{1H}$ and $\rm D_{2V}$ or $\rm D_{1V}$ and $\rm D_{2H}$ are observed, as shown in Fig. \ref{fig:BSM} (c).

A complete BSM is possible by using hyper-entanglement with only linear elements. Hyper-entanglement is a state that is entangled in more than one degree of freedom and the extra degrees of freedom allows for secondary interferometry, so that the remaining two Bell states can be distinguished \cite{kwiat1998embedded,walborn2003hyperentanglement,schuck2006complete}. If nonlinear elements are added, all four Bell states can be discriminated with a success probability of 100\% \cite{kim2001quantum}. Additional auxiliary degrees of freedom in hyper-entanglement can also be distinguished by the nonlinearity devices of~\cite{sheng2010complete}.

\subsubsection{Photonic Modulation Technologies}
\label{sec:encodingtech}
\begin{figure}[!htpb]
\begin{center}
\includegraphics[width=\columnwidth,angle=0]{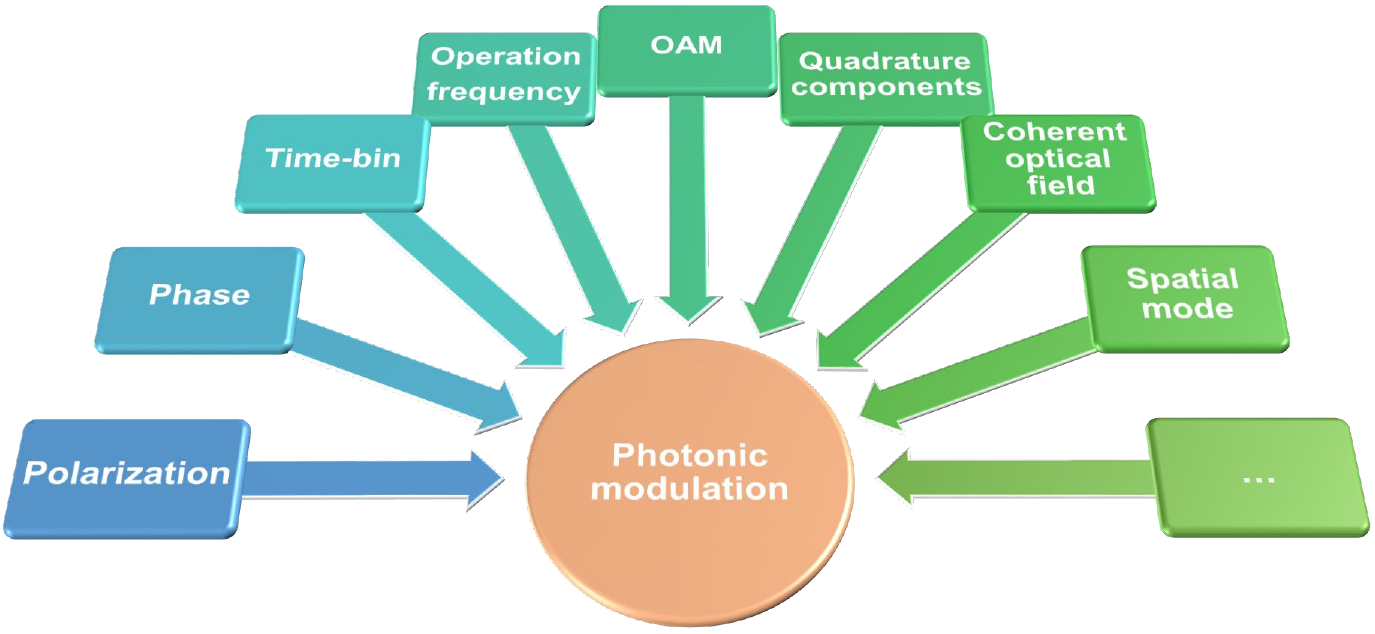}
\caption{Mapping qubits to physical resources using photonic modulation. OAM: orbital angular momentum.}
\label{fig:Photonicencoding}
\end{center}
\end{figure}

In QSDC, the information can be conveyed by different fundamental resources, as shown in Fig. \ref{fig:Photonicencoding}, namely polarization~\cite{zhang2017quantum,zhu2017experimental}, phase~\cite{sun2018design,qi2019implementation,pan2020experimental,sun2020toward,zhang2022realization}, time-bin~\cite{zhang2022realization}, operation frequency~\cite{hu2016experimental}, orbital angular momentum (OAM)~\cite{mi2015high}, quadrature components~\cite{pirandola2008quantum,pirandola2009confidential,cao2021continuous}, coherent optical filed~\cite{vazquez2021quantum}, and spatial mode~\cite{liu2012high}. Some of these have shown promising potential QSDC. In the following, we will briefly highlight these in the context of QSDC. Some new modulation techniques of QKD will also be mentioned, in order to pave the way for their QSDC counterparts in the future. We mainly focus our attention on the field of the discrete variables, while the quadrature components of light based solutions belong to the family of continuous variables~\cite{braunstein2005quantum,hosseinidehaj2018satellite}, which will be touched upon in less detail.

The \textbf{polarization} of a single-photon may be described by its state vector~\cite{bennett1998quantum,imre2013quantum}: the state of $|1\rangle$ stands for the vertically polarized photon $|\uparrow\rangle=|V\rangle$ also seen in the Bloch sphere of Fig. \ref{fig:BlochSphere}. Analogously, we have 
\begin{eqnarray}
|\rightarrow\rangle=|H\rangle=|0\rangle, \\
|\nearrow\rangle=|+\rangle,\\
|\searrow\rangle=|-\rangle,\\
|\circlearrowright\rangle=|R\rangle,\\
|\circlearrowleft\rangle=|L\rangle. 
\end{eqnarray}
Furthermore, for entangled states, we have $|\psi^{+}\rangle=\frac{1}{\sqrt{2}}\left(|\rightarrow\rangle_{A}|\uparrow\rangle_{B}+|\uparrow\rangle_{A}|\rightarrow\rangle_{B}\right)$. These states represent the information carrier of the classical bits in many quantum communications protocols. They also constitute a natural choice for experimental implementations~\cite{zhang2017quantum,zhu2017experimental}, because they may be readily generated by the polarization based modulation of a laser source \cite{townsend1998experimental} or polarizing multiple lasers~\cite{hughes2002practical,choe2018silica}\footnote{Note that this light source may raise some security concerns in quantum communication, because the laser sources cannot be exactly the same.}. Based on the characteristics of the quantum channels described in Section~\ref{sec:componentsandchannel}, we know that the polarization of photons is more suitable for FSO channels than for fiber, because the fiber channel tends to perturb the polarization of photons \cite{breguet1994quantum}. Furthermore, photon polarization based FSO transmission designed for satellites requires some reference frames for the accurate alignment of measurement bases \cite{bonato2006influence}, unless a specifically designed reference-frame-independent protocol is used \cite{laing2010reference}. At the measurement stage, the receiver can separate the orthogonal states and forward them to the single-photon detectors for determining, whether they are constituted by $|H\rangle$ or $|V\rangle$. The diagonally polarized states can also be readily identified in a similar way.

Phase encoding maps information onto the relative phase difference between two consecutive pulses. Typically, after modulation by an unbalanced Mach-Zehnder interferometer, the form of the weak coherent state becomes~\cite{lo2007security,tang2013source}
\begin{equation}
\left |\sqrt{\frac{\mu}{2}}e^{i\theta}\right \rangle_s\left |\sqrt{\frac{\mu}{2}}e^{i(\theta+\varphi )}\right \rangle_l,
\end{equation}
where $\mu$ is the average photon numbers, $\theta$ is the initial random phase, $\varphi $ is the modulated phase encoding information bits. Furthermore, the subscripts $s$ and $l$ denote the short arm and long arm of an unbalanced Mach-Zehnder interferometer. The discrete phases of $\varphi \in\{0, \frac{\pi}{2}, \pi, \frac{3\pi}{2}\}$ which are prepared by Alice as the initial states and then Bob randomly applies the phase shifts of either 0 or $\frac{\pi}{2}$ for performing demodulate measurement \cite{gobby2004quantum,muller1997plug,song2023practical} in BB84 QKD. This scheme may also be readily adapted for QSDC as detailed in~\cite{qi2019implementation,pan2020experimental}. 

The family of \textbf{time-bin} based methods convey the qubit states by relying on specific consecutive time intervals, written in the form of
\begin{equation}
|e\rangle=|\sqrt{\mu}\rangle_e|\sqrt{0}\rangle_l,~|l\rangle=|\sqrt{0}\rangle_e|\sqrt{\mu}\rangle_l,
\end{equation}
where the subscripts $e$ and $l$ denote the temporal modes early and late, respectively. It can be prepared by emitting a pulse which has to pass through an unbalanced interferometer. Thus the computational basis $|e\rangle$ indicates that the photon takes a short path and arrive early, while $|l\rangle$ represents the long arm, hence its photon arrives late. After preparation, if this time-bin based state goes through a nonlinear crystal, eventually a time-bin entanglement can be created \cite{marcikic2004distribution}. The phase and time-bin based qubits are more robust than the polarization based qubits, when performing quantum communication via fiber~\cite{zhang2022realization}. Hence the former are is popular in quantum communication.

\textbf{Operation frequency} based encoding applies the same unitary operation to a single-photon block periodically to encode secret messages, and the receiver of this single-photon block is capable of decoding the secret messages through the discrete time Fourier transform~\cite{hu2016experimental,gupta2023experimental}.

\textbf{Orbital angular momentum}~\cite{zhou2016orbital,erhard2018twisted} is another quantity that may be conveniently carried by a laser beam upon exploiting the so-called azimuthal angular dependence of $e^{-il\phi}$, where $l$ is the azimuthal index assuming an unbounded integer and $\phi$ is the azimuthal angle in the beam's cross-section \cite{allen1992orbital,djordjevic2016integrated}. For a given azimuthal index $l$, the orbital angular momentum has the discrete value of $L=l\hbar$, indicating that the beam carries $l\hbar$ amount of orbital angular momentum per photon. It has been demonstrated that orbital angular momentum states can be used for conveying information at a capacity beyond one bit per photon both for QSDC~\cite{mi2015high,jian2016efficient} and DQKD \cite{farman2018ping}. Additionally, a high-dimensional system `qudit' (a unit of information in a $N$ dimension space) typically has a superior security level in quantum cryptography protocols~\cite{cerf2002security,djordjevic2013multidimensional}. 

The QSDC scheme relying on \textbf{quadrature components} for encoding is commonly referred to as a continuous-variable protocol, which has the advantage of compatibility with existing classical communication infrastructure~\cite{qi2016simultaneous,pan2020simultaneous}, low cost, and high rate~\cite{hosseinidehaj2018satellite}. These continuous variable protocols typically employ Gaussian states~\cite{pirandola2008quantum,pirandola2009confidential,srikara2020continuous}, squeezed states\cite{paparelle2022implementation,paparelle2023practical}, and continuous-variable entangled states~\cite{marino2006deterministic,shapiro2014secure,zhuang2015ultrabroadband,shapiro2019quantum,chai2019novel,cao2021continuous} as information carriers for secure communication. Recently, the CV QSDC experiment demonstrations were given by Paparelle \textit{et al}.~\cite{paparelle2022implementation,paparelle2023practical}. There are two types of modulation methods. The information to be transmitted  can be converted into Gaussian random numbers and then mapped either to the position or momentum of the  quadrature components~\cite{cao2021continuous}, which is hence referred to as Gaussian modulation~\cite{laudenbach2018continuous}. Discrete modulation entails the discrete control of phase or quadrature amplitude based on the information to be transmitted~\cite{pirandola2008quantum,pirandola2009confidential,srikara2020continuous,paparelle2023practical,marino2006deterministic,shapiro2014secure,zhuang2015ultrabroadband,shapiro2019quantum,di2021two}.

In 2021, a protocol called quantum keyless private communication was proposed in~\cite{vazquez2021quantum}, which uses the quantum-domain version of on-off keying modulation to transmit information~\cite{vazquez2023quantum}, where the coherent state $|\alpha\rangle$ represents information bit 1 and the vacuum state $|0\rangle$ represents information bit 0. Single-photon detectors are used for detection. We refer to this as modulation based on \textbf{coherent optical fields.}

The \textbf{spatial-mode} is a frequently used degree of freedom exploited for information transfer, when combined with other degrees of freedom, it enables the design of high-capacity QSDC protocols~\cite{wang2011high,liu2012high,wu2020high}.

The family of \textbf{hybrid methods} combining different modulation techniques which complement each other is also often used for the sake of combining their different benefits. For instance, combined polarization-orbital angular momentum states are rotationally invariant and exhibit high robustness against spatial perturbations. Hence they are eminently suitable for mitigating the frame-misalignment problems encountered in free-space quantum communication~\cite{d2012complete,vallone2014free}. Modulation techniques relying on high-dimensional states that can be used for high-rate and high-security quantum communication are also available at the time of writing, but they are limited to short distances~\cite{djordjevic2013multidimensional,sit2017high}.

By contrast, there are many other mature modulation methods in QKD, but their feasibility in QSDC still needs to be studied. Examples of practical \textbf{phase-encoding} related QKD systems include the differential phase shift aided schemes of~\cite{inoue2002differential}. In differential phase shift aided QKD, the information bits are mapped to the phase difference between the adjacent pulses. It has been shown that it is feasible to realize a high-speed clock frequence of 10 GHz using the simple experimental setup of \cite{takesue2007quantum}, which achieves a 12.1 bit/s secure key rate over 200 km of optical fibre. In coherent one-way QKD, the bit string is carried by the coherent one-way pulse of $|0\rangle|\alpha\rangle$ for bit 0 and by $|\alpha\rangle|0\rangle$ for bit 1. Then the receiver can detect them with the aid of time-of-arrival measurements. In 2015, the coherent one way technique attained a long-distance record of 307 km for QKD~\cite{korzh2015provably}. Both the differential phase shift and the coherent one-way techniques use weak coherent pulses as the light source, but they are immune to photon number splitting attacks \cite{inoue2005robustness,stucki2005fast}.

A range of early contributions relied on single-sideband modulation, where the information bits are mapped to one of the two sidebands surrounding a central frequency \cite{merolla1999single,duraffourg2001compact,merolla2002integrated,guerreau2003long}. Bloch \textit{et al}.~\cite{bloch2007frequency} conceived a \textbf{frequency-encoding} QKD scheme, while its improved version was proposed by Zhang \textit{et al}. \cite{zhang2008frequency} a year later. 
In frequency-modulated QKD, Alice can generate four states by using the modulator, which can be formulated as follows,
\begin{eqnarray}
|+;1\rangle=\frac{1}{\sqrt{2}}|1\rangle_{\omega_0}+\frac{1}{2}|1\rangle_{\omega_0+\Omega}-\frac{1}{2}|1\rangle_{\omega_0-\Omega},\nonumber\\
|-;1\rangle=\frac{1}{\sqrt{2}}|1\rangle_{\omega_0}-\frac{1}{2}|1\rangle_{\omega_0+\Omega}+\frac{1}{2}|1\rangle_{\omega_0-\Omega},\nonumber\\
|+;2\rangle=|1\rangle_{\omega_0},\nonumber\\
|-;2\rangle=\frac{1}{\sqrt{2}}|1\rangle_{\omega_0+\Omega}-\frac{1}{\sqrt{2}}|1\rangle_{\omega_0-\Omega},
\end{eqnarray}
where $|n\rangle_{\omega}$ represents the number of photons in mode $\omega$. Those four states constitute a pair of bases given by $\{|+;1\rangle, |-;1\rangle\}$ and $\{|+;2\rangle, |-;2\rangle\}$, which is equivalent to the rectilinear and diagonal orthogonal basis sets of the BB84 QKD protocol~\cite{bennett1984quantum} and thus this modulation scheme may be readily used for implementing QKD. No unbalanced interferometers are required for frequency encoding, which has the advantage of avoiding stabilization. 

The status of these photonic modulation techniques in the context of QSDC is summarized in Table~ \ref{table:PEstatus}.

\begin{table}[]
\begin{footnotesize}
\begin{center}
\caption{Status of some popular photonic transmission techniques in the realization of QSDC.}
\begin{tabular}{|l|c|}
\hline
Technique                                                & Status                                           \\ \hline
Polarization                                              & $\surd$                                          \\ \hline
Phase                                                     & $\surd$                                       \\ \hline
Time bin                                            & $\surd$                                          \\ \hline
Operation frequency                                            & $\surd$                                          \\ \hline
Orbital angular momentum                                  & $\bigcirc$                                       \\ \hline
Quadrature components                                            & $\surd$                                          \\ \hline
Coherent optical field                                            & $\bigcirc$                                          \\ \hline
Spatial mode                                      & $\bigcirc$ \\ \hline
Hybrid methods & $\bigcirc$                                                
\\ \hline
Differential phase shift                                                      & ?                                                \\ \hline
Coherent one-way                                                       & ?                                                \\ \hline
Frequency                                                 & ?                                                \\ \hline
\multicolumn{2}{|p{8cm}|}{$\surd$\;\;There have already been experimental demonstrations.}                                 \\ 
\multicolumn{2}{|p{8cm}|}{$\bigcirc$ It has been proposed, but there is no experimental demonstration.} \\ 
\multicolumn{2}{|p{8cm}|}{$\;?$\;\;The feasibility is uncertain at the time of writing.}                                 \\ \hline
\end{tabular}
\label{table:PEstatus}
\end{center}
\end{footnotesize}
\end{table}

\section{Quantum Secure Direct Communication}
\label{sec:QSDC}
\subsection{Point-to-Point Communication Protocols}
\label{sec:PtoP}
\begin{figure}[!h]
\begin{center}
\includegraphics[width=\columnwidth,angle=0]{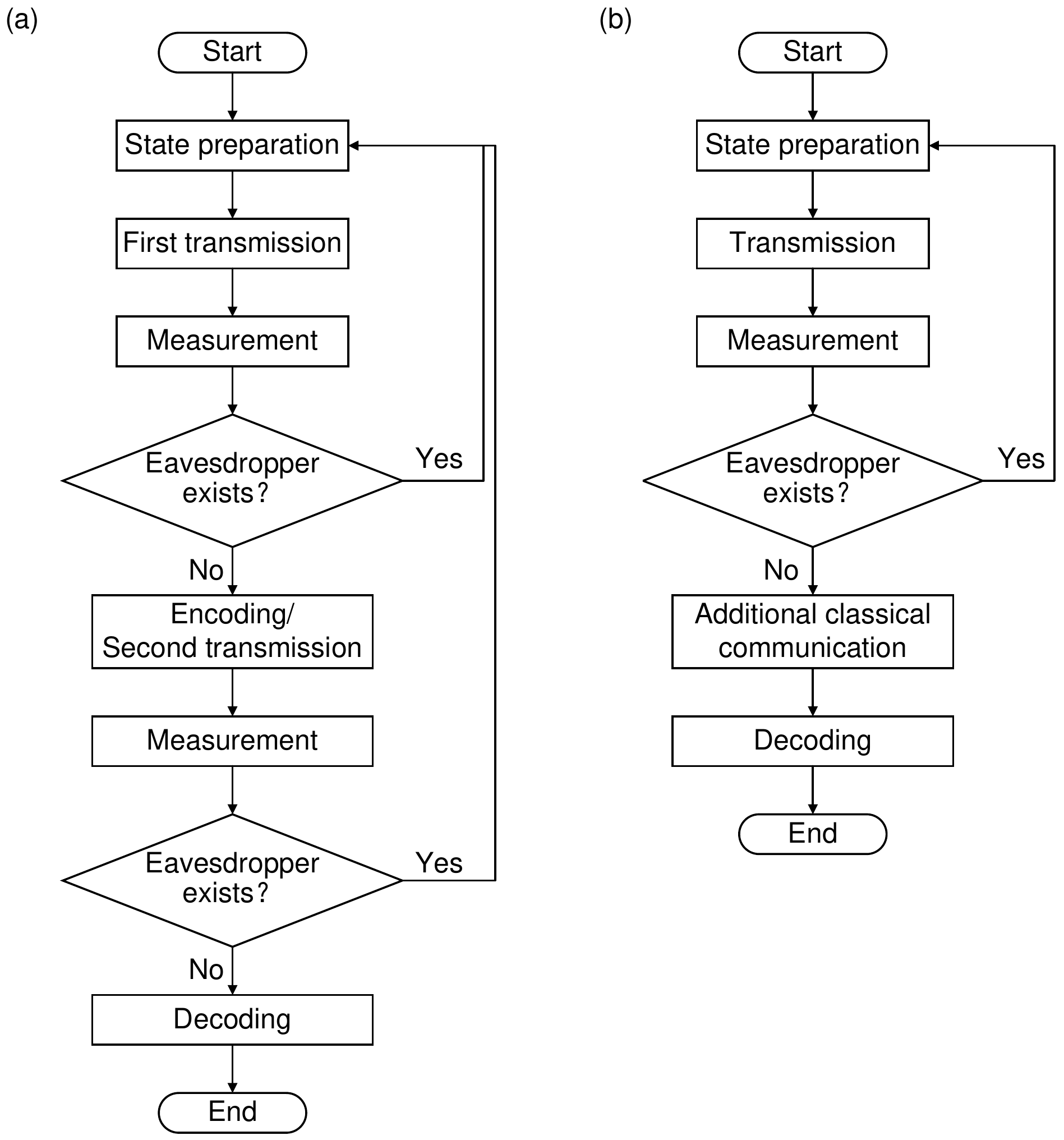}
\caption{The flow diagram of (a) the QSDC (high-capacity, two-step, DL04, high-dimension two-step, and MDI) and (b) DSQC protocols (continuous-variable protocol relying on two-mode squeezed states and Zhu-Xia-Fan-Zhang (ZXFZ) protocol).}
\label{fig:QSDCandDSQC}
\end{center}
\end{figure}

A point-to-point QSDC protocol is defined as a sequence of steps with associated quantum operations and rules that control the communications with the goal of securing communication between two users. In this section, we will highlight some of the typical QSDC protocols step by step and show how these protocols facilitate for two parties to transmit confidential messages directly through the quantum channel~\cite{long2002theoretically,deng2003two,deng2004secure,wang2005quantum,marino2006deterministic,zhou2020measurement}, rather than simply exchanging secret keys. 
We will introduce the QSDC protocol and DSQC protocol, which enable secure communication over a quantum channel. The processes of these protocols are illustrated in Fig.~\ref{fig:QSDCandDSQC}.

\subsubsection{High-Capacity QSDC Protocol~\rm{\cite{long2000theoretical,long2002theoretically}}}
\label{sec:High-capacity QSDC protocol}
\begin{figure}[!h]
\begin{center}
\includegraphics[width=\columnwidth,angle=0]{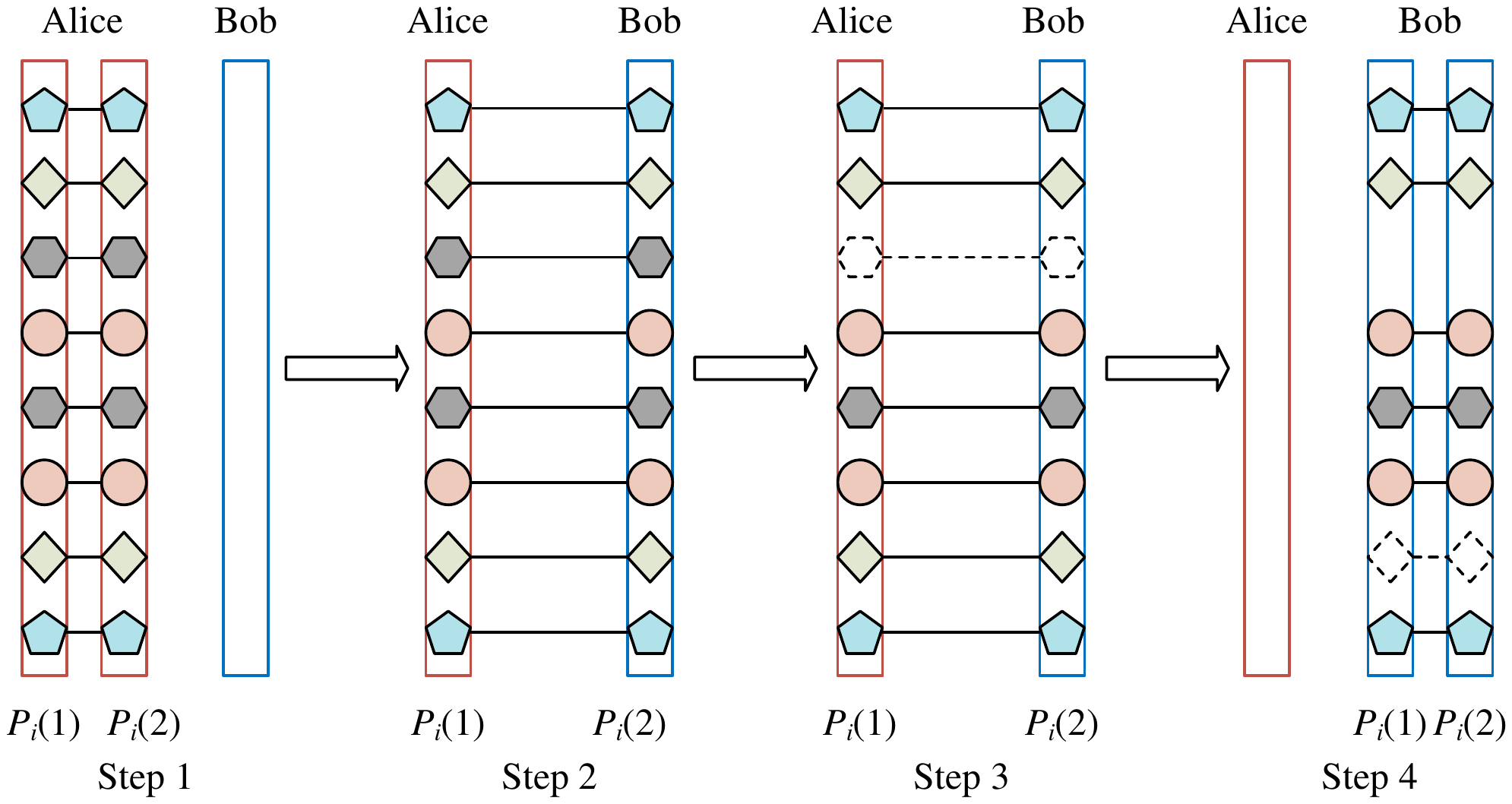}
\caption{Schematic illustration of the high-capacity QSDC protocol. Two symbols linked with a line are in a Bell-state. The pentagons, the rhombuses, the hexagons, and the circles represent the Bell state $|\phi^{+}\rangle_{AB}$, $\left |\phi^{-}\right \rangle_{AB}$, $\left |\psi^{+}\right \rangle_{AB}$, and $\left |\psi^{-}\right \rangle_{AB}$, respectively. The blank dashed symbols represent the state after Alice and Bob complete eavesdropping detection.}
\label{fig:HighCQSDC}
\end{center}
\end{figure}

Alice and Bob have the two-bit classical messages 00, 01, 10, 11 corresponding to $\left |\phi^{+}\right \rangle_{AB}, \left |\phi^{-}\right \rangle_{AB}, \left |\psi^{+}\right \rangle_{AB}, \left |\psi^{-}\right \rangle_{AB}$, respectively. An ordered set of $N$ EPR pairs is denoted by $\{[P_{1}\left ( 1 \right),P_{1}\left ( 2 \right)]$, $[ P_{2}\left ( 1 \right ),P_{2}\left ( 2 \right )]$, $\cdots,$$[P_{i}\left ( 1 \right ),P_{i}\left ( 2 \right )]$, $\cdots,$$[ P_{N}\left ( 1 \right ),P_{N}\left ( 2 \right )]\}$, where $P_{i}\left ( 1 \right )$ is the EPR partner particle of $P_{i}\left ( 2 \right )$ and vice versa. Then an ordered EPR partner particle sequence $[P_{1}\left(1\right), P_{2}\left(1\right), \cdots, P_{i}\left(1\right), \cdots,P_{N}\left(1\right)]$ is generated by taking one of the EPR partner particles, say $P_{i}\left(1\right)$ from each EPR pair $[ P_{i}\left ( 1 \right ),P_{i}\left ( 2 \right )]$. The order of these $N$ EPR pairs remains unchanged throughout the whole process of confidential message transmission. They carry out the following procedures \cite{long2002theoretically}, which are shown in Fig. \ref{fig:HighCQSDC}.

\begin{itemize}
\item \textbf{Step 1, state preparation}. Alice prepares an ordered set of $N$ EPR pairs $\{[P_{1}\left ( 1 \right),P_{1}\left ( 2 \right)]$, $[ P_{2}\left ( 1 \right ),P_{2}\left ( 2 \right )]$, $\cdots,$$[P_{i}\left ( 1 \right ),P_{i}\left ( 2 \right )]$, $\cdots,$$[ P_{N}\left ( 1 \right ),P_{N}\left ( 2 \right )]\}$ representing her eavesdropping check bits and her confidential messages to be transmitted to Bob. The check bits are randomly selected 00, 01, 10, and 11, and they are inserted into the messages. Alice then splits the EPR pair sequence into two halves, namely into an ordered EPR partner particle sequence: $[P_{1}\left(1\right)$, $P_{2}\left(1\right)$ $,\cdots,$ $P_{i}\left(1\right)$ $,\cdots,$ $P_{N}\left(1\right)]$, and into the corresponding EPR partner particle sequence: $[P_{1}\left(2\right)$, $P_{2}\left(2\right)$ $, \cdots,$ $P_{i}\left(2\right)$ $,\cdots,$ $P_{N}\left(2\right)]$.

\item \textbf{Step 2, first transmission}.  Alice sends one of the ordered EPR partner particle sequences, say $[P_{1}\left(2\right)$, $P_{2}\left(2\right)$ $,\cdots,$ $P_{i}\left(2\right)$ $,\cdots,$ $P_{N}\left(2\right)]$ to Bob and tentatively stores the other one in her quantum memory.

\item \textbf{Step 3, first eavesdropping detection}. Alice randomly chooses a sufficiently large fraction of the samples representing the check bits from the EPR partner sequence stored in her memory and performs measurement on these samples by randomly using either the Z-basis or the X-basis. Naturally, she will get the result of either 0 or 1. Again, the rest of the EPR partner sequence is stored by Alice as seen in Fig.~\ref{fig:HighCQSDC}. She then informs Bob through an authenticated classical channel - which may of course be mapped to another wavelength in the same wavelength division multiplex aided fiber link -  of the positions of the specific samples measured by her. Based on the information received from Alice, Bob measures the corresponding EPR sample particles in his hand. Then Alice and Bob publicly compare the results of their measurement to detect eavesdropping. If their results are the  `same'~\cite{long2000theoretical,long2002theoretically}, they conclude that there is no eavesdropping. If there is no eavesdropping, they proceed to the next step. Otherwise, they terminate the communication. This is the first eavesdropping detection opportunity during the transmission of $P_{i}\left(2\right)$. In Fig.~\ref{fig:HighCQSDC}, the pair of dashed hollow circles represent the checking qubits. 

\item \textbf{Step 4, second transmission, measurement, and second eavesdropping detection}. Alice sends the remaining EPR partner particle sequence $[P_{1}\left(1\right), P_{2}\left(1\right), P_{3}\left(1\right)\cdots, \cdots, P_{N}\left(1\right)]$ to Bob, which does not include the particles that have been choosen for detecting a potential eavesdropper. For instance, $P_{3}\left(1\right)$ in Fig.~\ref{fig:HighCQSDC} has been measured by Alice and thus was not transmitted. Bob performs the BSM on every EPR pair $[ P_{i}\left ( 1 \right ),P_{i}\left ( 2 \right )]$ in order to decode the confidential message and records the measurement results after receiving the rest of the sequence from Alice. The remaining check bits are announced by Alice, whose BSM results are selected to determine whether the QSDC process is successful. As shown in Fig.~\ref{fig:HighCQSDC}, the EPR pair $[ P_{7}\left ( 1 \right ),P_{7}\left ( 2 \right )]$ represents the second set of check qubits. If the error rate is deemed to be below a certain threshold, the remaining results of the BSM are deemed to represent the transmitted confidential messages. The second eavesdropping detection opportunity is included here for estimating the reliability of the communication.
\end{itemize}

To elaborate a little further, there are two eavesdropping detections, which  allows the QSDC protocol to guard against eavesdropping. On the one hand, an eavesdropper cannot steal the confidential messages without being detected, when she adopts the intercept-and-resend attack strategy. Eve has no access to both parts of the EPR pairs at the same time, since the ordered $N$ EPR pair sequence is sent from Alice to Bob in two staggered intervals. In order to obtain the other EPR partner particle sequence $[P_{1}\left(1\right), P_{2}\left(1\right), \cdots, P_{i}\left(1\right), \cdots,P_{N}\left(1\right)]$, Eve first has to intercept the EPR partner particle sequence $[P_{1}\left(2\right), P_{2}\left(2\right), \cdots, P_{i}\left(2\right), \cdots,P_{N}\left(2\right)]$ and then a fake particle sequence $[P^{*}_{1}\left(2\right), P^{*}_{2}\left(2\right), \cdots, P^{*}_{i}\left(2\right), \cdots,P^{*}_{N}\left(2\right)]$ of her has to be sent to Bob. However, this attack can be easily detected by the above first eavesdropping detection in Step 4, because Alice randomly chooses some EPR partner particles in her hand to perform measurements and asks Bob to do the same. Alice and Bob will discover that half of their measurement results are conflicting because there is no quantum correlation between $P_{i}\left(1\right)$ and $P^{*}_{i}\left(2\right)$. As a further measure, the malicious nature of direct measurement by Eve may be readily spotted by the second eavesdropping detection. If Eve applies direct measurement to the EPR partner particle sequence and resends it, a high error rate will be experienced in Step 4 for the reason that the Bell states will collapse.

This is the basic principle of QSDC which has the beneficial feature of high capacity, since the four legitimate states of the EPR pair can carry two classical bits of information. This represents a higher capacity than that of other protocols that make use of EPR pairs as their information carrier \cite{ekert1991quantum, bennett1992quantum,horodecki2009quantum}. Bob can decode the information directly without the exchange of classical bits. The protocols of~\cite{long2000theoretical,long2002theoretically} use block transmission of quantum states to prevent the leakage of confidential messages and detect eavesdropping via random sampling tests. The idea that quantum mechanics could be beneficially exploited for direct communication over quantum channels has evolved substantially further after its conception~\cite{long2000theoretical,long2002theoretically}, which was originally proposed for deterministic key distribution. As a next evolutionary step, let us now consider the following two-step QSDC protocol.

\subsubsection{Two-Step QSDC Protocol}
\label{sec:Two-step QSDC protocol}
\begin{figure}[!h]
\begin{center}
\includegraphics[width=\columnwidth]{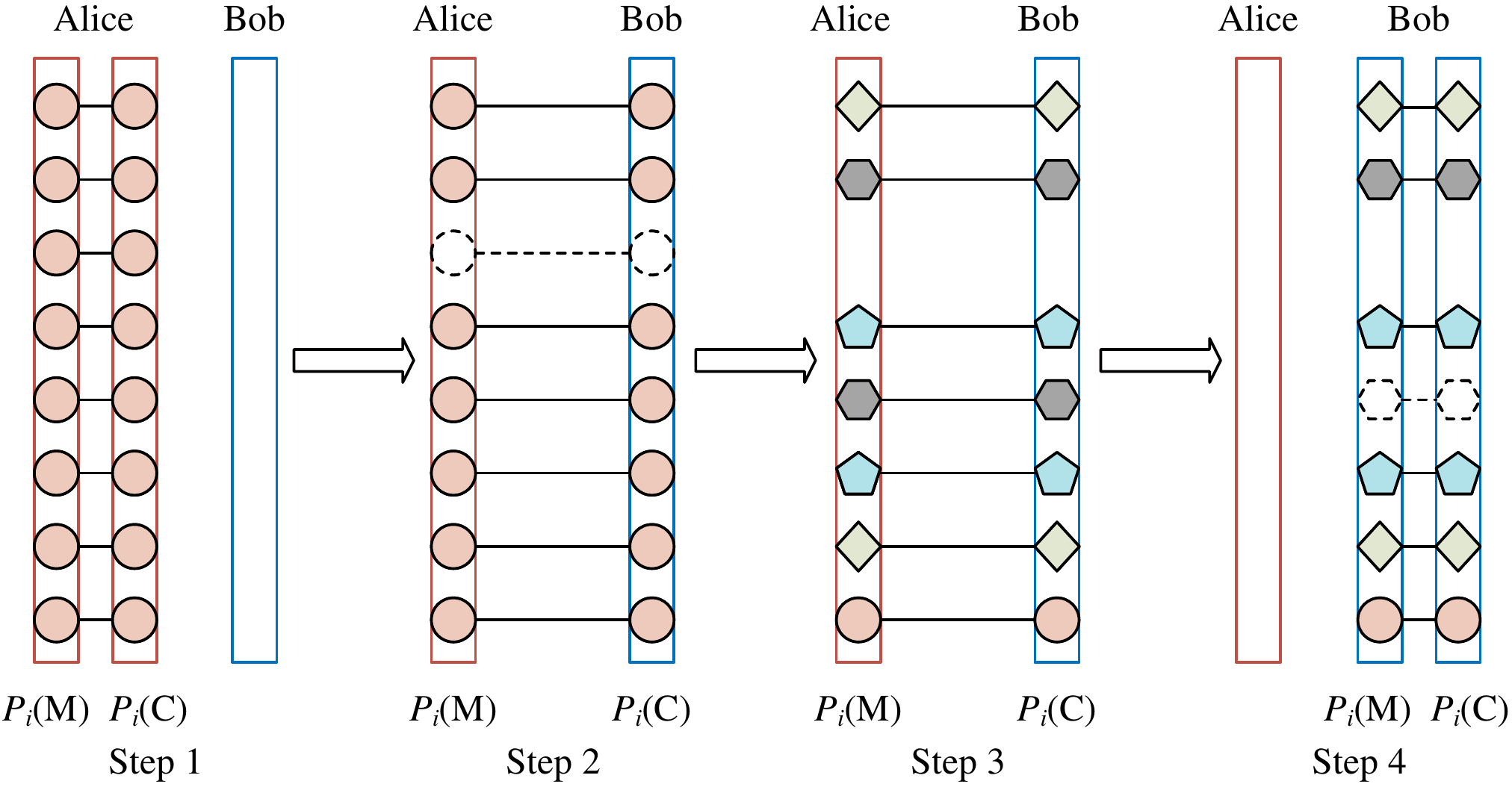}
\caption{The two-step QSDC protocol. Two hexagons linked with a line are in a Bell-state. The pentagons, the rhombuses, the hexagons, and the circles represent the Bell state $|\phi^{+}\rangle_{AB}$, $\left |\phi^{-}\right \rangle_{AB}$, $\left |\psi^{+}\right \rangle_{AB}$, and $\left |\psi^{-}\right \rangle_{AB}$, respectively. The blank symbols linked with a dashed line are the EPR pairs to do encoding-trick which is used for eavesdropping detection, it also in the one of the four Bell states while no secret message is carried.}
\label{TwostepQSDC}
\end{center}
\end{figure}

The two-step QSDC protocol depicted in Fig. \ref{TwostepQSDC} is described as follows \cite{deng2003two}.

\begin{itemize}
\item \textbf{Step 1, state preparation}. Alice and Bob agree on the specific mapping between the four Bell states $\left |\psi^{-}\right \rangle$, $\left |\psi^{+}\right \rangle$, $\left |\phi^{-}\right \rangle$, $\left |\phi^{+}\right \rangle$ and the two-bit classical information as 00, 01, 10, 11, respectively. Observe in Fig.~\ref{TwostepQSDC} that Alice prepares an ordered sequence of $N$ EPR pairs all in the state $\left |\psi\right \rangle_{\rm CM}$=$\left |\psi^{-}\right \rangle$=$1/\sqrt{2}\left(| 0\rangle_{\rm C}| 1\rangle_{\rm M}-| 1\rangle_{\rm C}| 0\rangle_{\rm M}\right)$, which is denoted by $[( P_{1}\left ( \rm C\right ),P_{1}\left ( \rm M\right ))$, $( P_{2}\left ( \rm C \right ),P_{2}\left ( \rm M\right ))$, $\cdots$, $( P_{i}\left ( \rm C \right ),P_{i}\left ( \rm M\right ))$, $\cdots$, $( P_{N}\left (\rm C \right ),P_{N}\left (\rm M \right ))]$. The subscript $i$ represents the index of the EPR pair in the sequence, while C and M represent the pair of particles in the EPR pair. Alice then divides the ordered sequence of $N$ EPR pairs into two EPR partner particle sequences. One of them is $[P_{1}\left(\rm C\right), P_{2}\left(\rm C\right),\cdots, P_{i}\left(\rm C\right), \cdots, P_{N}\left(\rm C\right)]$, which is called the checking sequence or C sequence for short. The other is the remaining EPR partner particle sequece $[P_{1}\left(\rm M\right), P_{2}\left(\rm M\right), P_{3}\left(\rm M\right), \cdots, P_{i}\left(\rm M\right), \cdots, P_{N}\left(\rm M\right)]$, which is termed as the message-coding sequence or M sequence, as seen in Fig.~\ref{TwostepQSDC}.
\item \textbf{Step 2, first transmission and first eavesdropping detection}. The C sequence $[P_{1}\left(\rm C\right)$, $P_{2}\left(\rm C\right)$, $P_{3}\left(\rm C\right)$, $\cdots$, $P_{N}\left(\rm C\right)]$ of Fig.~\ref{TwostepQSDC} is sent from Alice to Bob. Alice and Bob then detect eavesdropping through the following actions: (a) Bob randomly selects some EPR partner particles in the C sequence and tell Alice the position of these particles; (b) Bob randomly chooses one of the measurement bases $\{\sigma_{x},\sigma_{z}\}$ to measure the EPR partner particles selected; (c) Bob tells Alice which of the two measurement bases he has performed on each particles and additionally informs her of the outcome of his measurement; (d) Alice chooses the same measurement bases as Bob to measure the corresponding EPR partner particles in the M sequence. She will get the complete opposite results compared to Bob, provided that no eavesdropper contaminates the quantum channel: Alice gets 0 (1), Bob gets 1 (0). If the error rate is below the tolerance threshold, Alice and Bob conclude that there is no eavesdropper and proceed to next step. By contrast, in the presence of eavesdropping they curtail their communication.
\item \textbf{Step 3, information encoding}. As seen in Fig.~\ref{TwostepQSDC}, Alice performs one of the four unitary operations ($U_{0}$, $U_{1}$, $U_{2}$ and $U_{3}$) on each of the particles in the M sequence to encode her confidential messages. In the process of encoding, Alice has to apply an 'encoding-trick' to the M sequence. Explicitly, she randomly chooses some EPR partner particles in the M sequence as the samples to perform one of the four unitary operations, but no valuable payload information is mapped to them. Only Alice knows the position of these particles and she keeps it secret until the M sequence is transmitted to Bob. The number of these particles must be sufficiently high for estimating the error rate, and all the remaining particles are used for encoding confidential payload messages.
\item \textbf{Step 4, second transmission, measurement, and second eavesdropping detection}. Once Bob receives the M sequence, Alice tells him the position of the samples and the specific unitary operations applied to them. Bob performs the BSM on each and every EPR pair $[ P_{i}\left (\rm C \right ),P_{i}\left (\rm M \right )]$ to decode the confidential payload messages. By checking the measurement results, Bob will then get an estimate of the error rate within the current M sequence transmission. In fact, Eve is capable of perturbing the qubits, but cannot steal the confidential payload messages because she can only get one of the partner particles from an EPR pair in the second transmission. If the error rate of the Eve-checking pairs is reasonably low, Alice and Bob can then trust the process, and may proceed to correct the errors in the confidential payload messages using a classical-domain error correction method. Otherwise, they have to abandon this particular transmission session and go back to Step 1 of Fig.~\ref{TwostepQSDC}.
\end{itemize}
 
Let us now continue our journey through QSDC history by considering the DL04 protocol.

\subsubsection{Deng-Long 2004 (DL04) QSDC Protocol}
\label{sec:DL04 QSDC protocol}
\begin{figure*}[!h]
\begin{center}
\includegraphics[width=\textwidth,angle=0]{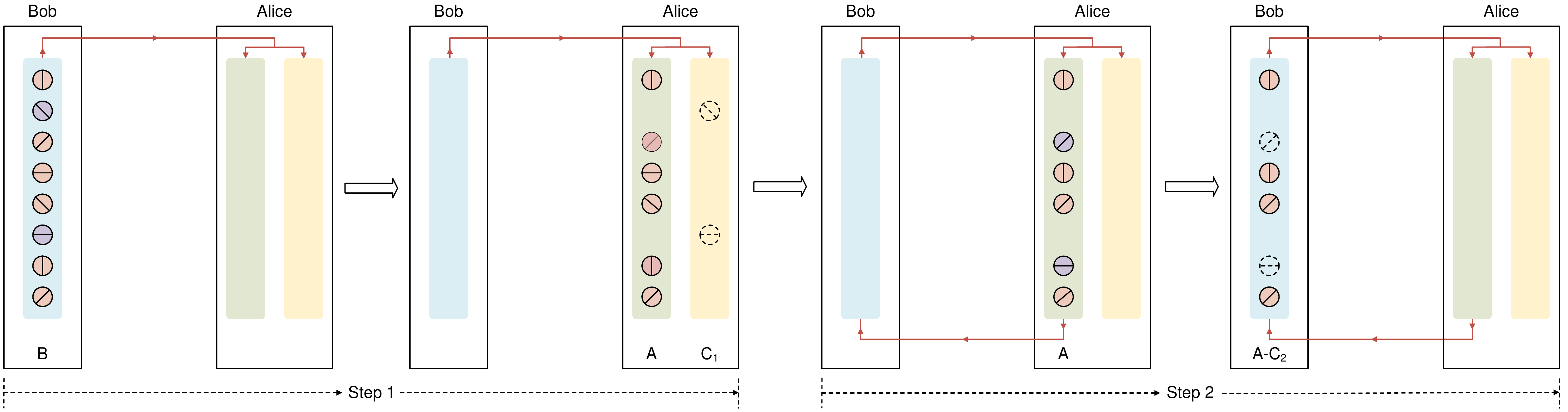}
\caption{Illustrating of DL04 QSDC protocol. The circles having vertical, horizontal, diagonal, and backslash lines correspond to the quantum state $|0\rangle$, $|1\rangle$, $|+\rangle$, and $|-\rangle$, respectively. To distinguish the Eve-checking samples from the confidential messages, the checking samples are denoted by the purple circles although they are also in one of the four states \{$|0\rangle, |1\rangle, |+\rangle, |-\rangle$\}.}
\label{fig:DL04}
\end{center}
\end{figure*}

As Fig. \ref{fig:DL04} shows, single photons are used as carriers of confidential messages in the DL04 protocol, which relies on the following two steps \cite{deng2004secure}.

\begin{itemize}
\item \textbf{Step 1, the secure postal pigeon sending stage}. As seen in Fig.~\ref{fig:DL04}, $N$ single photons are prepared by Bob and each of them is randomly mapped to one of the four states $\{\left |0\right \rangle, \left |1\right \rangle, \left |+\right \rangle, \left |-\right \rangle\}$, as shown in Table \ref{table:1}. These single photons are called B-batch photons, and they are transmitted directly to Alice after preparation. Upon receiving the batch of single photons, Alice and Bob check the presence of eavesdropping by the following actions: (a) Sufficient single photon samples are picked randomly from the B batch. This fraction is termed as the $\rm C_1$ batch, which is marked in gray in Table \ref{table:1}, leaving behind the other photons forming the B batch having a cardinality of $\rm A=B-C_1$;  (b) Alice randomly chooses either the measurement basis Z or X for measuring each photon in the $\rm C_1$ batch and then publishes both the position of these photons as well as the measurement bases applied to them and the measurement results; (c) Bob then calculates the error rate to estimate the probability of eavesdropping. More specifically, Bob compares the measurement results of Alice to his original quantum states to obtain the error rate, when Alice has chosen the same measurement basis as the preparation basis of Bob. If the error rate is lower than a pre-determined threshold, the transmission of the B batch through the quantum forward channel is considered to be secure and they proceed to the next step. Otherwise, the communication is aborted.
\item \textbf{Step 2, the message coding and postal pigeon returning stage}. Alice decides either to perform the operation $I=|0\rangle\langle0|+|1\rangle\langle1|$ to encode the information 0 or to perform the operation $U=i\sigma_{y}=|0\rangle\langle1|-|1\rangle\langle0|$ to encode the message 1\footnote{This is a process of quantum one-time pad, because it relies on inserting some decoy photons into the message photons and only Alice knows the positions of these decoy photons.}. Observe that the unitary operation $U$ acts as flipping the state in both measuring bases, hence we can obtain
\begin{eqnarray}
U|0\rangle=-|1\rangle,\; U|1\rangle=|0\rangle,
\end{eqnarray}
and
\begin{eqnarray}
U|+\rangle=|-\rangle,\; U|-\rangle=-|+\rangle.
\end{eqnarray}
This operation offers the option of deterministically detecting confidential messages for Bob. To guarantee the security of the second transmission, Alice has to randomly choose some photons in the A-batch as Eve-checking samples. We refer to them as the $\rm C_2$-batch and Alice maps random bits to them. These instances are marked in gray in Table \ref{table:1}. She publicly announces the positions of these photons and of the coded random bits after Bob receiving the returned photons. Armed with the knowledge of his preparation bases and original quantum states, Bob directly decodes the confidential payload messages and random bits by using the same preparation bases to measure the returned photons. Then the error rate is estimated to assess if there has been any eavesdropping attack from Eve.
\end{itemize}

\begin{table*}[!h]
\begin{footnotesize}
\begin{center}
\caption{DL04 QSDC protocol example corresponding to Fig~\ref{fig:DL04}.}
\begin{tabular}{|c|l|c|c|c|c|c|c|c|c|}
\hline
\multicolumn{10}{|c|}{Step 1}                                                                                                                                                                                                                                                                                                                                                                                                                                \\ \hline
                                            & Preparation bases                         & Z                                & X                                   & X                                                        & Z                                & X                                & Z                                                          & Z                                                        & X                                \\ \cline{2-10}
\multirow{-2}{*}{Bob}                       & Original quantum states (A-batch)          & $|0\rangle$                      & \cellcolor[HTML]{C0C0C0}$|-\rangle$ & $|+\rangle$                                              & $|1\rangle$                      & $|-\rangle$                      & \cellcolor[HTML]{C0C0C0}{\color[HTML]{000000} $|1\rangle$} & $|0\rangle$                                              & $|+\rangle$                      \\ \hline
                                            & Eavesdropping detection                   &                                  & X                                   &                                                          &                                  &                                  & X                                                          &                                                          &                                  \\ \cline{2-10}
\multirow{-2}{*}{Alice}                     & Quantum states detected                    &                                  & $|-\rangle$                         &                                                          &                                  &                                  & $|+\rangle$                                                &                                                          &                                  \\ \hline
\multicolumn{10}{|c|}{Step 2}                                                                                                                                                                                                                                                                                                                                                                                                                                \\ \hline
                                            & Encoding operations & I                                &                                     & I                                                        & U                                & U                                &                                                            & U                                                        & I                                \\ \cline{2-10}
                                            & Quantum sates after encoding (B-batch)     & \multicolumn{1}{l|}{$|0\rangle$} & \multicolumn{1}{l|}{}               & \multicolumn{1}{l|}{\cellcolor[HTML]{C0C0C0}$|+\rangle$} & \multicolumn{1}{l|}{$|0\rangle$} & \multicolumn{1}{l|}{$-|+\rangle$} & \multicolumn{1}{l|}{}                                      & \multicolumn{1}{l|}{\cellcolor[HTML]{C0C0C0}$-|1\rangle$} & \multicolumn{1}{l|}{$|+\rangle$} \\ \cline{2-10}
\multirow{-3}{*}{Alice}                     & Secret message bits or random number bits & 0                                &                                     & 0                                                        & 1                                & 1                                &                                                            & 1                                                        & 0                                \\ \hline
\multicolumn{1}{|l|}{}                      & Measurement bases & Z                                &                                     & X                                                        & Z                                & X                                &                                                            & Z                                                        & X                                \\ \cline{2-10}
\multicolumn{1}{|l|}{\multirow{-2}{*}{Bob}} & Decoding Results                          & 0                                & \multicolumn{1}{l|}{}               & {\cellcolor[HTML]{C0C0C0}0}                                                        & 1                                & 1                                & \multicolumn{1}{l|}{}                                      & {\cellcolor[HTML]{C0C0C0}1 }                                                       & 0                                \\ \hline
\end{tabular}
\label{table:1}
\end{center}
\end{footnotesize}
\end{table*}

By now the confidential payload messages have been transmitted directly over the quantum channel. In addition to the capability of detecting the presence of an  eavesdropper, the communicating parties must ensure that the secret messages are unlikely to be leaked to an eavesdropper before she is detected. Therefore, eavesdropping detection is necessary before mapping the secret messages to the single photons. Although the eavesdropper can intercept the quantum states carrying the confidential messages in the second transmission, she can only infer random results by measuring them, since she is unaware of the original quantum state. Alice encodes the confidential message in Step 2 just like in the one-time pad encryption. The quantum batch $\rm C_2$ is then inserted randomly into the secret message encoding sequence in this QSDC protocol, which encrypts the transmitted messages to ciphertext. A QSDC protocol relying on single photons was presented in~\cite{liu2012high}, where every qubit can carry 2 bits of information, as the transmitter can map messages both to  the polarization states and to the spatial-mode states of single photons completely independently.

\subsubsection{High-Dimensional Two-Step QSDC Protocol}
\label{sec:high-dimensional two-step QSDC protocol}
\begin{figure}[!h]
\begin{center}
\includegraphics[width=\columnwidth,angle=0]{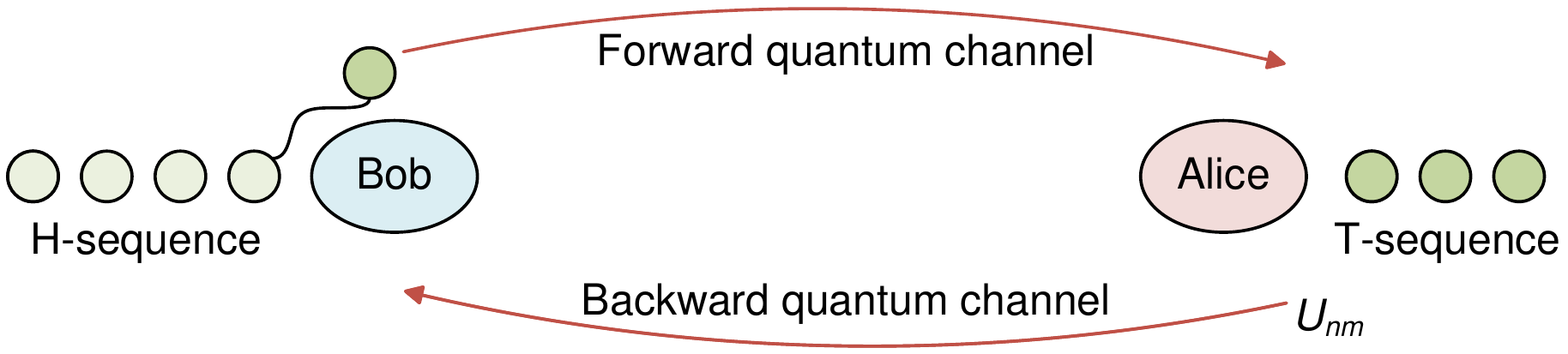}
\caption{Scheme showing principles involved in the high-dimension two-step QSDC of~\cite{wang2005quantum}. The connected circles represent an EPR pair. The bottle green circle is the EPR partner particle that belongs to what we refer to as T-sequence, while the light green one belongs to the H-sequence. $U_{nm}$ is the unitary transformation used for encoding.}
\label{fig:Hdtwostep}
\end{center}
\end{figure}

Consider a quantum system relying on a $d$-dimensional Hilbert space \cite{liu2002general}, where a set of maximally $d$-dimensional Bell states can be defined as follows
\begin{eqnarray}
|\Psi_{nm}\rangle=\sum_{j}e^{2\pi ijn/d}|j\rangle\otimes|j\oplus m\rangle/\sqrt{d},
\end{eqnarray}
where we have $n,\, m,\, j=0,\, 1,\,\cdots,\, d-1$, $j\oplus m=(j+m)\,$mod$\,d$. The unitary transformation of 
\begin{eqnarray}
U_{nm}=\sum_{j}e^{2\pi ijn/d}|j\oplus m\rangle\langle j|
\end{eqnarray}
can map the Bell state $|\Psi_{00}\rangle=\sum_{j}|j\rangle\otimes|j\rangle/\sqrt{d}$ to the Bell state $|\Psi_{nm}\rangle$, formulated as
\begin{eqnarray}
U_{nm}|\Psi_{00}\rangle=|\Psi_{nm}\rangle.
\end{eqnarray}
Now, let us describe the detailed steps of the high-dimensional two-step QSDC protocol of~\cite{wang2005quantum} with the aid of Fig.~\ref{fig:Hdtwostep}.

\begin{itemize}
\item \textbf{Step 1, state preparation}. Bob prepares a sequence of $d$-dimensional Bell states $|\Psi_{00}\rangle_{HT}$. The subscripts $H$ and $T$ serve as the labels of the two particles in the EPR pairs. Explicitly, $H$ represents the home particle of Bob while the $T$-particle will be transmitted by Alice to Bob and returned to Alice later. Bob selects one of the particles in each EPR pair to construct a partner particle queue, i.e., $[P_{1}(\rm H)$, $P_{2}(\rm H)$, $P_{3}(\rm H)$, $\cdots$, $P_{N}(\rm H)]$, termed as the home sequence or H-sequence for short. Thus the other new queue of $[P_{1}(\rm T)$, $P_{2}(\rm T)$, $P_{3}(\rm T)$, $\cdots$, $P_{N}(\rm T)]$ is composed of the remaining partner particles of each and every EPR pair. This queue may be referred to as the travel sequence or T-sequence. The subscript $N$ refers to the EPR pair index in the sequence.
\item \textbf{Step 2, first transmission and first eavesdropping detection}. Bob sends the T-sequence to Alice, and then they carry out the following substeps to finish the first eavesdropping detection. (a) Alice randomly selects one of several conjugate single-particle measurement bases to measure each of the sample particles, which are randomly picked from the $T$-sequence; (b) Alice then publishes the index of sample particles and also the choice of the measurement bases applied to them; (c) Bob applies measurements to the EPR partner particles of those that are Alice's sample particles; (d) both Alice and Bob disclose their own measurement results, hence Bob compares his measurement results to those of Alice to check whether or not an eavesdropper attack perturbs the quantum states. If their results are highly correlated, they continue to the next step. Otherwise, the first transmission of particles is insecure and they have to abandon this communication session and restart from Step 1.
\item \textbf{Step 3, information encoding}. Alice maps her secret message bits to the photons of the T-sequence with the aid of the unitary operation $U_{nm}$, but excludes the specific sample particles that have been chosen in Step 2. In addition, Alice and Bob have to detect the presence of the eavesdropper by comparing their appropriately selected particles for estimating the error rate. To do this, Alice randomly chooses some photons of the T-sequence for conveying random bits during the process of secret message encoding. She will not expose the position and the encoding bases $U_{mn}$ of these sample particles before Bob receives the T-sequence. Since the encoding bases of the secret message and random bits are $U_{mn}$, the eavesdropper does not know, which particles carry the message and which convey random bits. This is beneficial for confidentiality.
\item \textbf{Step 4, measurement}. Bob applies a joint BSM to every EPR pair after receiving the T-sequence from Alice. At this time, he has both the H-sequence and T-sequence again, except for the sample particles that have been used for the first eavesdropping detection.
\item \textbf{Step 5, second eavesdropping detection}. After Alice sends the index of the sample particles and the type of unitary operation in Step 4 as classical information to Bob, he carries out the second eavesdropping detection by combining his own measurement results. If the error rate is too high, Alice and Bob must abandon this transmission session and restart it from the beginning.
\end{itemize}

As discussed in Ref. \cite{bechmann2000quantum}, the high-dimensional two-step QSDC provides better security than that obtainable with the aid of two-dimensional Bell states \cite{deng2003two}. Furthermore, the protocol has the advantage of high capacity \cite{krenn2014generation}, where a particle can carry ${\rm log}_{2}d^{2}$ bits of classical information.

\subsubsection{Measurement-Device-Independent QSDC Protocol}
\label{sec:Measurement-device-independent QSDC protocol}
In theory, the quantum cryptographic protocols are unconditionally secure as guaranteed by the laws of quantum physics. However, the practical devices suffer from inevitable imperfections that can be exploited by the eavesdropper to infer some confidential information, especially when using single-photon detectors. For example, Huang \textit{et al}.~\cite{huang2018implementation} showed that QSDC systems may be compromised by detector blinding attacks \cite{lydersen2010hacking}. As a remedy, measurement-device-independent (MDI) protocols were proposed for protecting practical quantum cryptographic systems against the detector side channel attacks \cite{zhou2020measurement,niu2018measurement}. In the MDI protocol, the measurement-device is under the control of an untrusted party called Charlie who performs a BSM. Note that even if an adversary controls the measurement-device, he would not gain any useful information about the confidential message. Hence MDI protocols can remove all security loopholes from the measurement unit. More significantly, the realization of MDI protocols is entirely feasible at the time of writing. Recently, Gao \textit{et al}. \cite{gao2019long} proposed a long-distance MDI QSDC protocol by relying on  ancillary entangled sources, which were located in the middle of the link by adding an extra relay node.

\begin{figure}[!h]
\begin{center}
\includegraphics[width=6cm,angle=0]{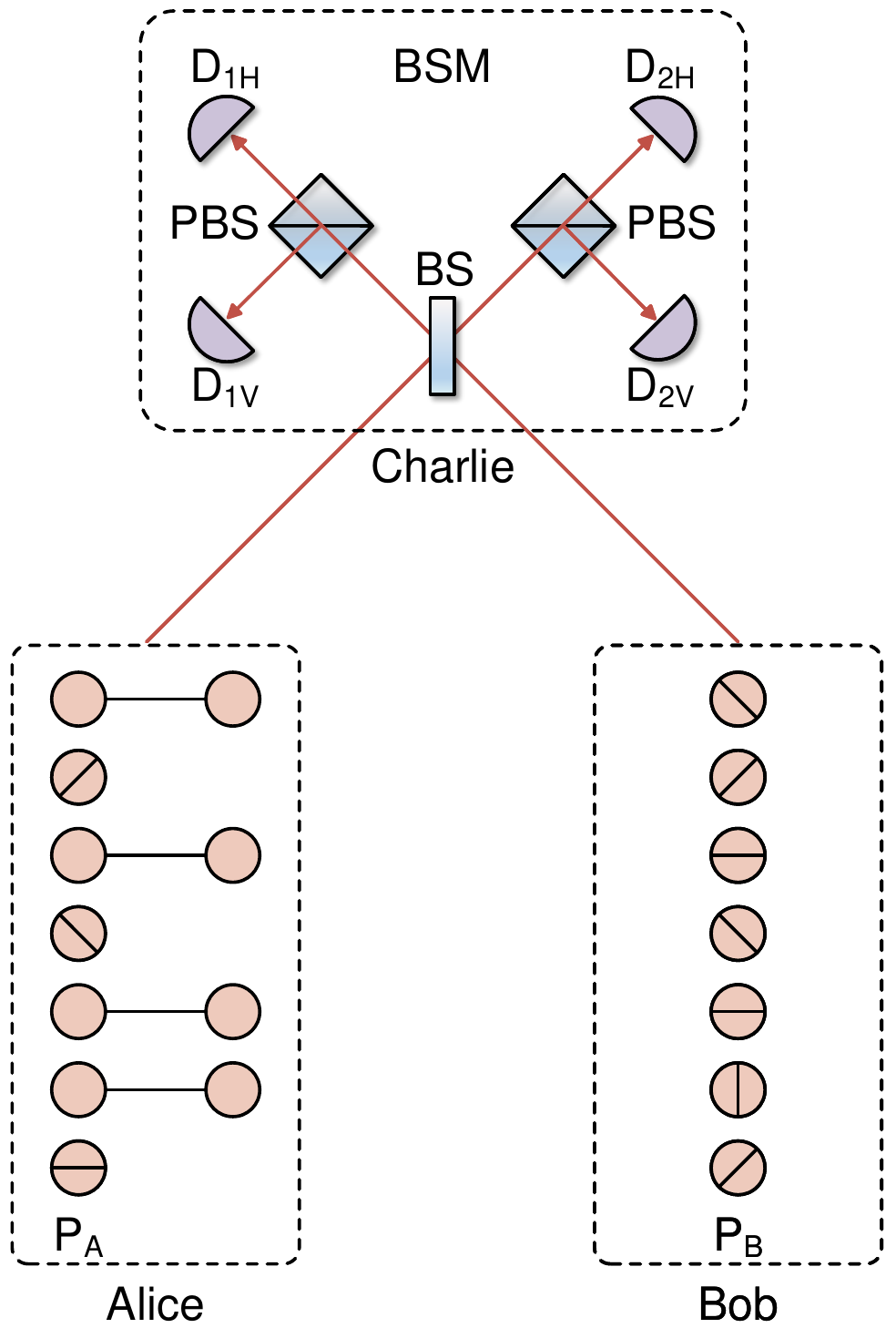}
\caption{The MDI QSDC protocol. Two circles linked with a line denote the Bell state $|\psi^{-}\rangle$. The circles having vertical, horizontal, diagonal, and backslash represent the single photons $|0\rangle$, $|1\rangle$, $|+\rangle$, and $|-\rangle$, respectively. BS: beam splitter, PBS: polarization beam splitter, D: single photon detector, BSM: Bell-state measurement.}
\label{fig:MDIQSDC}
\end{center}
\end{figure}

The measurement-device-independent QSDC protocol is illustrated in Fig. \ref{fig:MDIQSDC}, which relies on both single-photon states $\{|0\rangle, |1\rangle, |+\rangle, |-\rangle\}$ and Bell states $\{|\psi^{+}\rangle, |\psi^{-}\rangle, |\phi^{+}\rangle, |\phi^{-}\rangle\}$. It may be summarized in the following steps \cite{zhou2020measurement}.

\begin{itemize}
\item \textbf{Step 1, state preparation}. Alice prepares a queue of $N+t_{0}$ EPR pairs, all of which are in the state $|\psi^{-}_{12}\rangle$. Then the EPR pair sequence is separated into two parts: $\rm S_{Ah}$ and $\rm S_{At}$, each of which includes one of the particles in the EPR pair. She also generates a sequence of $t_{1}$ single photons, each randomly representing one of the four states  $\{|0\rangle, |1\rangle, |+\rangle, |-\rangle\}$. These single photons are then  randomly inserted into the EPR partner particle sequence $\rm S_{At}$. Therefore an ordered qubit sequence $P_{A}$ is prepared by Alice, as seen in Fig.~\ref{fig:MDIQSDC}. In the meantime, Bob produces a sequence of $N+t_{0}+t_{1}$ single photons, which are randomly in one of the four states $\{|0\rangle, |1\rangle, |+\rangle, |-\rangle\}$. This sequence is denoted by $\rm P_{B}$ in Fig.~\ref{fig:MDIQSDC}.
\item \textbf{Step 2, qubit transmission and measurement}. Alice sends the sequence $\rm P_{A}$ to an untrusted relay termed as Charlie and located in the middle, while Bob sends the sequence $\rm P_{B}$ to Charlie. Charlie then performs a BSM that projects the incoming qubits into a Bell state, and he publishes the measurement results. Once a partner particle from an EPR pair of $P_{A}$ and a single photon from $P_{B}$ are projected into a Bell state, the other partner particle of Alice is instantaneously collapsed into one of the four states $\{|0\rangle, |1\rangle, |+\rangle, |-\rangle\}$ with equal probabilities\footnote{This is actually a process of quantum teleportation carried out in a more complicated manner, in which Bob's state associated with a unitary operation $U_{T}$ ($U_{T}=I$ or $U_{T}=i\sigma_{y}$) is teleported to Alice. $U_{T}$ is only known by Bob.}, as shown in Table \ref{table:2}. The state after BSM can be deduced by Bob according to Table \ref{table:2}, but it is unknown to anyone else. For example, Alice's state is $|0\rangle_{1}$ for $|\psi^{-}_{12}\rangle|1\rangle_{3}$ if the BSM result of Charlie is $|\phi^{+}_{23}\rangle$.
\item \textbf{Step 3, eavesdropping detection}. To proceed further, Alice announces publicly the index and states of the $t_{1}$ single photons. Bob also publishes the state of the corresponding single photon in the sequence $P_{B}$. They then compare the BSM results, where they employ the same bases. The method of eavesdropping detection is identical to that in the MDI QKD \cite{lo2012measurement}. A BSM will project the incoming two photons that were  prepared by the same two bases into one of the two Bell states, formulated as,
\begin{eqnarray}
|+\rangle_{1}|+\rangle_{3}=\frac{1}{\sqrt{2}}\left(|\phi^{+}\rangle_{13}+|\psi^{+}\rangle_{13}\right),\\
|+\rangle_{1}|-\rangle_{3}=\frac{1}{\sqrt{2}}\left(|\phi^{-}\rangle_{13}-|\psi^{-}\rangle_{13}\right),\\
|0\rangle_{1}|0\rangle_{3}=\frac{1}{\sqrt{2}}\left(|\phi^{+}\rangle_{13}+|\phi^{-}\rangle_{13}\right),\\
|0\rangle_{1}|1\rangle_{3}=\frac{1}{\sqrt{2}}\left(|\psi^{+}\rangle_{13}+|\psi^{-}\rangle_{13}\right).
\end{eqnarray}
Charlie has a 50\% probability to obtain the other two Bell-states in the face of  eavesdropping attacks, so the error rate is increased. Nonethless, if the error rate remains below the maximum tolerable level, they proceed to the next step. Otherwise, they decide to abort the communication session.
\item \textbf{Step 4, confidential message encoding}. As her next action, Alice applies one of two unitary operators $\{I, i\sigma_{y}\}$ to the particles in her hand where the unitary operation $U_{m}=I$ represents 0 and the operation $U_{m}=i\sigma_{y}$ corresponds to 1. To provide an integrity check for the confidential message, $t_{0}$ EPR pair partner particles are randomly positioned to encode random bits. Then Bob publishes his preparation bases of the remaining single photons.
\item \textbf{Step 5, qubit transmission and measurement}. Alice sends the qubits to Charlie after encoding. Charlie measures the qubits by using the bases that Bob has published in Step 4 and announces the measurement results.
\item \textbf{Step 6, decoding and integrity check}. Bob decodes Alice's bits by combining his initial states with the measurement results of Step 2 and Step 5. Alice discloses the index of the $t_{0}$ EPR pair partner particles selected as well as the random bits mapped to them. If no perturbation is imposed by Eve and the direct communication is deemed to be secure, the error rate will be below the maximum tolerable threshold. Then Bob obtains the confidential payload message bits.
\end{itemize}

\begin{table}[!h]
\begin{footnotesize}
\begin{center}
\caption{The correspondence among Bob's state, Charlie's BSM result and Alice's state.}
\begin{tabular}{|c|c|cccc|c}
\hline
\multicolumn{2}{|c|}{\multirow{2}{*}{}}        & \multicolumn{4}{c|}{Charlie's BSM results}                                                                                                                             \\ \cline{3-6}
\multicolumn{2}{|c|}{}                         & \multicolumn{1}{c|}{$|\phi^{+}_{23}\rangle$} & \multicolumn{1}{c|}{$|\phi^{-}_{23}\rangle$} & \multicolumn{1}{c|}{$|\psi^{+}_{23}\rangle$} & \multicolumn{1}{c|}{$|\psi^{-}_{23}\rangle$} \\ \hline
\multirow{4}{*}{Bob's state} & $|0\rangle_{3}$ & $-|1\rangle_{1}$                             & $-|1\rangle_{1}$                             & $|0\rangle_{1}$                             & $-|0\rangle_{1}$                             \\ \cline{2-2}
                             & $|1\rangle_{3}$ & $|0\rangle_{1}$                              & $-|0\rangle_{1}$                             & $-|1\rangle_{1}$                             & $-|1\rangle_{1}$                             \\ \cline{2-2}
                             & $|+\rangle_{3}$ & $|-\rangle_{1}$                             & $-|+\rangle_{1}$                             & $|-\rangle_{1}$                              & $-|+\rangle_{1}$                             \\ \cline{2-2}
                             & $|-\rangle_{3}$ & $-|+\rangle_{1}$                             & $|-\rangle_{1}$                              & $|+\rangle_{1}$                              & $-|-\rangle_{1}$
                                                        \\ \cline{1-2}       \hline
\end{tabular}
\label{table:2}
\end{center}
\end{footnotesize}
\end{table}

The MDI QSDC protocol has the same security level as the MDI QKD protocol in Step 3 and the teleportation process of Bob's state is also secure after the eavesdropping detection. Firstly, Alice sends two kinds of photons to Charlie, but they have the same density matrix, $Tr_{1}\left(|\psi^{-}_{12}\rangle\langle\psi^{-}_{12}|\right)=I/2$ for the EPR pairs and $\frac{|0\rangle\langle0|}{4}+\frac{|1\rangle\langle1|}{4}+\frac{|+\rangle\langle+|}{4}+\frac{|-\rangle\langle-|}{4}=I/2$ for single photons. Hence Eve cannot differentiate between the EPR pair particles and single photons. Secondly, Bob knows the initial state $|q\rangle_{3}$, but it is sealed to others, so only he can infer the unitary encoding operation $U_{m}$ of Alice, even though both Alice and Charlie know the result of $U_{m}|q\rangle_{3}$.


\subsubsection{Marino-Stroud 2006 (MS06) DSQC Protocol}
\label{sec:MS06 DSQC protocol}
The DSQC protocol has also been extended to the continuous-variable (CV) domain related to infinite-dimensional Hilbert spaces \cite{braunstein2005quantum,hosseinidehaj2018satellite,marino2006deterministic,pirandola2008quantum,pirandola2009confidential}.
The continuous variable DSQC scheme uses the squeezing phase of a two-mode squeezed state, as detailed in~\cite{marino2006deterministic}. The characteristics of the two-mode squeezed state are eminently suitable for quantum direct communication and the associated communication protocol is shown in Fig. \ref{fig:CVDSQC}. The whole process can be divided into the following 6 steps:

\begin{figure}[!h]
\begin{center}
\includegraphics[width=\columnwidth,angle=0]{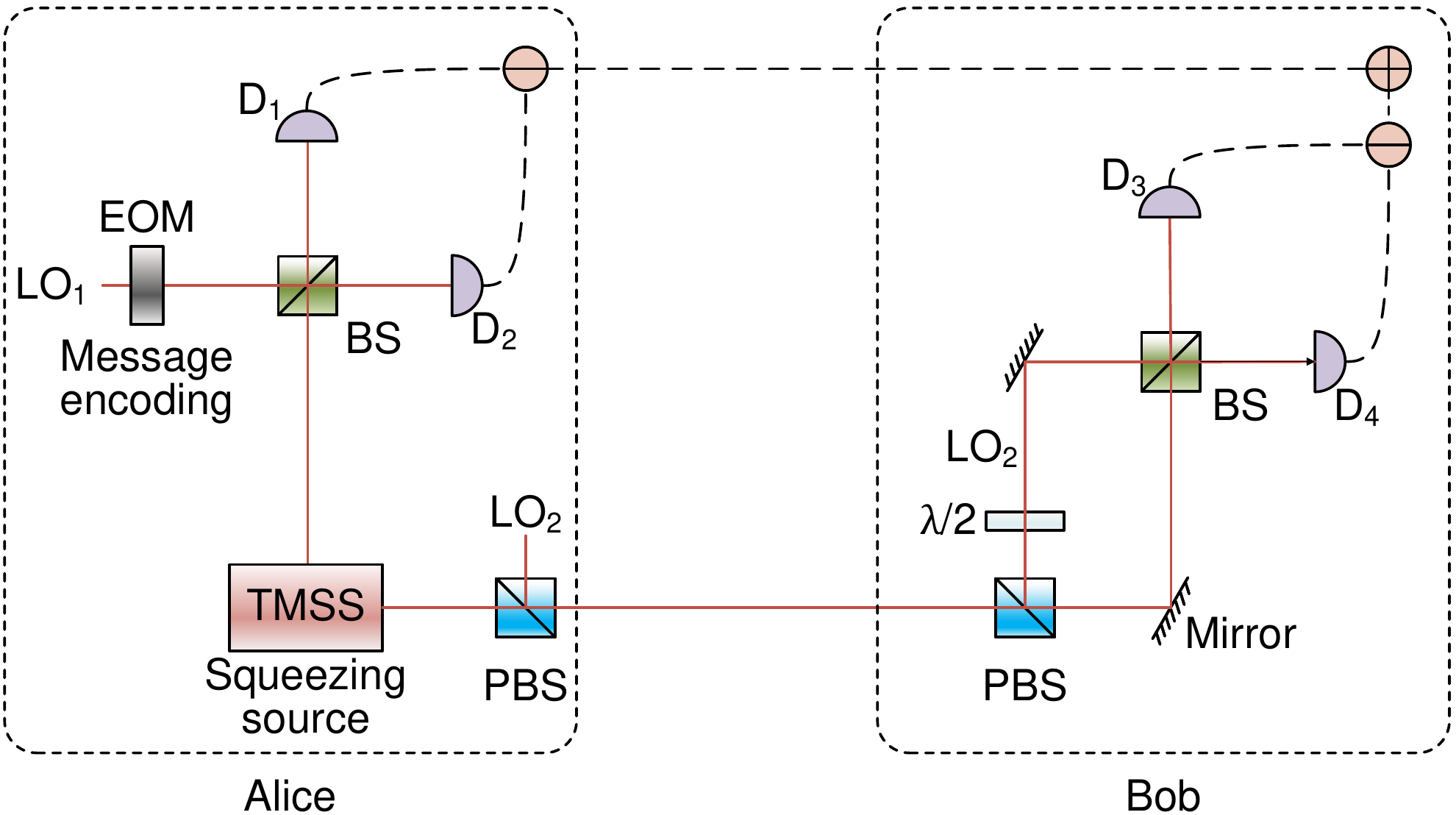}
\caption{Schematic of the CV DSQC protocol using two-mode squeezed states. LO: local oscillator, EOM: electro-optical modulator, BS: beam splitter, TMSS: two-mode squeezed state, PBS: polarizing beam splitter,  D: photodetector, $\lambda$/2: half wave plate.}
\label{fig:CVDSQC}
\end{center}
\end{figure}

\begin{itemize}
\item \textbf{Step 1, state preparation and transmission}. Alice generates and distributes the two-mode squeezed state, keeping one of the modes at her side and sending the other mode to Bob over a quantum channel as indicated by Fig.~\ref{fig:CVDSQC}. A coherent-state local oscillator $\rm{LO}_{2}$ is also sent to Bob by Alice together with the transmitted mode by using a polarizing beam splitter for combining them.
\item \textbf{Step 2, encoding}. Alice maps the confidential messages to her part of the two-mode squeezed state by imposing a phase shift on the $\rm{LO}_{1}$ and then carries out homodyne detection. She also inserts check bits at random time slots throughout the message encoding process, so that the communicating parties can detect eavesdropping.
\item \textbf{Step 3, measurement}. Bob splits the incoming beam in half at his polarizing beam splitter, which are then distributed to the separate parties, namely to $\rm{LO}_{2}$ and to the squeezed mode. Then, they are combined in a balanced beam splitter to perform homodyne detection and the phase of $\rm{LO}_{2}$ is kept constant during the measurement.
\item \textbf{Step 4, repeated transmission}. Alice and Bob repeat Step 1 to Step 3 until the secret messages have been completely encoded.  
\item \textbf{Step 5, eavesdropping detection}. Alice informs Bob of the time slot of her Eve-check bits and of the corresponding measurement results. Alice and Bob publicly compare the Eve-check bits to evaluate the error rate and confirm whether an eavesdropper is present. If the channel's integrity has been verified, they continue to the next step. Otherwise, they restart their communication session over a different quantum channel.
\item \textbf{Step 6, message decoding}. Alice sends the measurement results of the confidential messages to Bob. Once the two local measurement results are combined by Bob, he will have a signal whose variance fluctuates between two different levels, which represent the confidential information of Alice. Hence, Bob can retrieve the entire confidential message.
\end{itemize}

This protocol is immune to the intercept and resends attacks. Eve may  intercept the mode sent to Bob. If this occurs, then she can extract the information of squeezing degree and she can then resend a new mode at the same squeezing degree to Bob. This fraudulent action can be easily detected from the new mode sent by Eve, because it is not entangled with the mode retained by Alice. More explicitly, this will result in increased noise and no pair of distinct variances in the combined signal of Bob. As for the partial interception strategy, Eve can acquire part of the beam and combine it with the published measurement results of Alice to steal information. However, this will result in a reduction of the squeezing degree, making it easy to detect.

\subsubsection{ZXFZ DSQC Protocol}
\label{sec:ZXFZ DSQC protocol}
In 2003, a specific encryption scheme was invented~\cite{deng2003controlled}, which was subsequently adopted also to design a DSQC protocol by Zhu, Xia, Fan, and Zhang (ZXFZ)~\cite{zhu2006secure,wang2006quantum}. Below we summarize the ZXFZ DSQC protocol, which is based on a secret transmission order of EPR pairs \cite{zhu2006secure}. To elaborate, Alice and Bob exploit that the unitary operations of $U_{0}=I=|0\rangle\langle0|+|1\rangle\langle1|$, $U_{1}=\sigma_{z}=|0\rangle\langle0|-|1\rangle\langle1|$, $U_{2}=\sigma_{x}=|1\rangle\langle0|+|0\rangle\langle1|$, and $U_{3}=i\sigma_{y}=|0\rangle\langle1|-|1\rangle\langle0|$ are used for conveying two bits of confidential information 00, 11, 01, and 10, respectively.

\begin{itemize}
\item \textbf{Step 1, state preparation}. Alice prepares an EPR pair sequence, where $[P_{1}\left(1,1'\right), P_{2}\left(2,2'\right),\cdots,P_{i}\left(i,i'\right),\cdots,P_{N}\left(N,N'\right)]$, each pair is in the same state of $|\psi^{-}\rangle=\frac{1}{\sqrt{2}}\left(|0\rangle_{i}|1\rangle_{i'}-|1\rangle_{i}|0\rangle_{i'}\right)$. A sufficiently large subset is selected for the Eve-checking set (C set) and the rest of the EPR pairs serve as the confidential message set (M set). The basic elements of the C set or M set are EPR pairs, rather than being a single photon of an EPR pair. Alice uses the above four unitary operations for encoding the confidential message onto the M set, while random bits are mapped to the C set. Taking the C set as an example, Alice's random bits are (0100101101$\cdots$) and she chooses the first 50 EPR pairs as the C set. So she maps 01 to $P_{1}\left(1,1'\right)$, 00 to $P_{2}\left(1,1'\right)$, 10 to $P_{3}\left(3,3'\right)$,$\cdots$, by applying the four unitary operations to one of the particles in each EPR pair.
\item \textbf{Step 2, transmission}. Given a secret transmit order of particles, Alice sends these particles to Bob one by one. A partner particle in the EPR pair is taken as the minimal transmitted unit. For instance, Alice sends the particles in the order of $S_{1}(2)$,  $S_{2}(1)$, $S_{3}(51)$, $S_{4}(5')$, $S_{5}(2')$, $S_{6}(60)$, $S_{7}(10)$, $S_{8}(1')$, $\cdots$, $S_{j}(x)$, $\cdots$, $S_{k}(x')$, $\cdots$, $S_{2N}(y)$, where $S_{j}(i), (j\in\{1, 2, \cdots, 2N\}, i\in\{1, 2, \cdots, N\})$. This means that Alice sends the original particle $i$ at the $j-\rm{th}$ output index.
\item \textbf{Step 3, announcing correlation}. Bob confirms to Alice that he has recieved all the $2N$ particles. Alice then announces the exact quantum correlation of two particles that pertain to the C set over a public channel as exemplified by $S_{2}\sim S_{8}$, $S_{1}\sim S_{5}$, $\cdots$, $S_{j}\sim S_{k}$, $\cdots$.
\item \textbf{Step 4, eavesdropping detection}. Bob performs a BSM based on the pairs Alice has told him, and then the measurement results are published subsequently. Upon comparing the measurement results with her original information, Alice estimates the error rate and detects the presence or absence of eavesdroppers.
\item \textbf{Step 5, announcing correlation}. When Alice can ascertain that no eavesdropper is present, she classically informs Bob of the matching information of two particles in the M set. If Alice finds an unacceptably high error rate, she curtails the communication session and starts a new one from the Step 1.
\item \textbf{Step 6, measurement}. Finally, Bob decodes the confidential messages by performing the BSM.
\end{itemize}

The security of this protocol is ensured by the secret order of the particles. Even if Eve captures all the particles, she cannot glean any useful message without knowing the correct order. But this scheme becomes  insecure, if an eavesdropper steals the secret message by relying on Trojan horse attack strategies, hence it has been improved in Ref~\cite{li2006improving}.

\subsubsection{Other Protocols for QSDC or DSQC}
\label{sec:Other protocols for QSDC or DSQC}
As detailed above, numerous theoretical proposals have been conceived for QSDC or DSQC, in which the communication security is guaranteed by encrypting information using quantum states~\cite{he2006quantum,pirandola2008quantum,pirandola2009confidential,lum2016quantum} or by denying eavesdroppers access to the entirety of correlated quantum states~\cite{shapiro2019quantum,zhuang2015ultrabroadband}. To achieve confidential direct communication, Alice can also map the secret message to a coherent state $|\alpha_{M}\rangle$ and add a random amplitude $\alpha_{R}$ chosen from a Gaussian-distribution for ensuring that an encrypted quantum state of $|\alpha_{M}+\alpha_{R}\rangle$ is used for conveying confidential information along with random bits, which will be revealed to Bob for assisting his message decoding via the classical authentication channel \cite{pirandola2008quantum,pirandola2009confidential}. The technique of  quantum data locking \cite{lum2016quantum} provides a new way of realizing QSDC under the realistic practical assumption that Eve can only access quantum memories having limited coherence time. The secret messages are encoded onto a quantum state and locked by a random unitary operation applied to it. Then the locked quantum state containing messages is transmitted from Alice to Bob. Bob can unlock the original message by using an inverse unitary operation. The choice of the unitary operations between Alice and Bob depends on a pre-shared key, which could be generated using QKD. Both the protocols in \cite{pirandola2008quantum,pirandola2009confidential} and in \cite{lum2016quantum} required only the transmission of quantum states over a quantum channel once, thus they were less corrupted by channel impairments. By contrast, a pair of transmission is required in conventional QSDC protocols \cite{long2000theoretical,deng2003two,deng2004secure,wang2005quantum}.

Additionally, both the quantum illumination \cite{shapiro2014secure,zhuang2015ultrabroadband} and quantum low probability of intercept techniques  \cite{shapiro2019quantum} exhibit impressive potential for realizing QSDC at Gigabits per second communication rates only using single-wavelength operations over metropolitan-area fiber channels. In these two schemes, a spontaneous parametric down-conversion operation emits entangled signals and idler beams. Then the signal beam is transmitted to the transmitter of information for message encoding, while the idler beam is retained by the receiver of information. The receiver can recover transmitter's message by combining her idler beam and the returned signal beam as a benefit of their initial entanglement. 

The basic objective of the so-called quantum rebound capacity protocol~\cite{das2019quantum} is to have reliable communication from Alice to Bob that is protected private from the eavesdropper. In quantum private reading protocols~\cite{bauml2018fundamental,das2020entanglement}, the encoded confidential message can be reliably decoded by the reader, while a passive eavesdropper gains no information about it. Both of these could be regard as a variant of the QSDC protocol.

The above examples of QSDC or DSQC protocols include both discrete variable and continuous variable systems, proposing new techniques for realizing  point-to-point quantum communication, where confidential messages can be transmitted directly through the quantum link between a pair of legitimate users. The most striking contrast is in the additional classical communication step of Fig \ref{fig:QSDCandDSQC}. Explicitly, DSQC protocols need the transmitter's additional classical information for message decoding~\cite{sun2021deterministic}, whereas for QSDC there is no need to do this. This transmission from Step 6 of the continuous variable DSQC protocol relies on two-mode squeezed states, where Alice reveals her measurement results, and she also announces the matching information in Step 5 of the ZXFZ DSQC protocol.

While still relying on the basic principles of the earliest QSDC schemes, we can choose different physical entities to implement QSDC \cite{wang2005multi,wang2006multiparty,jin2006three,man2006quantum,lin2008quantum,chen2008controlled,chamoli2009secure,lu2010quantum,wang2011high,sun2012quantum,liu2013quantum,ren2013photonic,meslouhi2013quantum,zhuang2015ultrabroadband,mi2015high,jian2016efficient}. There are three foundamental features of a peer-to-peer QSDC protocol \cite{pan2023free}:
\begin{itemize}
\item (a) QSDC enables secure communication without the need for pre-distributed secret keys.
\item (b) The encoded information can be read out deterministically by the receiver without a basis reconciliation step, and hence in principle there is no additional classical bit exchange between the transmitter and the receiver, except for the process of eavesdropping detection and error rate estimation.
\item (c) Eve will be detected by the legitimate users in real-time.
\end{itemize}
Protocols that do not meet the above three characteristics are typically referred to as quasi-QSDC protocols~\cite{pan2023free}.

The most appealing feature of QSDC protocols is that a pair of legitimate users can prevent information leakage before eavesdropper detection. The eavesdropper can be detected by sampling the measurements of quantum cryptographic protocols, but information leakage cannot be avoided. To elaborate a little further, QKD is developed for transmitting a cryptographic key. If the eavesdropping detection of QKD discovers the presence of Eve, even if the encoded information has already been unveiled to Eve, the key is simply discarded. However, when the confidential message itself is transmitted directly, the legitimate users cannot throw away the message. Thus some new concepts were created: 

\textit{block-based transmission and transmission order rearrangement.} 

In block based transmission, Alice and Bob have to transmit a batch of quantum states and for the sake of preventing the leakage of a secret message they must carry out an Eve-check procedure to ascertain the absence or presence of an eavesdropper. By contrast, in case of order rearrangement Alice first encodes the confidential message and then reorganizes the position of the particles within a block, while keeping the order secret. However, the secret transmit order of particles may be announced publicly, provided that security of the channel has been confirmed. The above block transmission and order rearrangement methods are popularly used for constructing quantum communication protocols \cite{xia2007controlled,dong2011controlled,gao2013secure,ye2013quantum,li2006improving,cao2007multi,yadav2014two}.

\begin{figure*}[tp]
\begin{center}
\includegraphics[width=14cm,angle=0]{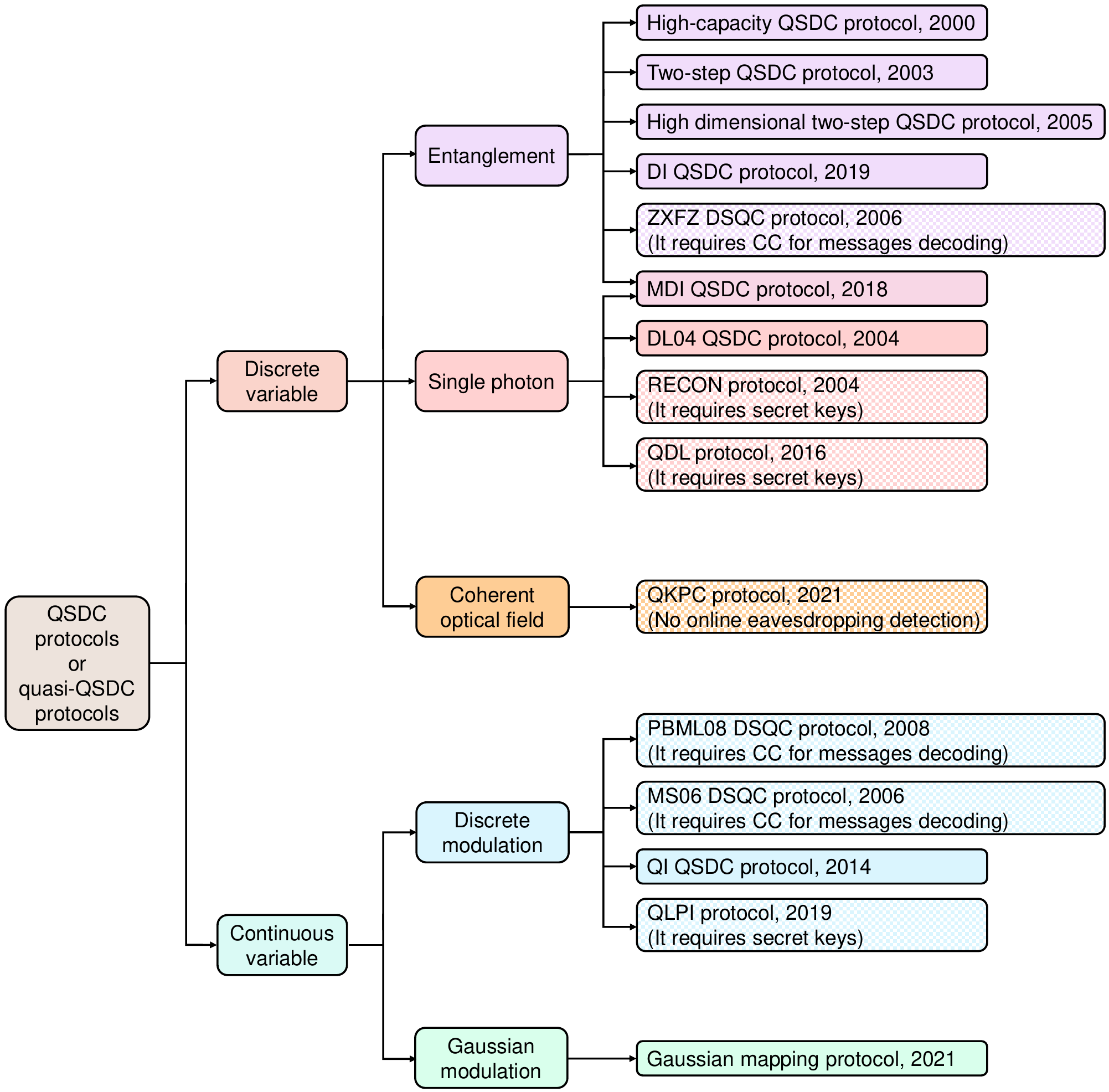}
\caption{The evolution of the QSDC protocols or quasi-QSDC protocols. This diagram is an extension based on Ref.~\cite{pan2023free}, which includes the following protocols: High-capacity QSDC protocol~\cite{long2000theoretical}, Two-step QSDC protocol~\cite{deng2003two}, High dimensional two-step QSDC protocol~\cite{wang2005quantum}, Device independent (DI) QSDC protocol~\cite{zhou2020device}, Zhu-Xia-Fan-Zhang (ZXFZ) DSQC protocol~\cite{zhu2006secure}, Measurement-device-independent (MDI) QSDC protocol~\cite{zhou2020measurement}, Deng-Long 2004 (DL04) QSDC protocol~\cite{deng2004secure},  Repeatable classical one-time-pad (RECON) protocol~\cite{deng2004repeatable}, Quantum data locking (QDL) protocol~\cite{lum2016quantum}, Quantum keyless private communication (QKPC) protocol~\cite{vazquez2021quantum}, Pirandola-Braunstein-Mancini-Lloyd 2008 (PBML08) protocol~\cite{pirandola2008quantum}, Marino-Stroud 2006 (MS06) protocol~\cite{marino2006deterministic}, Quantum illumination (QI) QSDC protocol~\cite{shapiro2014secure}, Quantum low probability of intercept (QLPI) protocol~\cite{shapiro2019quantum}, and Gaussian mapping protocol~\cite{cao2021continuous}. CC: classical communication. The protocol marked with grid-patterned color is referred to as quasi-QSDC.}
\label{fig:Protocols}
\end{center}
\end{figure*}

Following a step-by-step introduction of the protocol details, let us now compare the characteristics of various protocols. As illustrated in Fig.~\ref{fig:Protocols}, the protocols we have introduced can be categorized into discrete variable and continuous variable protocols based on the type of information carriers. Specifically, discrete variable protocols encompass entanglement-based, single-photon, and coherent light field protocols. On the other hand, when considering the different modulation techniques, continuous variable protocols are further divided into those based on discrete modulation and those based on Gaussian modulation.

Additionally, Table~\ref{tab:DVVsCV} presents a comparison between DV QSDC and CV QSDC, following the style outlined in Ref.~\cite{hosseinidehaj2018satellite}. It is noteworthy that in CV QSDC, we harness some pre-coding of the information~\cite{cao2021continuous}, and its information processing algorithms are considerably more complex than DV QSDC. Regarding the real-world performance of these solutions, such as their transmission rate, we will provide further details in the later sections.

\begin{table}[!h]
\caption{Comparison between DV QSDC and CV QSDC.}
\begin{center}
\begin{footnotesize}
\begin{tabular}{|m{1.5cm}|m{3cm}|m{2.95cm}|}
\hline
                                 &  DV QSDC &  CV QSDC \\ \hline
Information processing algorithm & $\bullet$ Simple                                                                                                             & $\bullet$ Complex \\ \hline
Light source                  &$\bullet$ Lack of genuine single-photon sources                                                              &$\bullet$ Relatively stable and mature light sources                                                                                               \\ \hline
Channel                        & \begin{tabular}[c]{@{}l@{}}$\bullet$ Fiber or free-space\\ $\bullet$ Photon loss results in \\ no response from the \\detector\end{tabular} & \begin{tabular}[c]{@{}l@{}}$\bullet$ Fiber or free-space\\ $\bullet$ Photon loss caused \\vacuum noise\end{tabular}                                           \\ \hline
Detection                    & \begin{tabular}[c]{@{}l@{}}$\bullet$ Single-photon detector\\ $\bullet$ Sensitive to background \\noise\end{tabular}                      & \begin{tabular}[c]{@{}l@{}}$\bullet$ Homodyne detection or \\heterodyne detection\\ $\bullet$ Having the ability to \\resist background \\light noise\end{tabular} \\ \hline
Rate     & $\bullet$ High data transmission rate in high-loss channels                                                                  & $\bullet$ High data transmission rate in low-loss channels                                                                                        \\ \hline
Cost                    & $\bullet$ Relatively high                                                                                                & $\bullet$ Relatively low                                                                                                                           \\ \hline
Compatibility              & $\bullet$ Need dedicated components                                                                                                                   & $\bullet$ Capable of better compatibility with existing telecom infrastructures \\ \hline
\end{tabular}
\end{footnotesize}
\label{tab:DVVsCV}
\end{center}
\end{table}

\subsection{Advances in Security Analysis}
\label{sec:Advancement in security analysis}
Classified by Eve's powers, three different types of attacks are commonly considered in QSDC's security analysis:
\begin{itemize}
\item \textbf{individual attack}: Eve prepares separate ancilla states, each of which is used as a probe to interact independently with qubits and these probes are measured one after the other;
\item \textbf{collective attack}: Eve independently attaches probes to each qubit, but she stores these ancilla states in the quantum memory to perform an optimal collective measurement at any later time;
\item \textbf{coherent attack}: Eve prepares a global ancilla state to be used as a probe for interacting with all qubits. Then the ancilla state is stored and collectively measured. This is the strongest class of attacks that Eve could carry out.
\end{itemize}
Individual attacks are the only ones that are feasible with the aid of existing quantum technologies, while both collective attacks and coherent attacks rely on quantum memories having long coherence time and coherent quantum operations associated with high fidelity, which are unavailable at the time of writing. However, despite the unavailability of quantum memory, the eavesdropper is assumed to have full access to the quantum channel and have all quantum technologies - some of which are unavailable at the time of writing - at her disposal, while operating without violating the laws of quantum mechanics. Clearly, sustaining confidentiality even under these worst-case assumptions would make the security of quantum communication `unconditional'. Additionally, it is assumed that Eve can monitor the classical authentication channel, but cannot tamper with it. A quantum communication scheme cannot be generally regarded as being information-theoretically secure until it is proven to be secure against coherent attacks~\cite{gisin2002quantum}. 

The security proof of QSDC protocols is still considered to be work in progress. The challenge in this context is that Eve can attack the qubits traveling along the two-way channel. Similarly, the above three different types of attacks can also be considered in the security analysis QSDC. The security analysis of the initial QSDC protocols focused on tackling individual attacks was given in~\cite{deng2003two}, followed by that of collective attacks in~\cite{qi2019implementation}. 

QSDC protocols can be changed to the DQKD protocols, when dispensing with block transmissions, that is, when the number of qubits $N$ is reduced to 1. The paradigmatic examples of DQKD are the Ping-Pong protocol of~\cite{bostrom2002deterministic} and the LM05 protocol of~\cite{lucamarini2005secure}. In \cite{lucamarini2005secure}, Bob randomly produces a photon polarized in one of the four states $\{|0\rangle, |1\rangle, |+\rangle, |-\rangle\}$ and he transmits it to Alice. When this particle reaches Alice, she measures it with a probability of $c$ with the objective of eavesdropping detection (checking mode), or she uses it for conveying a secret key bit with probability $1-c$ (encoding mode). After encoding the particle, it is sent back to Bob. The above steps are repeated  until all key bits are transmitted. Then Bob can deterministically access the secret key as well as an estimate of the error rate experienced during transmission over the forward quantum channel (Bob-Alice) after Alice revealing the index of Eve-checking bits. Alice will also publish some part of the key bits for estimating the error rate in the backward quantum channel (Alice-Bob). If entanglement is used, one particle of the entangled state will travel in a  forward-backward manner for supporting the information flow \cite{bostrom2002deterministic}. There is no block based transmission in DQKD protocols, so they are similar to the DL04 or to the two-step QSDC protocol when $N=1$. The security of DQKD protocols has been analyzed in~\cite{lin2009eavesdropping,lu2011upper,lu2011unconditional,fung2012quantum,lu2013two,beaudry2013security,lucamarini2014quantum,shaari2014checking,lu2015two,henao2015practical,lu2019ambiguous,krawec2022security,laurenza2020dense}. Some of them are analyzed under the scenario of Alice and Bob using QSDC protocols to distribute the cryptographic key, but they are appropriate for the security analysis of QSDC. To elaborate a little further, diverse  individual attack strategies are discussed in Ref. \cite{lucamarini2014quantum} and the upper bound of the amount of information stolen by Eve has also been given. A pair of legitimate communication parties can also benefit from encountering a scenario in that Eve attacks both the forward and the backward channel, because these correlated attacks may be more easily detected \cite{shaari2014checking}. Eve can only access the confidential information probabilistically by combining a photon number splitting attack with methods of state discrimination, even if a weak coherent pulse source is adopted \cite{lin2009eavesdropping,lu2019ambiguous}. In 2011, Lu \textit{et al}. \cite{lu2011unconditional} proved that the DL04 protocol is secure against collective attacks, when the secret key is transmitted by QSDC while relying on idealized perfect devices. In their proof, Eve attaches separable ancilla states to each qubit and applies a unitary operation to the joint state. She keeps the ancilla states in a quantum memory until receiving the qubit from Alice after encoding. In order to infer the maximal possible amount of information from a secret message, the optimal measurement is performed by combining the encoded state and her ancilla state. The joint state of Bob's initial state, Alice's encoded state, and Eve's attack were used for estimating the maximal possible amount of information that Eve can extract. On this basis, Lu \textit{et al}. \cite{lu2013two} assumed that Eve controls the detector and performs measurements by exploiting the measurement bases Bob has passed to her. The results showed that all detector-side-channel attacks are futile in the channel of Alice-Bob. One can also employ QSDC based on single-photon Bell-state measurements to resist detector-side-channel attacks~\cite{li2020quantum}. Moreover, Beaudry \textit{et al}. \cite{beaudry2013security} and Henao \textit{et al}. \cite{henao2015practical} also gave the security proof of two-way protocol.

\begin{figure}[!h]
\begin{center}
\includegraphics[width=\columnwidth,angle=0]{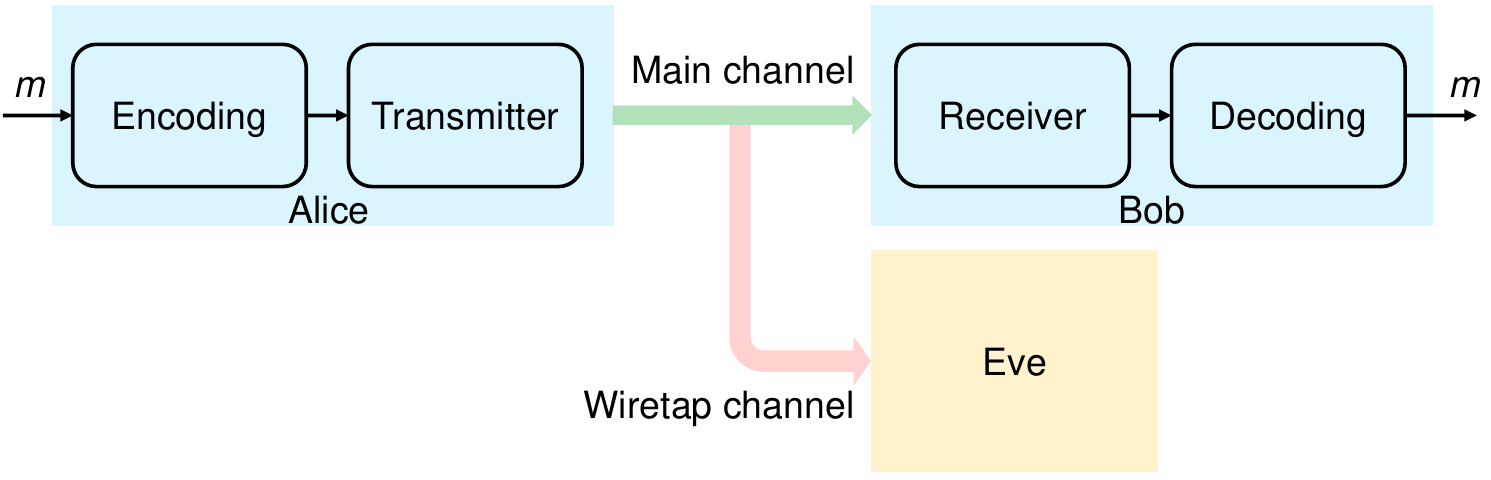}
\caption{The wiretap channel mode. $m$ is the transmitted message.}
\label{fig:Wiretapchannel}
\end{center}
\end{figure}

The security of practical QSDC systems may be analyzed from an information theoretical point of view by relying on Wyner's wiretap channel theory~\cite{wyner1975wire,boche2022semantic}, as shown in Fig.~\ref{fig:Wiretapchannel}. The mutual information $I\left(A:B\right)$ quantifies the information rate at which Bob can reliably receive from Alice via the main channel. By comparison, the quality of the eavesdropper's channel termed as the wiretap channel and the maximum attainable information rate of a malicious Eve is given by $I\left(A:E\right)$. In the framework of taking collective attacks into consideration, the lower bound of the secrecy capacity of the DL04 protocol can be expressed as~\cite{qi2019implementation,wu2019security,pan2020experimental}
 \begin{eqnarray}
C_{s}^{\rm DL04}&\!\!\!=\!\!\!&\max_{\left\{p\right\}}\left\{I\left(A:B\right)-I\left(A:E\right)\right\}\nonumber\\
&\!\!\!=\!\!\!&Q_{\rm Bob}\left [1-h\left(e\right) \right ]-Q_{\rm Eve}h\left(e_{x}+e_{z}\right)\nonumber\\
&\!\!\!=\!\!\!&Q_{\rm Bob}\left [1-h\left(e\right)  -gh\left(e_{x}+e_{z}\right) \right],
\end{eqnarray}
where $Q_{\rm Bob}$ is the reception rate of Bob, $Q_{\rm Eve}$ is the maximum rate Eve can access the qubits, $e$ is the QBER of second transmission, $e_{x}$ and $e_{z}$ is the QBER under X-basis and Z-basis of the first eavesdropping-check, $g$ is the gap between $Q_{\rm Eve}$ and $Q_{\rm Bob}$, and finally $h(x)=-x{\rm log}_{2}x-(1-x){\rm log}_{2}(1-x)$ is the binary entropy function. By contrast, the lower bound of the secrecy capacity of the two-step protocol is given by~\cite{wu2019security}
 \begin{eqnarray}
C_{s}^{\rm TW}=Q_{\rm Bob}\left [2-h_4\left(\textbf{e}\right)\right]-Q_{\rm Eve}\left [h\left(e_{x}\right)+h\left(e_{z}\right)\right],
\end{eqnarray}
where $h_4\left(\textbf{e}\right)$ is the `four-array' Shannon entropy and $\textbf{e}$ is error distribution. Furthermore, the secrecy capacity of MDI QSDC was given by~\cite{niu2020security,li2023single2,sun2023one}
 \begin{equation}
C_{s}^{\rm MDI}=Q_{\rm Bob}\left [1-h\left(e\right)\right]-Q_{\rm Eve}h\left(e_{u}\right),
\end{equation}
where $e_{u}$ is the QBER during the first eavesdropping detection and $e$ is the QBER of second transmission. Therefore, the secrecy capacity limits the maximum rate at which Alice can securely and reliably convey confidential messages to Bob through the quantum channel under the guarantee that Eve has no useful information about the confidential message.

The security metrics of the three protocols have been extensively investigated~\cite{wu2019security,niu2020security}. A crucial parameter in this context is the tolerable QBER threshold, below which the protocol exhibits a non-zero secrecy capacity. Previous findings suggest that the two-step protocols exhibit a higher tolerance for errors compared to the DL04 protocol~\cite{wu2019security,beaudry2013security,shaari2015finite}, while the DL04 protocol outperforms the MDI protocol in this respect~\cite{niu2020security}.  Moreover, high-dimensional entangled protocols can further enhance the tolerable QBER, with an increasing improvement observed, as the dimensionality increases~\cite{patra2023dimensional}.

A promising solution capable of increasing the channel capacity using masking was proposed in~\cite{long2021drastic}, where 
Eve's effective reception rate is limited to $Q_{\rm Eve}=Q_{\rm Bob}$, namely $g=1$. In addition, the finite-length security analysis of QSDC is currently under investigation~\cite{wu2022quantum}. In order to obtain the real-life secrecy capacity of QSDC, some of its practical influencing factors are starting to be taken into account, such as the detector efficiency mismatch~\cite{ye2021generic}, side-channel effects~\cite{ye2021generic}, source imperfections~\cite{ye2021generic,sun2023one}, and so on. Generally, the performance of QSDC or DSQC in achieving information-theoretic security using quantum states can be mathematically proven. However, the security proofs are still evolving, and Table~\ref{tab:securityanalysis} summarizes the progress in the security analysis of the typical communication protocols introduced in Section~\ref{sec:PtoP}.

\begin{table}[tp]
\begin{footnotesize}
\begin{center}
\caption{Security analysis progress of some representative protocols.}
\begin{tabular}{|m{2cm}|m{5.8cm}|}
\hline
Protocols                          & Progress \\ \hline
High-capacity QSDC protocol             & Security proof against the collective attack within a finite-length setting has been provided in~\cite{wu2022quantum}.     \\ \hline
Two-step QSDC protocol                 &Security proof against the collective attack within a finite-length setting has been provided in~\cite{wu2022quantum}.          \\ \hline
High dimensional two-step QSDC protocol & Security proof against the collective attack within a finite-length setting has been provided in~\cite{wu2022quantum}.             \\ \hline
DI QSDC                                 &Security analysis against the collective attack in the asymptotic regime have been provided in~\cite{zhou2020device,zhou2023device,roy2023device}. \\ \hline
ZXFZ DSQC protocol                     & Security analysis against some specific attacks has been provided in~\cite{zhu2006secure}. \\ \hline
MDI QSDC protocol                    &Security proofs against the collective attack in the asymptotic regime have been provided in~\cite{niu2020security,li2023single2,sun2023one}.                   \\ \hline
DL04 QSDC protocol                  &Security proof against the collective attack within a finite-length setting has been provided in~\cite{wu2022quantum}.          \\ \hline
RECON protocol                     & Security analysis against the optimal individual attack                                     has been provided in~\cite{deng2004repeatable}. \\ \hline
QKPC protocol                        & The private capacity has been calculated quantificationally in~\cite{vazquez2021quantum}.                      \\ \hline
PBML08 DSQC protocol                   &Security analysis against the individual attack                                        have given in~\cite{pirandola2008quantum,pirandola2009confidential}. \\ \hline
MS06 DSQC protocol                   & Security analysis against some specific attacks                                             has given in~\cite{marino2006deterministic}. \\ \hline
QI QSDC protocol                  & Security proofs against the collective attack in the asymptotic regime have been provided in~\cite{shapiro2014secure,zhang2013entanglement}.  \\ \hline
QLPI protocol                  & Security proof against the collective attack in the asymptotic regime have been presented in~\cite{shapiro2019quantum}. \\ \hline
Gaussian mapping protocol            & Security proof against the collective attack in the asymptotic regime has been provided in~\cite{cao2021continuous}.\\ \hline
QDL protocol                      &Secure against an attack situated between individual attack and collective attack~\cite{lupo2015quantum}. \\ \hline
\end{tabular}
\label{tab:securityanalysis}
\end{center}
\end{footnotesize}
\end{table}

\subsection{The Cryptographic Applications of Point-to-Point QSDC Protocols}
\label{sec:The cryptographic applications of point-to-point QSDC protocols}
The blueprint of managing security in communication has been proposed in~\cite{niemiec2013management}, where end-users of communications networks utilize specific quantum communication systems having different security levels. QSDC constitutes an important fundamental communication protocol capable of supporting high-level security. Hence, numerous quantum cryptographic solutions have been derived from QSDC, as seen in Table \ref{cryptographictasks1}. Based on the DL04 QSDC protocol, a single photon is harnessed by Alice and Bob during their one-way \cite{xin2006secure} or two-way transmission \cite{shi2010quantum}, where both of them deduce their secret messages after the announcement of measurement results. The role of Alice and Bob is symmetric. Explicitly, when Bob receives the sequence of $M$ photons, as shown in Fig.\ref{TwostepQSDC} of the two-step QSDC protocol, Alice and Bob hold half of every EPR pair in the state $\left |\psi^{-}\right \rangle$. They both are able to encode their secret messages and speak to each other (dialogue) \cite{pathak2013elementsbook}. This makes it natural to apply QSDC for the design of the quantum dialogue protocols of~\cite{ye2013quantum,zheng2014quantum}. 

\begin{table*}[]
\begin{footnotesize}
\begin{center}
\caption{An overview of the studies on different types of quantum cryptographic protocols based on QSDC.}
\begin{tabular}{|m{3cm}|m{3.5cm}|m{7cm}|m{2cm}<{\centering}|}
\hline
Primitive QSDC protocol               & Different types of quantum cryptographic tasks                         & \centering Design objective & Related references (studies) \\ \hline
DL04                                  &  \multirow{2}{*}{Quantum dialogue}                                 & To realize bidirectional QSDC, in which both legitimate users are the sender of the secret message as well as the receivers, and their secret message can be exchanged simultaneously. & \cite{xin2006secure}, \cite{shi2010quantum}                        \\ \cline{1-1} \cline{4-4}
Two-step                              &                                                                        &                                                                                                                                                                                                                                                     &\cite{zheng2014quantum}, \cite{gao2010two}                         \\ \hline
DL04                                  & \multirow{3}{3.5cm}{Quantum secret sharing of secure direct communication} & A dealer wants to share his secret message directly with a group of agents, but the secret message can only be obtained by all the agents if they collaborate. & \cite{zhang2005multiparty1,deng2005improving}, \cite{han2008multiparty}, \cite{du2012quantum}                          \\ \cline{1-1} \cline{4-4}
Two-step                              &                                                                        &                                                                                                                                                                                                                                                     & \cite{zhang2005multiparty} \\ \cline{1-1} \cline{4-4}
High-dimension two-step               &                                                                        &                                                                                                                                                                                                                                                     & \cite{wang2008multiparty}                       \\ \hline
QSDC with GHZ states \cite{lee2006quantum}            & \multirow{3}{3.5cm}{Quantum authentication}                               & Verifying the identity of communication participants to prevent a malicious eavesdropper from pretending to be a legitimate user, where a sender can simultaneously transmit a secret message over the quantum channel to the receiver.                    & \cite{lee2006quantum}                         \\ \cline{1-1} \cline{4-4}
Two-step                              &                                                                        &                                                                                                                                                                                                                                                     & \cite{yang2013quantum}                       \\ \cline{1-1} \cline{4-4}
DL04                                  &                                                                        &                                                                                                                                                                                                                                                     & \cite{yu2013authentication}                        \\ \hline
Three-party QSDC with GHZ states \cite{jin2006three} & \multirow{3}{3.5cm}{Quantum sealed-bid auction}                            & Allowing all bidders to submit their own bids, where the auctioneer makes all the bids public and determines the winning bidder. The honesty of auction must be pledged, and no malicious bidders can collide with the auctioneers. & \cite{naseri2009secure}, \cite{yang2009improved}, \cite{zhao2010secure},  \cite{luo2013loophole}                        \\ \cline{1-1} \cline{4-4}
Two-step                              &                                                                        &                                                                                                                                                                                                                                                     & \cite{zhang2010quantum}, \cite{wen2014attacks}                     \\ \cline{1-1} \cline{4-4}
DL04                                  &                                                                        &                                                                                                                                                                                                                                                     & \cite{zhang2018economic}                         \\ \hline
Two-step                              & Quantum steganography                                                  & Embedding the secret message into another innocent-looking quantum carrier for secure transmission of the secret messages.                                                                                                                  & \cite{qu2010novel}, \cite{qu2011quantum2}, \cite{xu2013high}                        \\ \hline
Two-step                              & Quantum watermarking                                                   & Quantum watermarking is utilized to embed owner identification into quantum multimedia, which is difficult to remove.                      & \cite{fatahi2012quantum}, \cite{mo2013quantum}                       \\ \hline
High-dimension two-step               & Quantum covert channel                                                 & To send secret messages over a covert channel which is established within the normal quantum channel.                                                                                                        & \cite{shu2013novel}                        \\ \hline
Two-step                              & Quantum anonymous ranking                                              & To allow users to attend a privacy-preserving ranking activity whereby each of the participants involved can anonymously infer his/her ranking information, but cannot get that of others.                                     & \cite{huang2014ranking}                         \\ \hline
DL04                                  & Quantum broadcast communication                                        & Transmitting secret messages from a sender to a dynamically changing group of receivers, where only the authenticated users can decode the relevant information and others obtain nothing.                                                        & \cite{yan2013quantum,cao2016cryptanalysis}                         \\ \hline
Two-step                              & Quantum signature                                                      & To guarantee the security of digital signatures for two participants, so that the message cannot be forged by the receiver or a possible attacker.                                                                       & \cite{yoon2014quantum}                         \\ \hline
Two-step                              & \multirow{3}{3.5cm}{Quantum key agreement}                               & To permit each participant to equally contribute to the generation of a shared key, which cannot be determined fully by any of the parties alone. Hence others cannot get the key through illegal means.                              & \cite{huang2014quantum}, \cite{sun2016efficient}                          \\ \cline{1-1} \cline{4-4}
Three-party QSDC with GHZ states \cite{jin2006three} &                                                                        &                                                                                                                                                                                                                                                     & \cite{zeng2016multiparty}, \cite{chou2018dynamic}                        \\ \hline
\end{tabular}
\label{cryptographictasks1}
\end{center}
\end{footnotesize}
\end{table*}

In 2005, the new concept of quantum secret sharing under QSDC was presented in Ref. \cite{zhang2005multiparty1}, where the advantages of both QSDC and quantum secret sharing are combined. This allows Alice to transmit a secret message to different agents, where no single person is capable of  reconstructing the complete original message, unless the users cooperate. Such protocols can be classified based on their information carriers. Some of them are based on single photons \cite{zhang2005multiparty1,deng2005improving,han2008multiparty,du2012quantum} and rely on the character of `DL04'. Others employ entangled states \cite{zhang2005multiparty,wang2008multiparty} and exhibit a `two-step' or `high-dimensional' character. In the protocols using a single photon, a batch of $N$ initial single photons is prepared by the first agent, and then the next agents apply the unitary operations\footnote{The unitary operation set is different from that of the encoding operation set in the QSDC protocol. For instance, an additional Hadamard gate operator is brought into the original encoding operation set. It is carefully picked by the agents for avoiding that a single agent intercepts and recovers the whole message independently.} to each and every photon to encrypt them. These photons will be transmitted to Alice after the last agent completes the encryption. Then eavesdropping detection and the  encoding of secret messages is carried out followed by return to the last agent. If all agents act in concert, all of them can acquire the secret message by applying their respective preparation bases and encryption operations. However, the process is slightly different, when it comes to entanglement-based protocols, where Alice first prepares an EPR photon pair sequence according to her secret message and randomly inserts some checking photon pairs into it. A partner EPR particle sequence is encrypted by the agents alternately, and Alice sends the retained sequence to the last agent for their cooperation to decrypt the secret message in the event of having no eavesdroppers.

QSDC schemes associated with authentication are composed of two parts: one of them is for an authentication process and the other is for direct communication\cite{lee2006quantum,yang2013quantum,yu2013authentication,sun2012quantumidentification}. These protocols are able to resist man-in-the-middle attacks. Commencing from the principles of three-party QSDC associated with GHZ states \cite{jin2006three}, the authors of  \cite{naseri2009secure,yang2009improved,zhao2010secure,luo2013loophole} put forward the concept of quantum sealed-bid auction. The auctioneer, say Alice, prepares a set of $M$ groups of $n$-particle GHZ states and then she distributes each $n$-particle GHZ state to $n$ bidders. These bidders encode their own bid information and return the corresponding particles to Alice for an $n$-particle GHZ-basis measurement. After this, the winner of the auction will be revealed to all bidders. Of course, both EPR pairs \cite{zhang2010quantum,wen2014attacks} and single-photon schemes \cite{zhang2018economic} can also be used as the basic resources of quantum sealed-bid auction.

By combining the two-step QSDC protocol and ping-pong protocol, the quantum steganography philosophy was is proposed in~\cite{qu2010novel,qu2011quantum2}. Explicitly, a pair of users employ block transmission to prevent information disclosure. Accordingly, first Bob prepares a large number of entangled states. Then the first layer of secret information is transmitted in the same way as in the ping-pong protocol, but auxiliary Bell states are introduced for capacity improvement. Then two adjacent BSM results are picked by Alice to act as the second secret information, which can be read out by the process of entanglement swapping as detailed in~\cite{qu2010novel,qu2011quantum2}. This  is how the hidden secret message is embedded. As a further advance, another `hidden rule' was conceived based on the tensor product of Bell states and unitary transformations for employment in quantum steganography~\cite{xu2013high}. Inspired by quantum steganography, the hidden layer of secret messages was also exploited for quantum watermarking \cite{fatahi2012quantum,mo2013quantum}. The covert quantum channel is established by changing the secret message encoding rule of QSDC, which can also be used for the transmission of confidential messages \cite{shu2013novel}.

In the protocol of quantum domain anonymous ranking, each user can acquire the correct rankings of his/her data, but nobody else can infer it \cite{huang2014ranking}. An ordered sequence of $N$ two-qudit entangled pairs is grouped into a pair of sequences and one of the sequences will visit the station of every users one by one to register their data, while the other sequence is kept at its source. According to the measurement results announced in the final stage, each of the $N$ users can get anonymously the ranking of his/her own data. Decoy qudits are required in some random positions of the photon sequence for detecting eavesdroppers and hence to secure the quantum channel. In \cite{yan2013quantum,cao2016cryptanalysis}, the decoy photons are randomly inserted into the sequence of photons of the GHZ states similarly to the procedure of quantum one-time pad in DL04, where the decoy photons are prepared for authenticating users, while the GHZ states can be used for QSDC broadcasting to $N$-receivers.

In \cite{yoon2014quantum}, the authors proposed a quantum version of the process of arbitrated signature between two users and a trust center. Accordingly, the users sign the public message by using a pre-shared key\footnote{The key is generated by QKD, thus this protocol assumes that the key is secure.} and the unitary operation of a quantum search algorithm \cite{grover1996fast,long2001grover}. Then the signed message qubits are transmitted by two-step QSDC. By this means, the message receiver can confirm that the transmitter signed the message legitimately, while the attacker cannot identify the signature and cannot forge it, because QSDC secures the quantum channel. Similarly, if the particles of each entangled states are held by two different users \cite{huang2014quantum} or even more than two users \cite{sun2016efficient,zeng2016multiparty,chou2018dynamic}, then every one in the group has the right to modify a key by applying his/her operation to the qubits in hand and the final key is jointly determined by all members.

As seen from the literature, QSDC is indeed capable of high-security communication. Thus QSDC is eminently suitable for a wide variety of quantum cryptographic tasks, which require a direct transport of deterministic information over the quantum channel. Both block transmission and the random decoy photon insertion techniques constitute powerful countermeasures against eavesdropping attacks.
 
\subsection{Networking Schemes}
\label{sec:Networking schemes}
\begin{figure}[!h]
\begin{center}
\includegraphics[width=\columnwidth,angle=0]{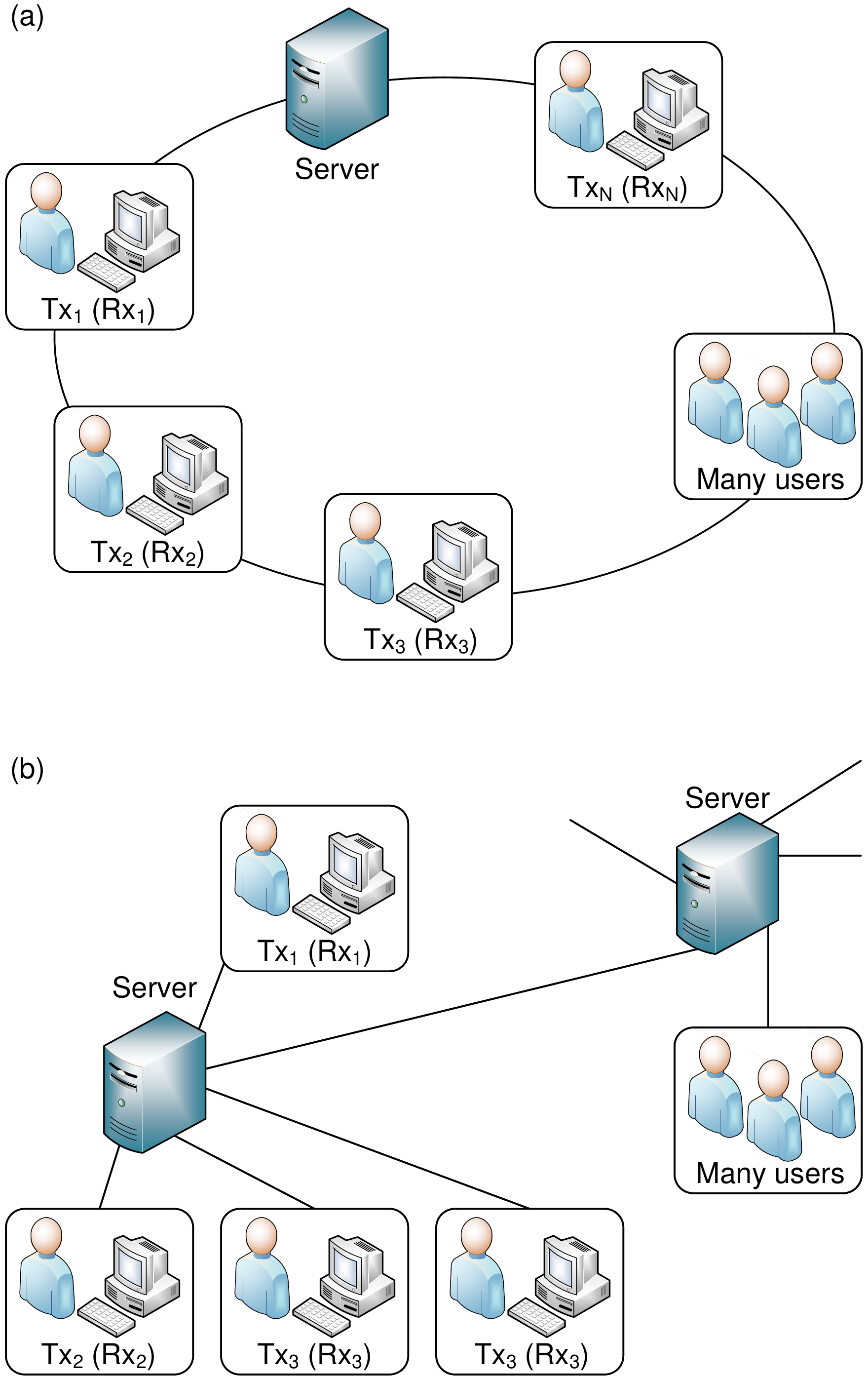}
\caption{Multi-user network configurations for QSDC. (a) loop-configuration, (b) star-configuration.}
\label{fig:Qnet}
\end{center}
\end{figure}

The topology of a hypothetical any-to-any multi-user QSDC system can be of a loop or star structure as shown in Fig. \ref{fig:Qnet}, similar to QKD networks \cite{phoenix1995multi,razavi2012multiple}. The nodes assume one of three roles: the server, the transmitter (Tx) and receiver (Rx). The server prepares qubits and receives classical requests from all the users on the network. It responds to them over the quantum channel and the classical channel in a different slot. Controllable switches are used for constructing the quantum link between a pair of communicating parties~\cite{niu2022qnus}.

\begin{figure}[!h]
\begin{center}
\includegraphics[width=\columnwidth,angle=0]{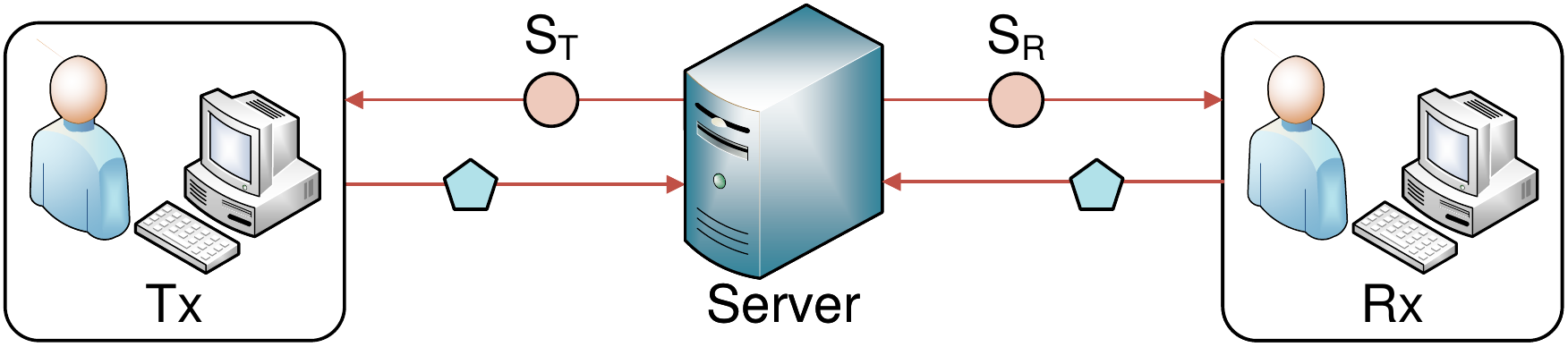}
\caption{The subsystem of the QSDC network. The pair of circles represent the state $|\psi^{-}\rangle_{TR}$, while the pair of pentagons denote the state $|\phi^{+}\rangle_{TR}$.}
\label{fig:QSDCnet}
\end{center}
\end{figure}

A multi-user QSDC network can be reduced to that seen in Fig. \ref{fig:QSDCnet}. If two users are not in the same loop or branch, the server of the loop supporting a $\rm Tx$ or $\rm Rx$ will carry out the tasks of qubit preparation, while the other servers will offer the quantum link for their communication. In this case, we assume that only a single subsystem can use the same quantum channel and classical channel simultaneously \cite{fu2002theoretical,deng2006quantum}. The server prepares a set of EPR pairs in the same quantum state $|\psi^{-}\rangle_{TR}$, and divides them into two parts, $\rm S_{T}$ and $\rm S_{R}$. The $\rm S_{T}$ sequence is composed of all the particles marked by $T$ and $R$ in every EPR pair $|\psi^{-}\rangle_{TR}$, respectively. They will be distributed to two different users. Then a subset of particles is selected randomly to detect eavesdroppers by applying the measurement basis $\sigma_{x}$ or $\sigma_{z}$ to them, similarly to the process in the two-step QSDC protocol~\cite{deng2003two}. If the transmission of qubits is deemed to be secure, the transmitter maps its secret message onto the particles of the sequence $\rm S_{T}$ by applying one of the unitary operations $\{U_{0}, U_{1}, U_{2}, U_{3}\}$, and then randomly picks some particles for the next eavesdropping detection action. The receiver also uses the sequence $\rm S_{R}$ for conveying information, where all bit values are randomly set to 0 or 1. The transmitter and receiver transmit the sequence $\rm S_{T}$ and $\rm S_{R}$ respectively back to the server. The server applies the BSM to each and every EPR pair received and publishes the outcomes given by $U_{A}=U_{T}\otimes U_{R}$. Then the receiver checks the security and deduces the transmitter's secret messages by applying $U_{T}=U_{A}\otimes U_{R}$. At this stage communication between the network users is completed.

\begin{figure*}[!h]
\begin{center}
\includegraphics[width=16cm,angle=0]{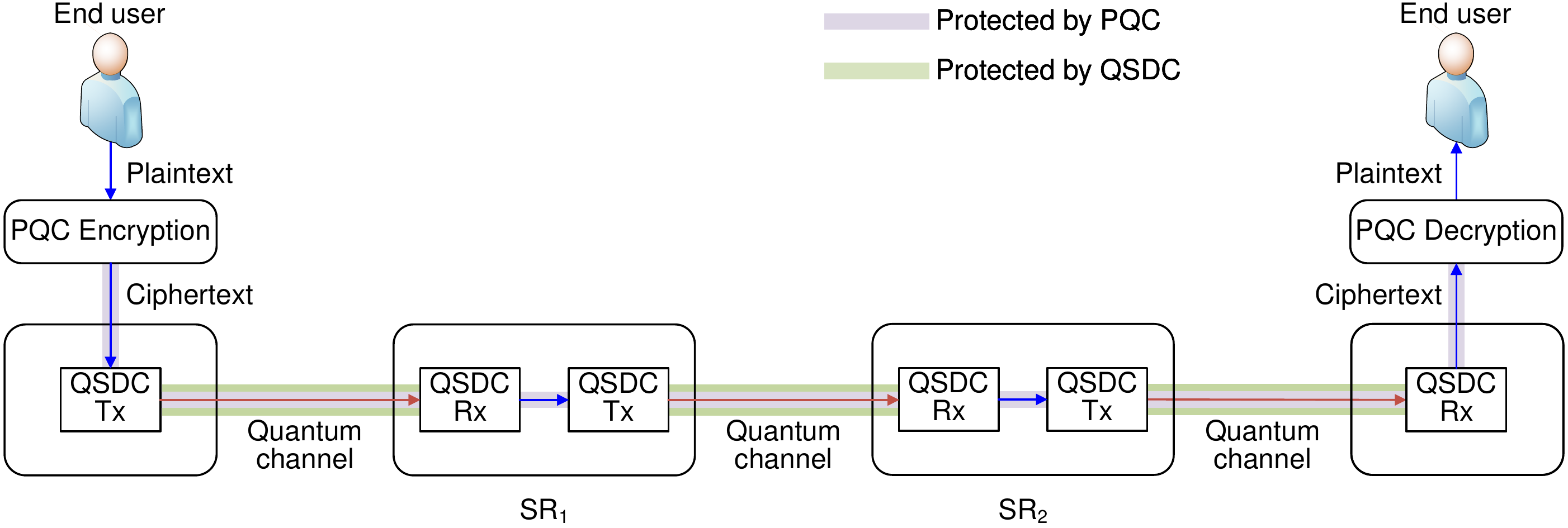}
\caption{A secure repeater network. PQC, post-quantum cryptography; SR, secure repeater.}
\label{fig:SecureRepeater}
\end{center}
\end{figure*}

Diverse QSDC networks relying on entanglement~\cite{deng2006quantum,xi2006quantum,deng2007quantum}, on single photon based regimes~\cite{fu2007economical}, and on hyperentanglement~\cite{bin2011bidirectional,hong2014quantum,kim2022security} have been conceived, providing an important step forward in terms of achieving any-to-any multi-users QSDC connectivity. Furthermore, the authentication process or identity verification between a quantum server and the users can be validated by entangled EPR pairs and controlled NOT gates~\cite{farouk2015generalized}. Some of the QSDC features have also been introduced into classical optical virtual private networks and into quantum virtual private networks with the objective of enhancing the security of passive optical networks~\cite{gong2013novel}. To fit into the operational mobile communication framework and allow telecom companies to provide secure communication, a controlled bidirectional QSDC protocol based on the properties of GHZ-states was invented and applied in mobile networks~\cite{chou2014quantum}. Some techniques of the network layer, such as quantum multiple access techniques and routing have also been considered in multi-user QSDC networks~\cite{wang2013quantum,zarmehi2016controlled}. Indeed, we may view the overall communication networks as a hybrid one relying on a quantum and a classical channel, where the confidential messages are transmitted directly over the quantum channels and the classical channels simply assist in eavesdropping detection.

Constrained by the capabilities of the state-of-the-art technologies at the time of writing, quantum communication relies on classical trusted nodes in networking applications~\cite{elliott2004darpa,poppe2008outline,chen2010metropolitan,sasaki2011field,wang2014field,cao2022single}, which have to be hosted in secure premises. Unless these premises are protected from potential eavesdroppers, the information security at the relay nodes potentially faces challenges. The secure repeaters of the near future will have to utilize QSDC and post-quantum cryptography for hop-by-hop relaying, while ensuring reliable and secure information transfer even in the presence of realistic eavesdropping-infested quantum channels, while protecting the information security at the classical relay nodes using post-quantum cryptography~\cite{long2022evolutionary}. Hence again, the relay nodes are not required to be trusted, as shown in Fig.~\ref{fig:SecureRepeater}. This dual protection scheme solves a major challenge in quantum communication and networking. This approach enhances the transmission distance of QSDC, potentially supporting large-scale secure networking applications, while additionally promoting the organic fusion of quantum communications and post-quantum cryptography. A secure relay has been experimentally characterized by combining 10 kilometers of optical fiber and short-distance free-space transmission for supporting relay-based image transmission~\cite{long2022evolutionary}. The experimental tests also indicate that the delay at the relay is relatively small~\cite{wang2023experimental}. This near-term quantum network supports both connection-oriented and connectionless network protocols~\cite{zhang2022connection,li2022connection,xiao2023connectionless} in classical networks. A seven-stage evolutionary roadmap of constructing a perfectly quantum Internet based on secure relays has been proposed in~\cite{long2022evolutionary}, but again, the ultimate solution hinges on the introduction of fully-fledged quantum repeaters~\cite{van2013designing,muralidharan2016optimal,cacciapuoti2019quantum,azuma2023quantum,zhang2023hybrid
}.

\subsection{Experimental Progress}
\label{sec:Experimental progress}
Substantial efforts have been invested into realizing DQKD~\cite{ostermeyer2008implementation,chen2016experimental,cere2006experimental,kumar2008two,khir2012implementationfreespace,khir2012implementation,qi2021loophole}, which has the potential of preparing the groundwork for QSDC experiments. At the time of writing QSDC has evolved from its theoretical protocol development phase to experimental demonstrations over the past few years. Both the DL04~\cite{hu2016experimental,qi2019implementation} and the two-step QSDC protocol~\cite{zhang2017quantum,zhu2017experimental} have been realized by dedicated experimentalists. Below, the associated experimental results are reviewed in a little more detail. Table~\ref{EDQSDC-table1} and Table~\ref{EDQSDC-table2} provide a summary of the most representative QSDC experiments (N.A. represents not available). These experiments were conducted to demonstrate and test a range of QSDC protocols in real-world conditions.

\begin{table*}[!h]
\begin{footnotesize}
\begin{center}
\caption{Summary of representative QSDC Experiments based on single photon.}
\begin{tabular}{|l|m{2.0cm}|m{2.0cm}|m{2.0cm}|m{2.0cm}|m{2.0cm}|m{2.0cm}|}
\hline
\multicolumn{1}{|l|}{Group \& year}                & Hu \textit{et al}., 2016~\cite{hu2016experimental}  & Qi \textit{et al}., 2019 \cite{qi2019implementation} & Sun \textit{et al}., 2020~\cite{sun2020toward} & Pan \textit{et al}., 2020~\cite{pan2020experimental} &Zhang \textit{et al}., 2022~\cite{zhang2022realization} & Liu \textit{et al}., 2022~\cite{liu2022fiber}      \\ \hline
Protocol                             & \multicolumn{6}{c|}{DL04}              \\ \hline
Information carrier & \multicolumn{6}{c|}{Single photon}         \\ \hline
Encoding                                  & Operation frequency              & Phase & Phase & Phase & Phase and time-bin & Phase       \\ \hline
Wavelength                                & 1550 nm               & 1550 nm       & 1550 nm & 1550 nm  & 1550 nm  &1550 nm\\ \hline
Channel                                   & Fiber      & Fiber & Fiber & Free space & Fiber &Fiber   \\ \hline
Repetition rate                & 10 MHz                & 1 MHz    & 1 MHz      & 16 MHz           & 50 MHz            &50 MHz \\ \hline
Distance                     & N.A.                  & 1.5 km    & 18.5 km     & 10 m          & 100 km& 5 km        \\ \hline
Error rate                                & N.A.                   & 0.6\%        & 0.96\%         & 0.49\%$\pm $0.27\%  & 2.5\% & 0.42\%$\pm $0.05\%                           \\ \hline
Rate                          & 4 kbps                & 50 bps    & 100 bps     & 500 bps        & 0.54 bps  & 3.43 kbps            \\ \hline
\end{tabular}
\label{EDQSDC-table1}
\end{center}
\end{footnotesize}
\end{table*}

\begin{table*}[!h]
\begin{footnotesize}
\begin{center}
\caption{Summary of representative QSDC Experiments based on entanglement. ITU, international telecommunication Union.}
\begin{tabular}{|l|c|c|c|}
\hline
\multicolumn{1}{|l|}{Group \& year}                & Zhang \textit{et al}., 2017 \cite{zhang2017quantum}   & Zhu \textit{et al}., 2017 \cite{zhu2017experimental}   & Qi \textit{et al}., 2021~\cite{qi202115}    \\ \hline
Protocol   & \multicolumn{3}{c|}{Two-step}                 \\ \hline
Information carrier &   \multicolumn{3}{c|}{Entangled photon}   \\ \hline
Encoding        & Polarization  & Polarization     & Polarization \\ \hline
Wavelength     & 795 nm                & 1549.32 nm    & 30  ITU channels
 \\ \hline
Channel     & Free space            & Fiber & Fiber  \\ \hline
Repetition rate  & N.A.            & N.A.               & N.A. \\ \hline
Distance      &N.A.               & 0.5 km        & 40 km        \\ \hline
Error rate      & 10\%         & N.A. & 0.13\%  \\ \hline
Rate       & 2.5 bps               & N.A.  & 1 kbps \\ \hline
Fidelity   & 90\%               & 91\%, 88\%     & >95\%        \\ \hline
\end{tabular}
\label{EDQSDC-table2}
\end{center}
\end{footnotesize}
\end{table*}

The information carriers inevitably suffer from the impairments of the quantum channel, such as its thermal effects, nonlinearities and dispersion. Therefore, quantum error correction codes \cite{wen2010one,zamir2018secure} and the decoherence-free subspace technique \cite{li2015fault,su2010robust,gao2020free} have been developed for protecting QSDC. In 2016, Hu \textit{et al}. \cite{hu2016experimental} proposed the so-called `single-photon frequency coding' technique for the DL04 protocol and characterized it experimentally. This scheme is different from the method relying on frequency-coded optical qubits (see Section \ref{sec:encodingtech}). The repetitive operations are applied to a single-photon block for conveying a secret message, formulated as~\cite{hu2016experimental}
\begin{eqnarray}
{\rm Operation}\!=\!\begin{cases}
&\!\!U\!=\!|0\rangle\langle1|-|1\rangle\langle0|,\;\sin(2\pi f\tau_{i}+\delta)>0, \\
&\!\!I\!=\!|0\rangle\langle0|+|1\rangle\langle1|,\;\;\sin(2\pi f\tau_{i}+\delta)<0,
\end{cases}
\end{eqnarray}
where $f$ is the modulation frequency, $\tau_{i}$ is the arrival time of a photon, and $\delta$ is the initial phase of each signal. The receiver of this single-photon block is capable of decoding the secret message through the discrete time Fourier transform. Hence, we refer to this modulation method as operation frequency based encoding in Section \ref{sec:encodingtech}. In contrast to the original DL04 protocol~\cite{deng2004secure}, the secret message bits are mapped to individual photons, where the spectrum of the single-photon block carries the secret messages here. Even though not all the photons can be detected by the receiver owing to the optical loss and the limited quantum efficiency of a single photon detector, the reliability and secrecy of information transmission would not be affected. However, the signal-to-noise ratio experienced in the frequency spectrum would be decreased. Another advantage of this method is the ability to support wideband communication. Assuming that the frequency range that can be exploited spans from $f_{\rm min}$ to $f_{\rm{max}}$ and that the channel spacing is $f_{b}$, the number of frequency channels becomes $N_{c}=(f_{\rm max}-f_{\rm min})/f_{b}+1$. The effective degree of freedom is given by~\cite{hu2016experimental} $N_{\rm{max}}=N_{c}!/[r!\left(N_{c}-r\right)!]$,  which is the number of possible combinations of $r$ frequency components from a set of $N_{c}$ frequency channels. This implies that ${\rm log}_2N_{\rm max}$ bits of information can be carried by a block of single photons, and the information transmission rate is $I=(1/T_{\rm{span}}){\rm log}_{2}N_{\rm max}$, where $T_{\rm{span}}$ is the time span of a single-photon block.

\begin{figure*}[htbp]
\centering
\subfigure[Operation frequency encoding \cite{hu2016experimental}. Reprinted with permission from Ref.~\cite{hu2016experimental}~\copyright~Springer Nature.]{
\begin{minipage}[t]{0.48\linewidth}
\centering
\includegraphics[width=3.44in,height=1.85in]{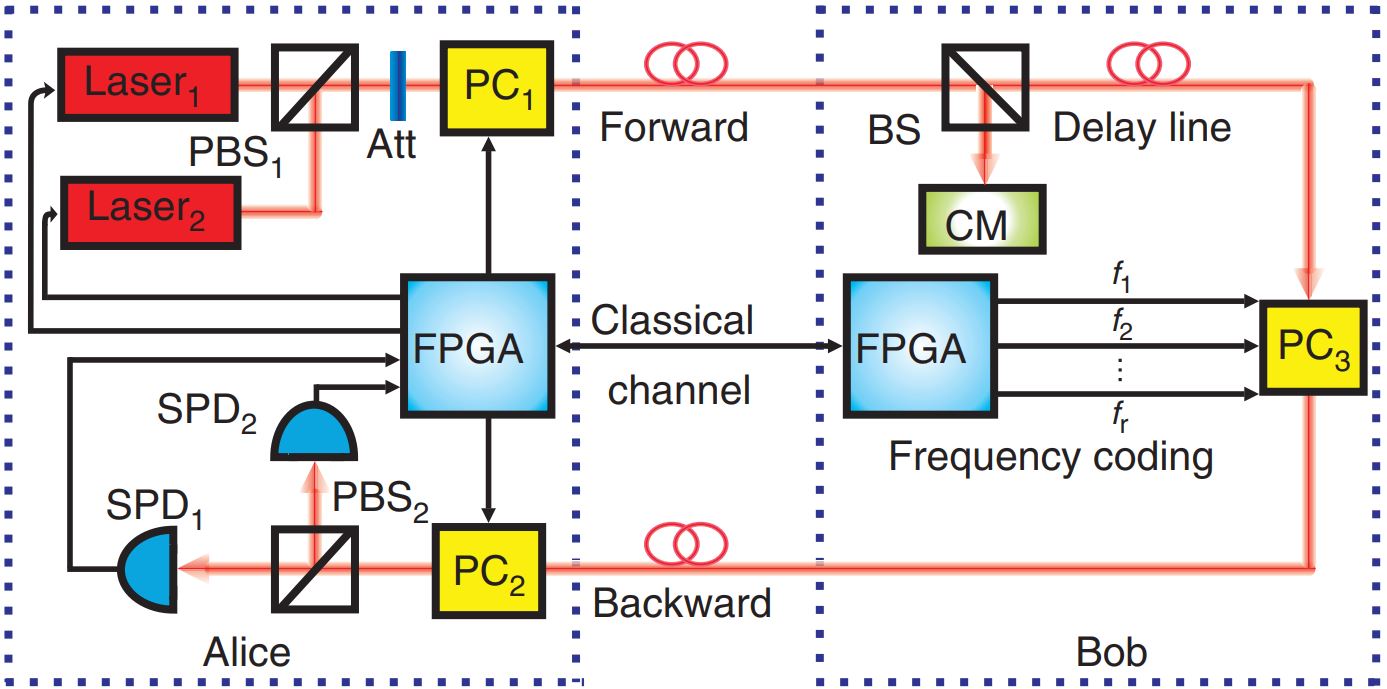}
\end{minipage}%
}\quad
\subfigure[QSDC system over 1.5 km fiber channel~\cite{qi2019implementation}. Reprinted with permission from Ref.~\cite{qi2019implementation}~\copyright~Springer Nature.]{
\begin{minipage}[t]{0.48\linewidth}
\centering
\includegraphics[width=3.44in,height=1.85in]{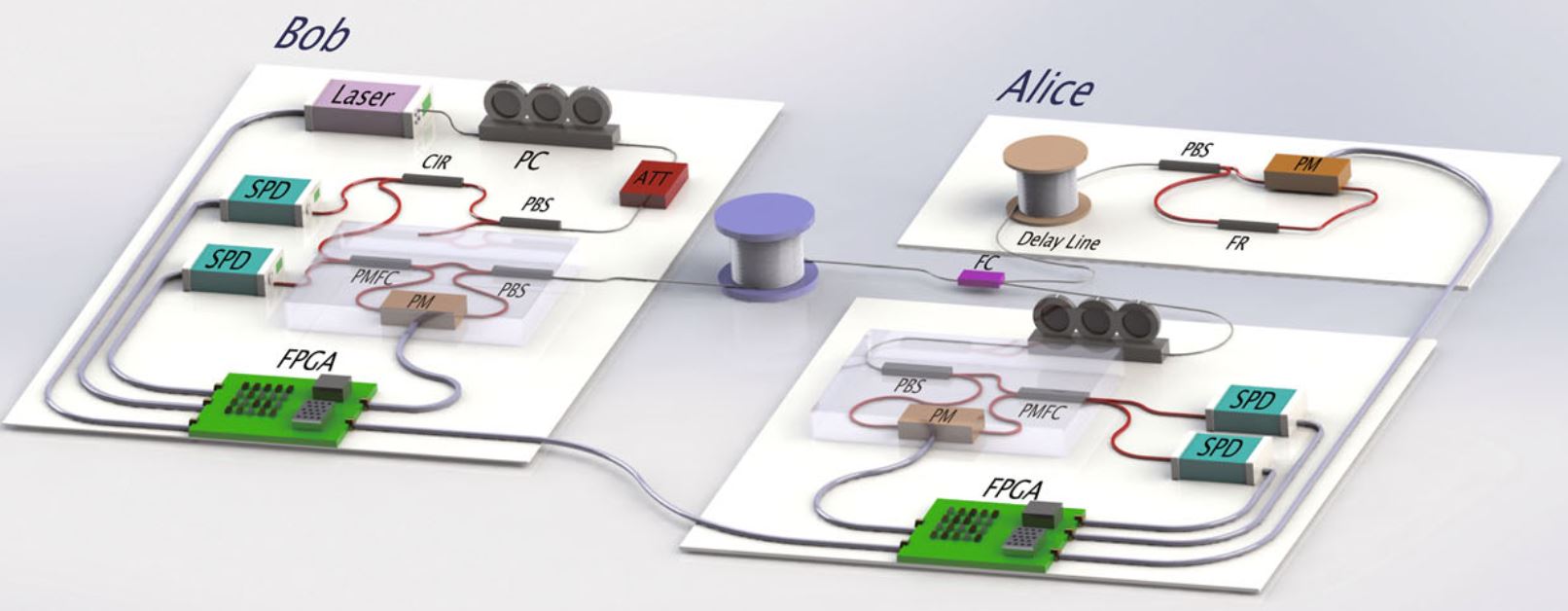}
\end{minipage}%
}%

\subfigure[Quantum-memory-free coding~\cite{sun2020toward}. Reprinted with permission from Ref.~\cite{sun2020toward}~\copyright~The IEEE.]{
\begin{minipage}[t]{0.47\linewidth}
\centering
\includegraphics[width=3.45in,height=1.88in]{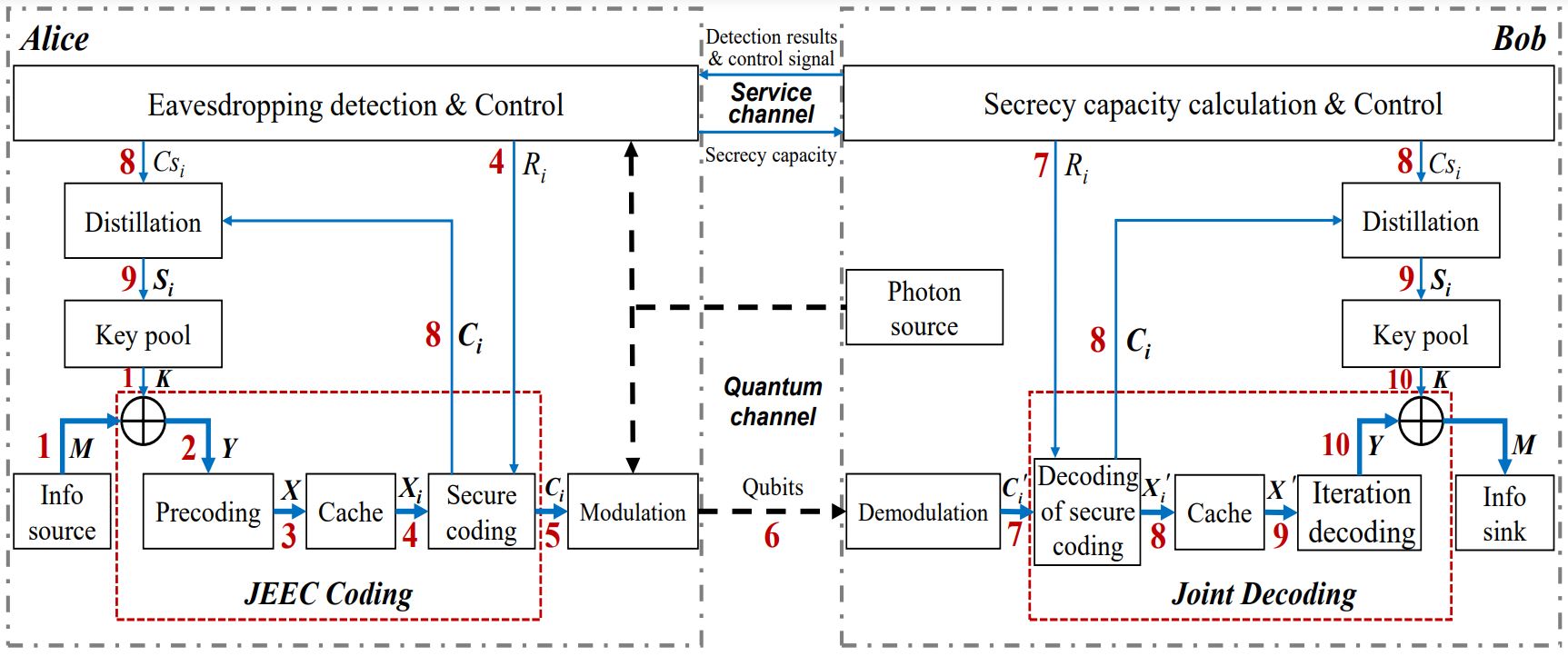}
\end{minipage}
}\quad%
\subfigure[Free-space QSDC system~\cite{pan2020experimental}. Reprinted with permission from Ref.~\cite{pan2020experimental}~\copyright~The Optical Society.]{
\begin{minipage}[t]{0.47\linewidth}
\centering
\includegraphics[width=3.45in,height=1.88in]{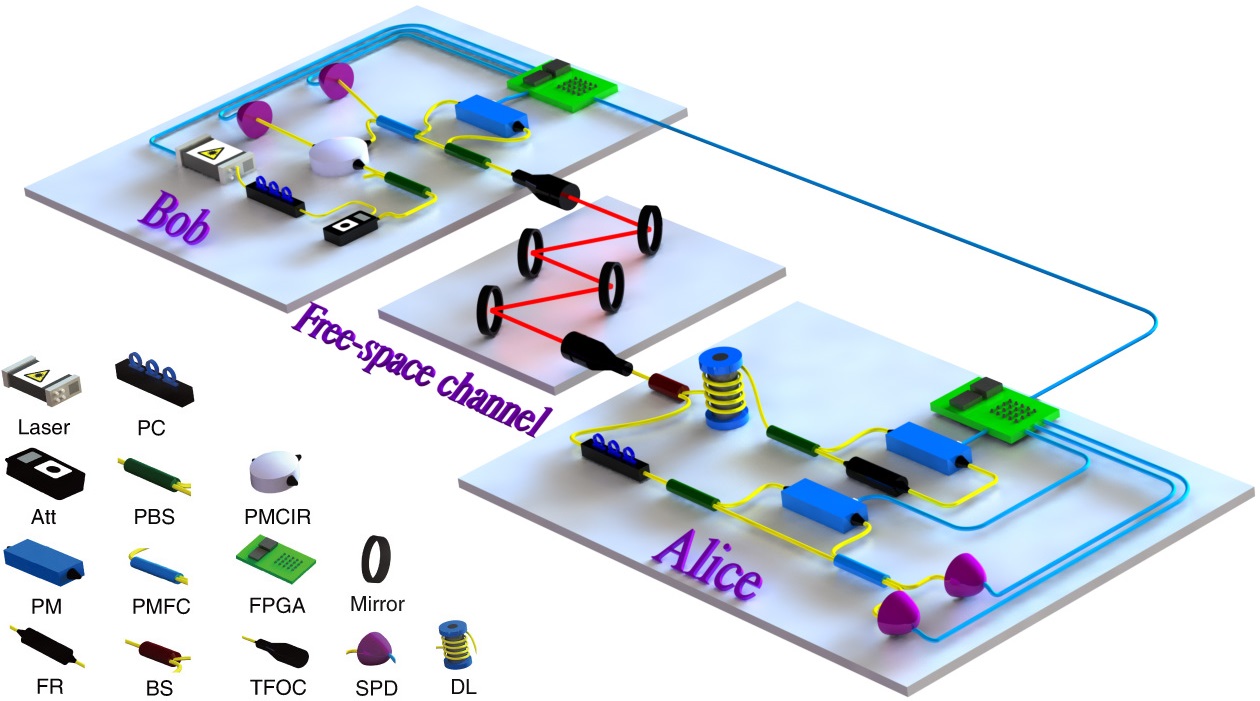}
\end{minipage}
}%

\subfigure[QSDC over 100 km fiber~\cite{zhang2022realization}. Reprinted with permission from Ref.~\cite{zhang2022realization}~\copyright~Springer Nature.]{
\begin{minipage}[t]{0.47\linewidth}
\centering
\includegraphics[width=3.45in,height=1.88in]{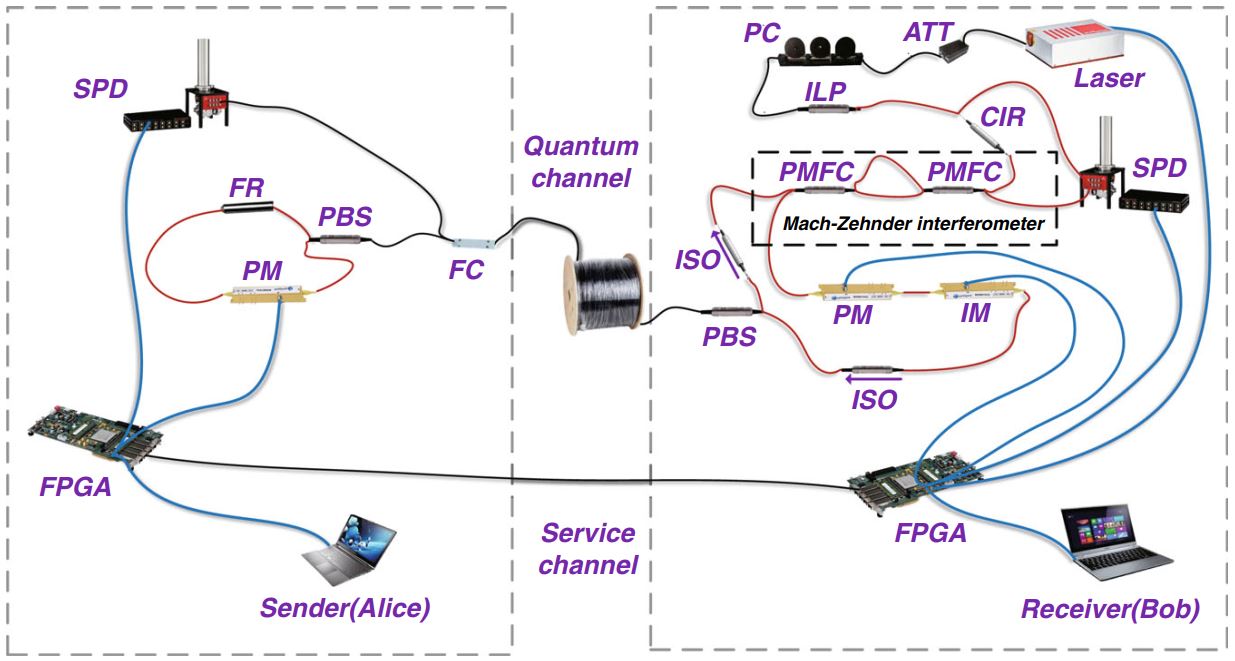}
\end{minipage}
}\quad
\subfigure[QSDC without active polarization compensation~\cite{liu2022fiber}. Reprinted with permission from Ref.~\cite{liu2022fiber}~\copyright~Springer Nature.]{
\begin{minipage}[t]{0.47\linewidth}
\centering
\includegraphics[width=3.45in,height=1.88in]{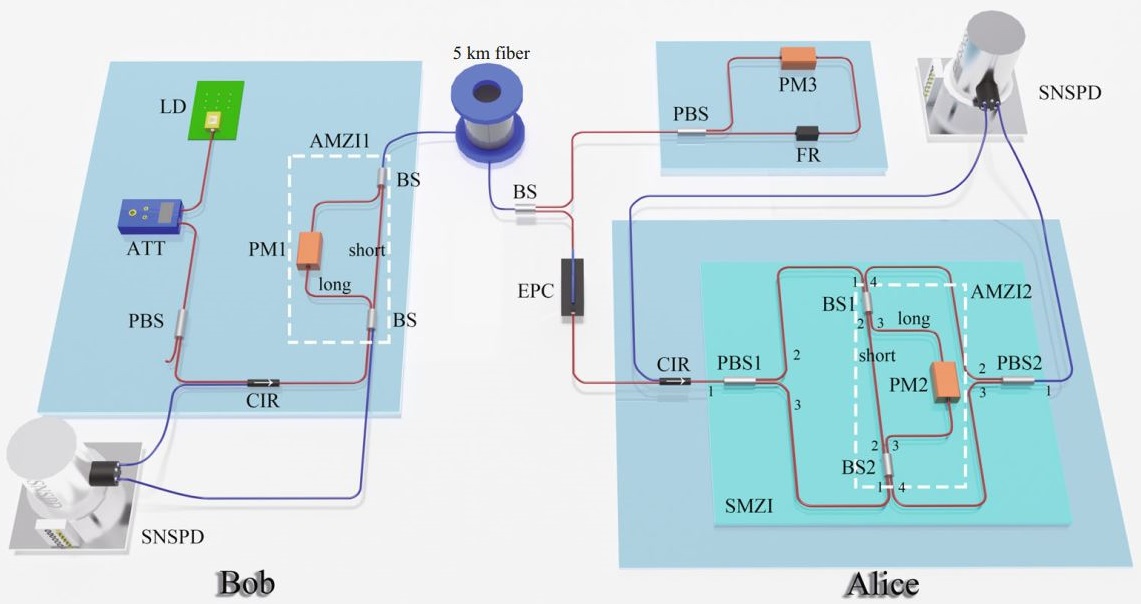}
\end{minipage}
}%
\centering
\caption{Experimental progress of DL04 QSDC protocol. In the figures:  Att, attenuator; AMZI, asymmetric Mach-Zehnder interferometer; BS, beam splitter; CIR, circulator; CM, control mode; DL, delay line; EPC, electronic polarization controller; FC, fiber coupler; FPGA, field programmable gate array; FR, Faraday rotator; ILP, in-line polarizer; IM, intensity modulator; ISO, isolator; LD, laser diode; PBS, polarization beam splitter; PC, polarization controller; PM, phase modulator; PMCIR, polarization-maintaining circulator; PMFC, polarization maintaining filter coupler; SMZI, Sagnac-Mach-Zehnder interferometers; SNSPD, superconducting nanowire single photon detector; SPD, single-photon detector; TFOC, triplet fiber-optic collimator. \label{fig:DL04Experiment}}
\end{figure*}

Fig.~\ref{fig:DL04Experiment} (a) shows the fiber-optic QSDC system designed by Hu \textit{et al}~\cite{hu2016experimental}. Their system had an operating frequency range spanning from 25 to 400 kHz and a channel spacing of 25 kHz. So there are 16 frequency bands and an information transmission rate of 4 kbps was achieved. Both the common intercept-resend attack of~\cite{gisin2002quantum} and photon-number-splitting attack of~\cite{huttner1995quantum} were considered in this experiment, when calculating the number of secure information bits versus the communication distance. 

Qi \textit{et al}.~\cite{qi2019implementation} described the practical QSDC of Fig. \ref{fig:DL04Experiment} (b), including both its optical and electronic parts. Motivated readers are referred to~\cite{qi2019implementation} for the detailed portrayal of this practical circuit. The optical link is set up elaborately by using asymmetric Mach-Zehnder interferometers, which manage the relative phase between two pulses \cite{marand1995quantum}. The upper optical components of Fig. \ref{fig:DL04Experiment} (b) at Alice' station are used for the message encoding and for the `postal pigeon' returning stage of the DL04 protocol~\cite{deng2004secure}, while eavesdropping detection relying on random sampling is carried out by the other one. 
The QSDC system of Fig.~\ref{fig:DL04Experiment} (b) was also concatenated with a low-density parity check coding scheme \cite{qi2019implementation} for enhancing its performance. A practical QSDC experiment conducted over 1.5 km fiber attained a secure communication rate of 50 bps at a QBER of 0.6\%, and both pictures as well as audio were successfully transmitted by this system. Nonetheless, the rate vs. distance performance of the system requires further improvement.

The full implementation of typical QSDC protocols \cite{long2002theoretically,deng2003two,deng2004secure,wang2005quantum,zhu2006secure,zhou2020measurement} requires block-based transmission, where a large number of quantum states have to be processed, which requires quantum memory. However, at the time of writing quantum memory has a rather limited coherence time. A compelling solution is to use an ingenious coding method to reduce the reliance on quantum memory \cite{sun2018design,sun2020toward}, which was hence termed as quantum-memory-free scheme. 
 
In quantum-memory-free QSDC~\cite{sun2018design,sun2020toward}, the forward error correction codeword is divided into several data frames for transmission, as seen in Fig.~\ref{fig:DL04Experiment} (c). Alice extracts a secure sequence from the previously sent data frame and encrypts the current secret message by using this sequence to obtain the ciphertext. Then, she utilizes the channel's secrecy capacity estimated during the previously sent data frames as the upper limit of the encoding rate representing the maximum normalized throughput for transmitting the current data frame, encoding the ciphertext into a forward-error-correction codeword. This system achieved an information transmission rate of 100 bps over an 18.5 km optical fiber channel. 

Pan \textit{et al}.~\cite{pan2020experimental} experimentally demonstrated single-photon based QSDC in a free-space channel for transmission over a distance of 10 m at an information transmission rate of 500 bps in a laboratory tabletop experiment. They have opted for a signal wavelength of 1550 nm, leveraging the advantages of the compatibility with atmospheric transmission windows, enhanced resilience against sunlight, and compatibility with fiber optic communication devices. At the time of writing, the majority of free-space quantum communication implementations rely on polarization encoding. However, polarization encoding is unsuitable for fiber-based optical networks due to the birefringence effect\footnote{Birefringence refers to the phenomenon of double refraction that occurs when light propagates through an anisotropic medium. Birefringence is induced by variations in manufacturing impurities or environmental conditions within the optical fiber, leading to the rotation of the polarization state of light as it propagates through the fiber.}, which adversely affects the transmission of polarized quantum states. By contrast, phase encoding and time-bin based encoding are well-suited for fiber networks. Hence, investigating the feasibility of phase encoding and time-bin based encoding in free-space QSDC~\cite{pan2020experimental,long2022evolutionary,pan2023free} or quantum cryptography~\cite{jin2019genuine,chen2020field,li2023free} will be beneficial for streamlining direct fiber/free-space interfaces. The performance improvement of free-space QSDC ultimately leads to satellite-to-ground communication~\cite{wang2021transmission,mendes2023optical} and the establishment of a global quantum network.

In 2022, the communication distance of QSDC was extended to 100 km~\cite{zhang2022realization}. Hence QSDC is indeed capable of supporting an entire metropolitan area, providing secure communication. This implementation employed a modified DL04 QSDC protocol~\cite{zhang2022realization}, in which the time-bin states were used for eavesdropping detection and the phase states were used for information transmission. As a result, the QBER during eavesdropping detection was low, thereby enhancing the secrecy capacity of QSDC. Additionally, the round-trip transmission of phase states was observed to exhibit self-compensating characteristics. A powerful forward error correction code was designed for mitigating the bit loss issues inherent in QSDC. Indeed, harnessing high-performance classical error correction codes at extremely low signal reception rates emerged as a pivotal research focus~\cite{sun2020toward,sun2021design}. As a further advance, Liu \textit{et al}.~\cite{liu2022fiber} constructed a robust optical fiber based QSDC system successfully operating without active polarization compensation, which also reduced the QBER during eavesdropping detection.
 
\begin{figure*}[htbp]
\centering
\subfigure[QSDC with the quantum memory~\cite{zhang2017quantum}. Reprinted with permission from Ref.~\cite{zhang2017quantum}~\copyright~The American Physical Society.]{\includegraphics[width=0.48\linewidth]{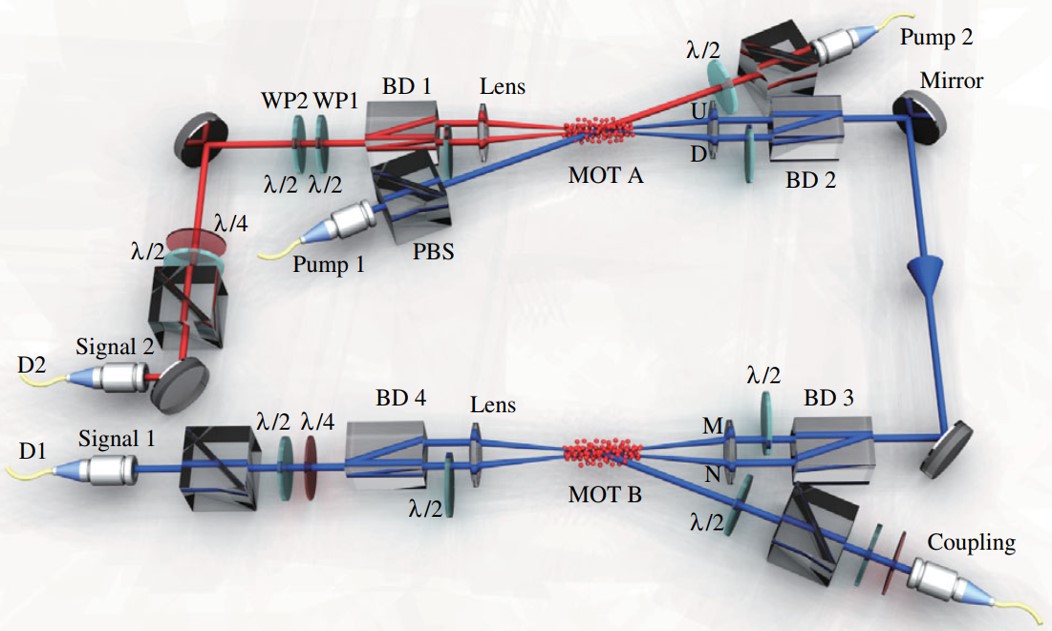}}
\quad
\subfigure[Entanglement-based long-distance QSDC~\cite{zhu2017experimental}. Reprinted with permission from Ref.~\cite{zhu2017experimental}~\copyright~Elsevier.]{\includegraphics[width=0.48\linewidth]{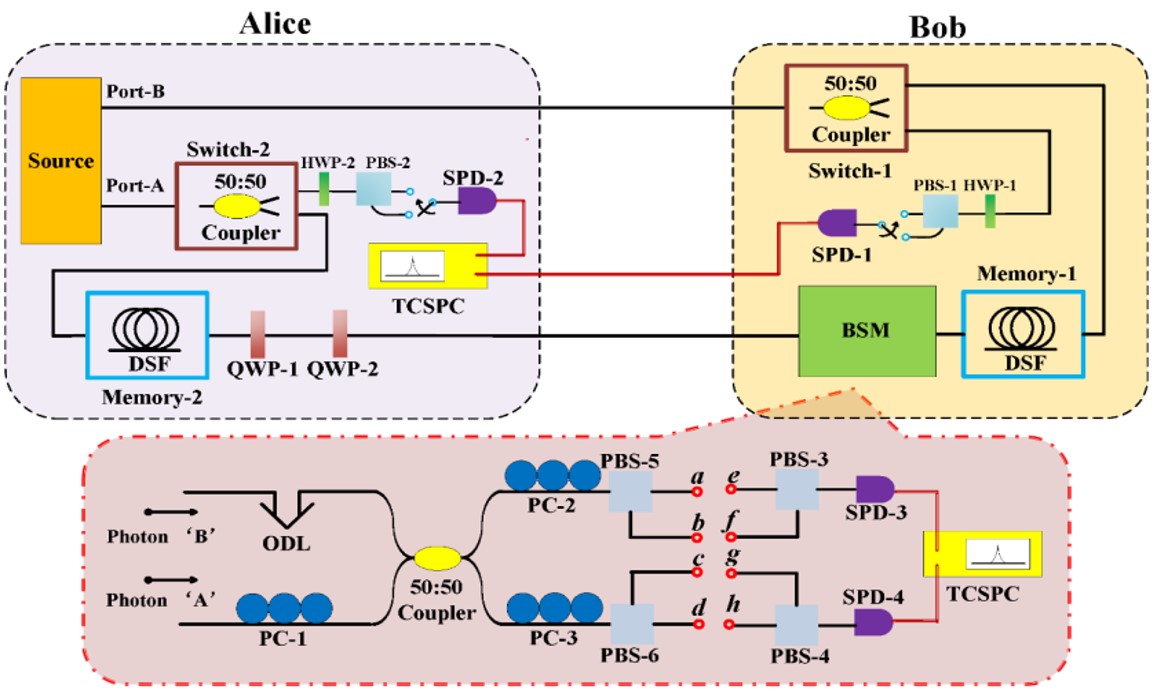}}

\subfigure[A 15-user QSDC network~\cite{qi202115}. Reprinted with permission from Ref.~\cite{qi202115}~\copyright~Springer Nature.]{\includegraphics[width=0.48\linewidth]{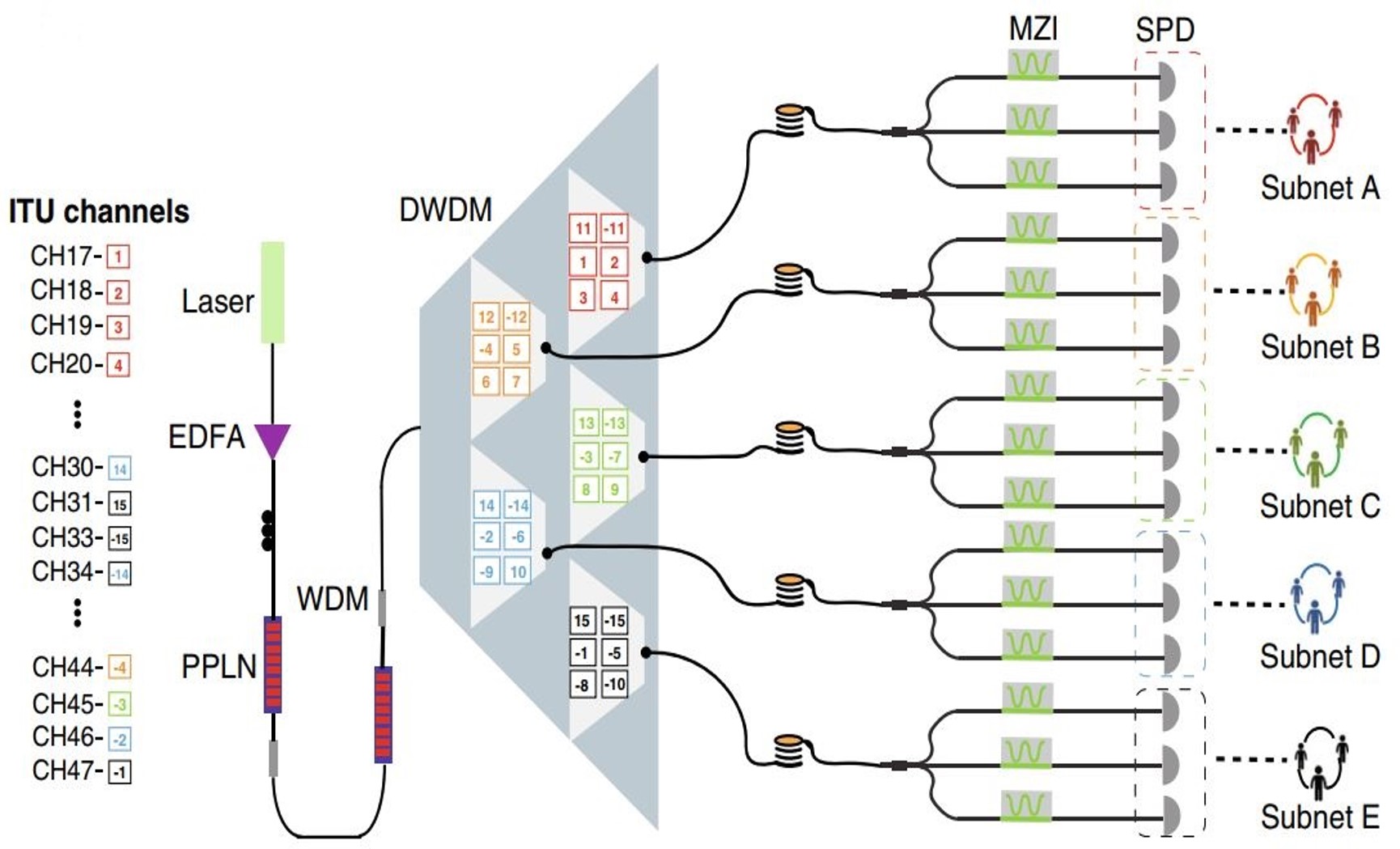}}
\centering
\caption{Demonstration of Two-step QSDC protocol. In the figures: BD1, BD2, BD3 and BD4, beam displacer; D1 and D2, single photon detector; DWDM, dense wavelength division multiplexing; DSF, dispersion shifted fiber; EDFA, erbium-doped fiber amplifier; $\lambda/2$, half-wave plate; HWP, half wave plate; ITU, International Telecommunication Union; MOT A and MOT B, magneto-optical trap; MZI, Mach–Zehnder interferometer; ODL, optical delay line; PBS, polarization beam splitter; PC, polarization controller; PPLN, periodically  poled  lithium  niobate; $\lambda/4$, quarter-wave plate; SPD, single-photon  detector; TCSPC, time-correlated single photon counting; U, D, M, and N, path; WDM, wavelength division multiplexing; WP1 and WP2, half-wave plate.\label{fig:EntExperiment}}
\end{figure*}

Experimentalists often store the photons by using a delay line, such as a fiber loop \cite{pittman2002single,ostermeyer2008implementation} and optical fiber delay line \cite{chen2016experimental,hu2016experimental,zhu2017experimental,qi2019implementation} as a simple design alternative. As a further advance, Zhang \textit{et al}. \cite{zhang2017quantum} opened the door for storing the entangled photons in QSDC using state-of-the-art atomic quantum memory. The detailed components of the QSDC system relying on quantum memory are shown in Fig. \ref{fig:EntExperiment} (a). The hybrid atom-photon entangled state is generated in MOT A, while Alice delivered a partner particle of entangled state to MOT B for storage, in order to wait for the other encoded particle. Then the information is decoded with the aid of density matrix reconstruction. A bit rate of 2.5 bps was achieved at the error rate of 10\% in this quantum-memory-assisted QSDC system.

Additionally, another practical challenge in the implementation of entanglement-based QSDC is to perform high-fidelity Bell-state discrimination. Zhang \textit{et al}. \cite{zhang2017quantum} applied so-called quantum state tomography~\cite{nielsen2002quantum} to discriminate the polarization of entangled states instead of applying the BSM mentioned above. Zhu \textit{et al}. \cite{zhu2017experimental} replaced the quantum state tomography by the BSM using only linear optical elements. In their experiment the Bell states $|\psi^{+}\rangle$ and $|\psi^{-}\rangle$ were discriminated successfully with a fidelity of 88\% and 91\%, respectively. Fig. \ref{fig:EntExperiment} (b) shows their experimental setup, where the communicating parties are linked by 0.5 km of optical fiber. The entangled photons were generated by spontaneous four-wave mixing \cite{zhu2016fiber}, and this fiber-based source is conveniently compatible with the transmission medium. Recently, Qi \textit{et al}.~\cite{qi202115} experimentally demonstrated a 15-user QSDC network based on entanglement distribution, achieving an information transmission rate of 1 kbps between any pair of users separated by a distance as high as 40 km.

Let us now focus our attention on QSDC experiments based on novel theoretical protocols. For example, Lum \textit{et al.}~\cite{lum2016quantum} demonstrated the feasibility of QSDC based on quantum data locking\footnote{Quantum data locking is a quantum communication primitive that enables encryption of a long message using a shorter secret key, ensuring information-theoretical security against adversaries possessing constrained quantum memory.} in 2016, achieving a successful data packet reception rate of 99.5\%. In 2019, Massa \textit{et al.}~\cite{massa2019experimental} experimentally characterized a two-way communication protocol~\cite{Del2018two}, where Alice transmitted a 100-bit code of a black-and-white image to Bob through a quantum channel, while Bob simultaneously transmitted 100 random bits to Alice.

\begin{figure*}[htbp]
\centering
\subfigure[Secure communication via QI \cite{shapiro2014secure}. Reprinted with permission from Ref.~\cite{shapiro2014secure}~\copyright~ Springer Nature.]{
\begin{minipage}[t]{0.48\linewidth}
\centering
\includegraphics[width=3.45in,height=1.7in]{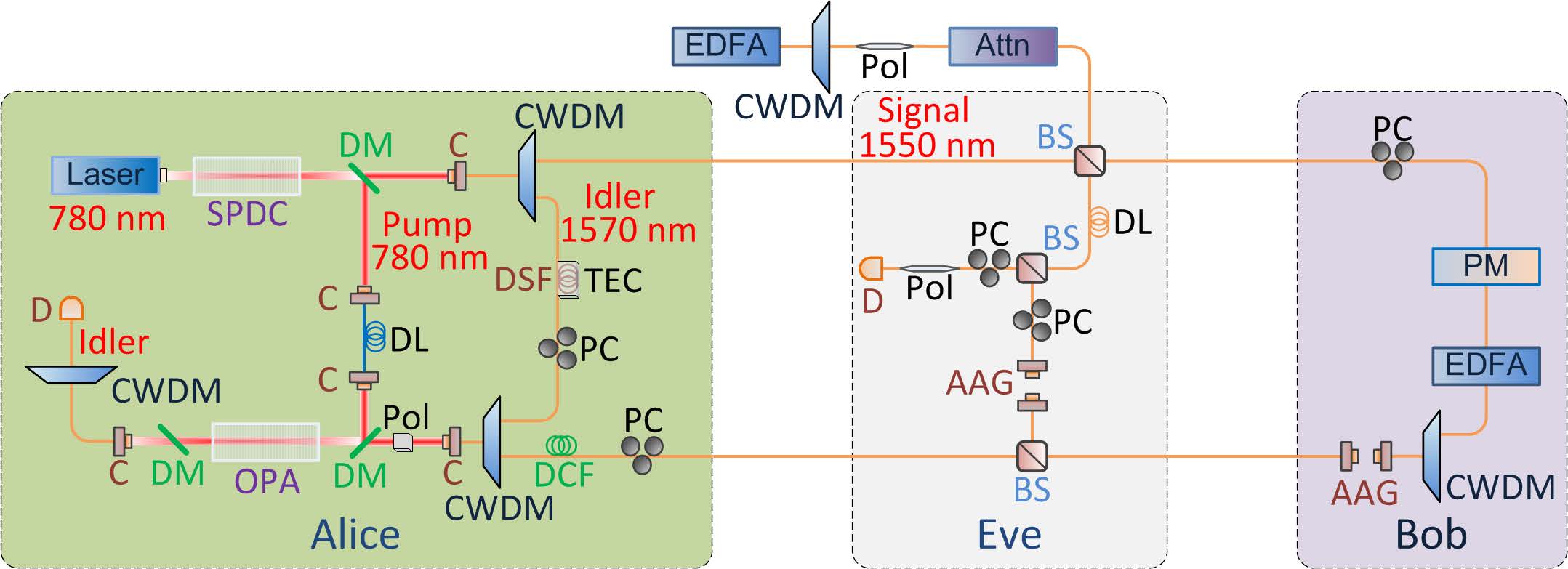}
\end{minipage}%
}\quad%
\subfigure[QSDC relying on QLPI~\cite{shapiro2019quantum}. Reprinted with permission from Ref.~\cite{shapiro2019quantum}~\copyright~The Optical Society.]{
\begin{minipage}[t]{0.48\linewidth}
\centering
\includegraphics[width=3.45in,height=1.88in]{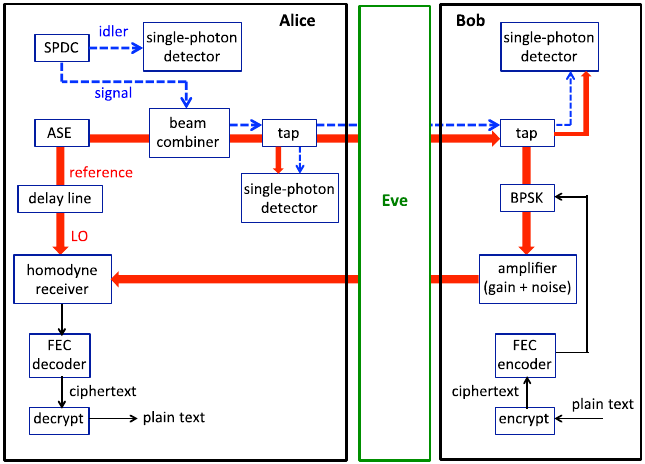}
\end{minipage}%
}%

\subfigure[Gaussian mapping based QSDC~\cite{cao2023realization}. Reprinted with permission from Ref.~\cite{cao2023realization}~\copyright~The AAAS.]{
\begin{minipage}[t]{0.47\linewidth}
\centering
\includegraphics[width=3.45in,height=1.88in]{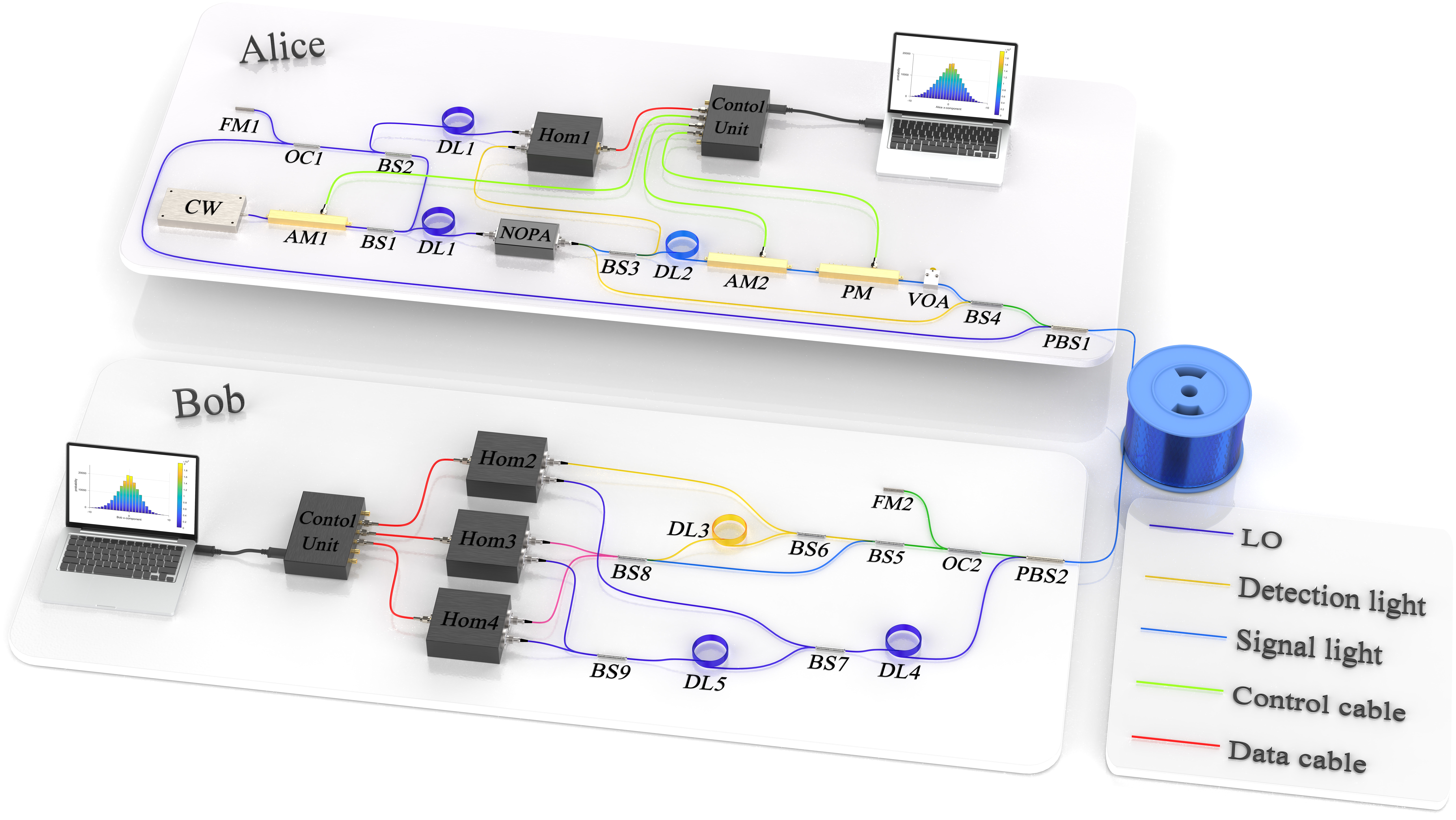}
\end{minipage}
}\quad%
\subfigure[QSDC relying on squeezed state~\cite{paparelle2023practical}. \copyright~Paparelle \textit{et al}. arXiv, 2023.]{
\begin{minipage}[t]{0.47\linewidth}
\centering
\includegraphics[width=3.45in,height=1.88in]{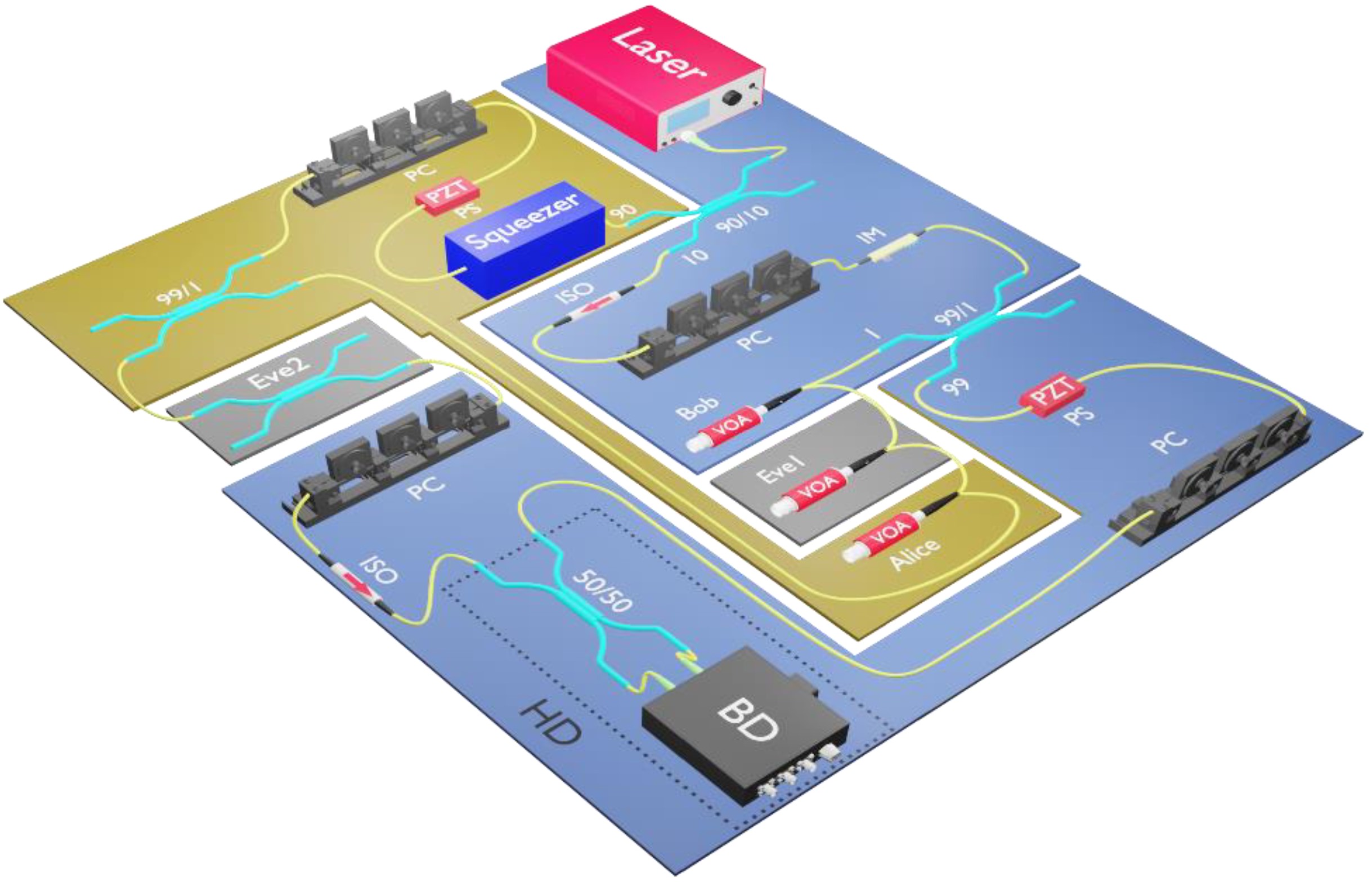}
\end{minipage}
}%
\centering
\caption{Experimental progress of CV QSDC protocol. In the figures: BD, balanced detector; 99/1, 90/10, 50/50, BS, beam splitter; AAG, adjustable air gap; AM, amplitude modulator; ASE, amplified spontaneous emission; Attn, attenuator; BPSK, binary phase shift keying; C, collimator; CW, 1550-nm continuous-wave light source; CWDM, coarse wavelength-division multiplexer; D, detector; DCF, dispersion-compensating fiber; DL, delay line; DM, dichroic mirror; DSF, dispersion-shifted fiber; EDFA, erbium-doped fiber amplifier; FEC, forward error correction; FM, faraday mirror; HD, Hom, homodyne detector; IM, intensity modulator; ISO, isolator; LO, local oscillator; NOPA, nondegenerate optical parametric amplifier; OC, optical circulator; OPA, optical parametric amplifier; PC, polarization controller; PBS, polarization beam splitter; PM, phase modulator; Pol, polarizer; PS, phase shifter; PZT, piezoelectric; SPDC, spontaneous parametric downconverter;  TEC, thermoelectric cooler; VOA, variable optical attenuator. \label{fig:CVQSDCExperiment}}
\end{figure*}

The aforementioned experimental progress falls within the domain of discrete variable QSDC. Concurrently, continuous variable QSDC is also transitioning from theory to experiments. This includes quantum-secured direct communication relying on squeezed states~\cite{paparelle2023practical} and continuous variable entanglement~\cite{shapiro2014secure,cao2023realization,nirala2023information}. The underlying theoretical protocols~\cite{zhuang2015ultrabroadband,cao2021continuous,srikara2020continuous} employed for these experiments exhibit two-way transmission of quantum states, which is similar to the DL04 protocol~\cite{deng2004secure} and to the two-step protocol~\cite{deng2003two}. Table~\ref{tab:CVQSDCExperiments} provides a summary of some representative CV QSDC experiments. Their general architecture is shown in Fig.~\ref{fig:CVQSDCExperiment}. These experiments are still at their `desktop stage', but they demonstrate tangible advantages at high data transmission rates in the context of short to medium distances. Additionally, the two-way protocol of~\cite{marshall2016continuous} exhibits substantial potential in cloud-based computing networks of the near future.

\begin{table*}[!h]
\begin{footnotesize}
\begin{center}
\caption{Summary of representative CV QSDC experiments.}
\begin{tabular}{|l|l|l|l|l|}
\hline
Group \& year & Shapiro \textit{et al}.~\cite{shapiro2014secure}, 2014 & Shapiro \textit{et al}.~\cite{shapiro2019quantum}, 2019 & Cao \textit{et al}.~\cite{cao2023realization}, 2023 & Paparelle \textit{et al}.~\cite{paparelle2023practical}, 2023 \\ \hline
protocol & QI & QLPI & Gaussian mapping& CV DL04 \\ \hline
Information carrier & \multicolumn{3}{c|}{CV entanglement} & Squeezed state \\ \hline
Encoding& Binary phase shift keying & Binary phase shift keying & Gaussian mapping & Discrete modulation \\ \hline
Wavelength & 1570 nm \& 1550nm & 1570 nm \& 1550nm & 1550 nm & 1560 nm \\ \hline
Channel & Fiber & Fiber & Fiber & Fiber \\ \hline
Repetition rate & N.A. & N.A. & 1 MHz & N.A. \\ \hline
Distance & N.A. & N.A. & 5 km & N.A. \\ \hline
Rate & 500 kbps & N.A. & 4.08$\times10^5$ bps & N.A. \\ \hline
\end{tabular}
\label{tab:CVQSDCExperiments}
\end{center}
\end{footnotesize}
\end{table*}

At the time of writing many of the grave QSDC impediments have indeed been overcome, despite relying on off-the-shelf optoelectronic devices. An information rate of a few dozens of kbps has been achieved over several tens of kilometers. These impressive achievements are expected to be followed by more practical high-performance implementations in the near future.

\section{Open Challenges and Future Research}
\label{sec:challenges and future works}
\begin{figure}[htpb]
\begin{center}
\includegraphics[width=\columnwidth,angle=0]{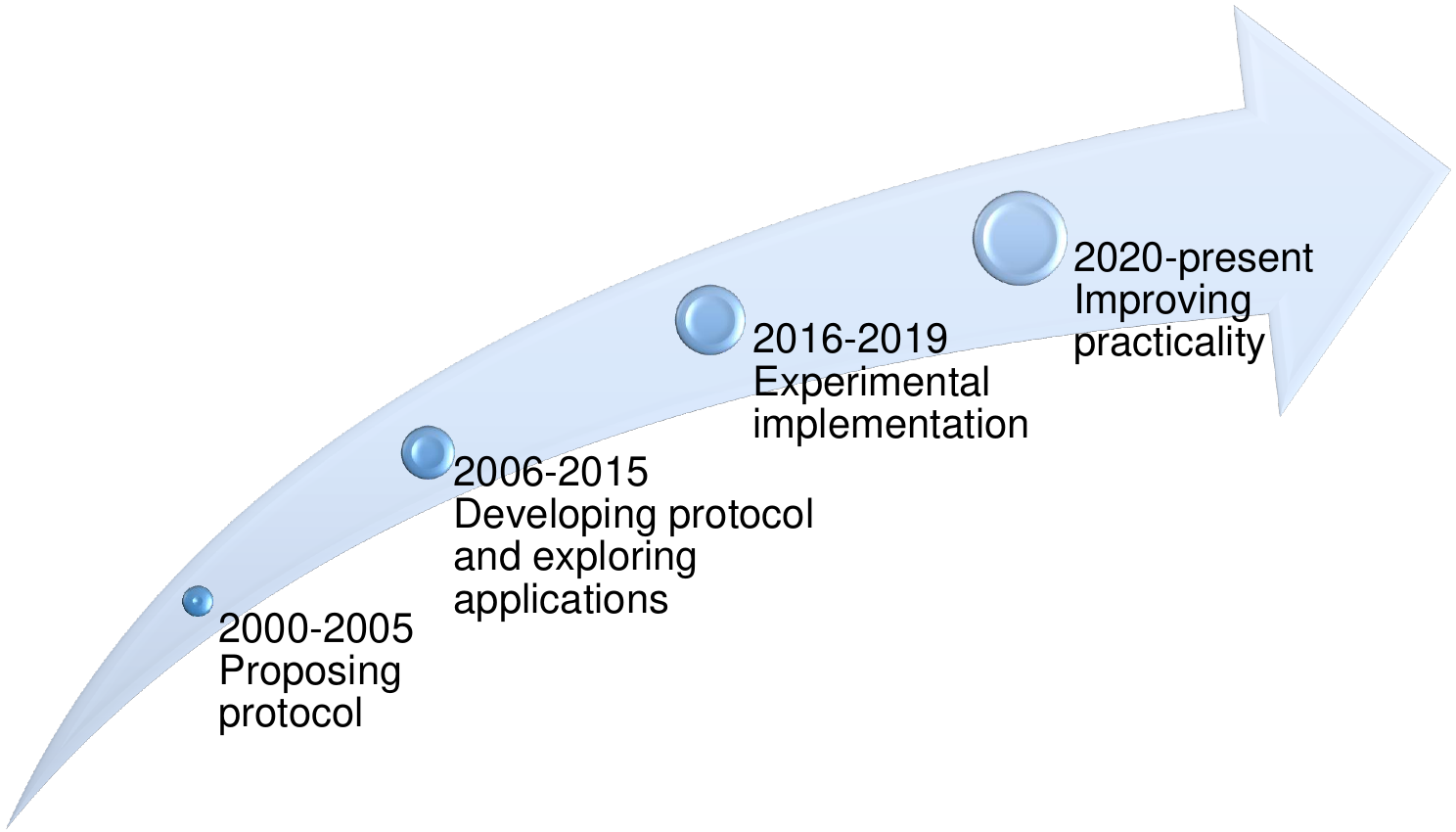}
\caption{Development and outlook of QSDC.}
\label{fig:Outlook}
\end{center}
\end{figure}

Looking back over the twenty year history of QSDC, several stages of development appear prominently before our eyes, as seen in Fig. \ref{fig:Outlook}. From these the following lessons can be derived.

Starting from 2000 when the QSDC philosophy was conceived, researchers have endeavored to construct point-to-point communication protocols. Then the early QSDC protocols were further developed for solving diverse quantum cryptographic tasks, including quantum dialogues, quantum signatures, and quantum steganography, which opened the way for the extensive application of QSDC. The security of QSDC is provable in theory: the QSDC protocols are information-theoretically secure. Recent engineering efforts completed the proof-of-principle experiment and took a step toward practical field-operation. Nonetheless, numerous open problems and challenges exist, hence substantial further research is needed.

\subsection{Designing QSDC Protocols}
\label{sec:Designing QSDC protocol}
As it transpires from the literature we have reviewed, QSDC protocols have mainly been studied in the discrete variable domain. But, the continuous-variable schemes are potentially more compatible with current telecom equipment, and have the advantage of cost-effective detectors, as well as higher rates than discrete-variable protocols. Hence there have been some initial QSDC studies in the continuous-variable domain \cite{he2006quantum,marino2006deterministic,pirandola2008quantum,pirandola2009confidential}. However, all of them belong to the family of DSQC or quasi-QSDC~\cite{pan2023free}, because additional communication is a prerequisite for the final information decoding, which reduces the efficiency of communication. Thus, one of the open problems is that of designing the family of continuous-variable QSDC protocols, whose receiver can detect information without the assistance of additional communication. Traditional QSDC protocols require round-trip transmission of quantum states. Designing one-way QSDC protocols would be beneficial both in terms of reducing channel losses and system complexity. The rate vs. distance performance has to be improved by harnessing novel physical resources~\cite{lucamarini2018overcoming}, for breaking the theoretical limits~\cite{pirandola2017fundamental}. In the transmission of confidential information over quantum channels with noise, loss, and eavesdropping, excellent forward error correction codes are required~\cite{pan2023free}, such as polar codes~\cite{zhou2022appending}, low density parity check codes~\cite{qi2019implementation}, and so on. A particularly promising technique is to design error correction codes for QSDC schemes.

\subsection{Security Proof of QSDC}
\label{sec:Security proof of QSDC}
Historically speaking, the security proof of BB84 QKD has been carried out from a range of different perspectives \cite{shor2000simple,koashi2006unconditional,renner2008security,koashi2009simple}. The security proof QSDC still requires further research. One of the challenges is that the two-way transmission is vulnerable to attacks by quantum hackers. The security analysis of imperfect devices is also urgently needed. The secrecy capacity considering finite-length effects and practical system parameters has to be determined.

\subsection{Experimental Implementations of QSDC}
\label{sec:QSDC experimental implementation}
Optimizing the parameters of devices and improving the performance of  practical QSDC systems represents ongoing challenges on the experimental side. Intuitively, the communication distance of single-photon QSDC may approach a half of QKD's distance at the same information rate, bearing in mind that QSDC requires two-way block based transmission. Combining quantum memory with QSDC is conducive to the further improvement of the communication distance and in support of QSDC with block based transmission \cite{zhang2017quantum}. Furthermore, the investigation of free-space optical QSDC would contribute to the future implementation of satellite-based QSDC networks. Chip based QSDC will result in a more stable, compact, and portable practical system~\cite{luo2023recent}.

\subsection{Hybrid QSDC-Classical Network}
\label{sec:Hybrid QSDC-classical network}
In closing we note that most of the cryptographic tasks considered operated in the quantum domain. However, it is also feasible to integrate QSDC into high-security classical communication network~\cite{wang2022quantum}. Specially, the secure repeater network is compatible with the existing Internet~\cite{long2022evolutionary}, but additional efforts are needed to design its architecture and establish the interface with the classical network. The classical-cryptography assured imperfect device QSDC proposed in~\cite{pan2023free} combines classical cryptographic techniques with QSDC using existing technology, enabling confidential communication at an improved security level.

\section{Conclusions and Lessons Learned}
\label{sec:Con}
Following our tour of quantum signal processing with a view to inspire a community-wide effort in the interest of filling the open challenges detailed in the previous section, we conclude by listing a few crisp lessens learned:
\begin{itemize}
\item The main benefit of communicating in the quantum-domain is that eavesdropping may be detected, which is not the case in the classical domain. Hence if a QKD process is perturbed by an eavesdropper, any further proceedings are curtailed and the key-negotiation is recommenced. Once the key is determined, it may be used exactly in the same way as in classic encryption.
  
\item In contrast to the family of QKD solutions, which constitute a family of pure secret key-negotiation protocols, QSDC supports secure communication without requiring a cryptographic key for encryption and decryption.

\item Dispensing with a secret key is possible, because in QSDC the confidential messages are directly embedded into the quantum system and transmitted between the communicating parties via a quantum channel.

\item The QSDC protocols are also eminently suitable for diverse cryptographic tasks, and a large number of cryptographic protocols beyond QSDC have been constructed.

\item The recent experimental progress was reviewed in Section~\ref{sec:Experimental progress} highlighting the associated technological challenges. It is clear that the performance of QSDC in terms of its information rate and communication distance is still very limited at the time of writing.

\item In conclusion, there are significant untapped opportunities and numerous open problems for a strongly interdisciplinary research community to solve, including numerous open problems in quantum information and communication theory, in quantum physics, in numerous aspects of quantum engineering. These include, but are no means limited to quantum error mitigation, quantum coding, quantum channel modelling and transmission techniques.

  \item A compelling direction is to further explore the potential of wireless QSDC both in the context of free-space optical satellite communication and terrestrial scenarios relying on both visible-light communications and potentially even on THz wireless communications. 

\item {\bf Join this exhilarating research momentum valued Colleague!}

\end{itemize}

\bibliographystyle{IEEEabrv}
\normalem

\end{document}